\documentclass[aps,twocolumn,superscriptaddress,preprintnumbers,pre]{revtex4}
\usepackage{xr-hyper}
\usepackage{mathrsfs, hyperref}
\usepackage{amssymb, amsbsy, amsmath, latexsym, dsfont, array, layout, graphics,mathrsfs,braket,amsfonts,amsthm,
  amssymb,graphicx,subfigure, youngtab,color,bm,mathtools,braket,verbatim,url,cleveref,natbib,hypernat,extarrows,inputenc,multirow,bbm,dsfont}
  
\usepackage[usernames,dvipsnames]{xcolor}
\colorlet{RED}{red}
\usepackage{graphicx,psfrag}
\usepackage{amsfonts}
\usepackage[figuresright]{rotating}  
\usepackage{amssymb}
\usepackage{amsmath}
\usepackage{subfigure}
\usepackage{multirow}
\usepackage{tabularx}
\usepackage[resetlabels]{multibib}
\newcites{supp}{Supplemental References}
\usepackage{amssymb, amsbsy, amsmath, latexsym, dsfont, array, layout, graphics,mathrsfs,braket,amsfonts,amsthm,
  amssymb,graphicx,subfigure,dcolumn, youngtab,color,bm,mathtools,braket,verbatim,url,cleveref,natbib,hypernat,extarrows,inputenc,multirow}
\usepackage[usernames,dvipsnames]{xcolor}
\graphicspath{ {./images/} }

\newcommand{\xw}[1]{{\color{black} #1}}
\usepackage{mathrsfs, hyperref}
\usepackage{amssymb, amsbsy, amsmath, latexsym, dsfont, array, layout, graphics,mathrsfs,braket,amsfonts,amsthm,
  amssymb,graphicx,subfigure, youngtab,color,bm,mathtools,braket,verbatim,url,cleveref,natbib,hypernat,extarrows,inputenc,multirow,bbm}
\usepackage{graphicx}
\usepackage{dcolumn}
\usepackage{bm}
\usepackage[usernames,dvipsnames]{xcolor}
\usepackage[normalem]{ulem}
\usepackage[T1]{fontenc}
\graphicspath{ {./images/} }
\allowdisplaybreaks 

\newcommand{\splitatcommas}[1]{%
\begingroup 
\begingroup\lccode`~=`, \lowercase{\endgroup 
\edef~{\mathchar\the\mathcode`, \penalty0 \noexpand\hspace{0pt plus 1em}}%
}\mathcode`,="8000 #1%
\endgroup 
} 

\begin{document}
\title{Symmetry-based classification of exact flat bands in single and bilayer moir\'e systems}
\author{Siddhartha Sarkar}\thanks{These two authors contributed equally}
\affiliation{%
Department of Physics, University of Michigan, Ann Arbor, MI 48109, USA
}
\author{Xiaohan Wan}\thanks{These two authors contributed equally}
\affiliation{%
Department of Physics, University of Michigan, Ann Arbor, MI 48109, USA
}
\author{Shi-Zeng Lin}
\email{szl@lanl.gov}
\affiliation{%
Theoretical Division, T-4 and CNLS, Los Alamos National Laboratory, Los Alamos, New Mexico 87545, USA
}
\affiliation{%
Center for Integrated Nanotechnologies (CINT), Los Alamos National Laboratory, Los Alamos, New Mexico 87545, USA
}
\author{Kai Sun}
\email{sunkai@umich.edu}
\affiliation{%
Department of Physics, University of Michigan, Ann Arbor, MI 48109, USA
}
\begin{abstract}
We study the influence of spatial symmetries on the appearance and the number of exact flat bands (FBs) in single and bilayer systems with Dirac or quadratic band crossing points, and 
systematically classify all possible number of exact flat bands in systems with different point group symmetries.
We find that a maximum of 6 FBs can be protected by symmetries, and show an example of 6 FBs in a system with QBCP under periodic strain field of $\mathcal{C}_{6v}$ point group symmetry. All known examples of exact FBs in single and bilayer systems fall under this classification, including chiral twisted bilayer graphene, and new examples of exact FBs are found. We show the construction of wavefunctions for the highly degenerate FBs, and prove that any such set of FBs are $\mathds{Z}_2$ nontrivial, where all WFs polarized on one sublattice together have Chern number $C = 1$ and WFs polarized on the other sublattice together have $C = -1$. These bands also satisfy ideal non-Abelian quantum geometry condition.
We further show that, just like in TBG, topological heavy fermion description of the FBs with higher degeneracy is possible 
as long as the Berry curvature distribution is peaked around a point in the BZ.

\end{abstract}
\maketitle
\textit{Introduction.}--The interplay between topology and strong interaction makes topological flat bands(FBs) ideal for nontrivial states such as fractional Chern insulators, high $T_c$ superconductors etc. One well known example is twisted bilayer graphene (TBG), where the interference between Dirac points (DPs) in the two layers give rise to FBs, at some ``magic'' angles, whose nontrivial topology is a result of nontrivial Berry phase winding of the constituent DPs of the two layers. Interestingly, Tarnopolosky et al.~\cite{tarnopolsky2019origin} showed that, in the chiral limit of TBG, where the AA tunneling is set to zero, the middle two bands at the Fermi level become exactly flat at certain magic angles. They also showed that the exact wave-functions(WFs) of these two bands are complex analytic functions, analogous to WFs of the lowest Landau level (LLL) on a torus~\cite{haldane1985periodic}. Just like LLL WFs, the analytic WFs of chiral TBG also carry Chern number $C=\pm 1$. Since then, exact FBs were shown to appear in continuum models of a number of different systems at chiral limit such as, twisted bilayer checkerboard lattice (TBCL)~\cite{li2022magic}, TBG with spatially alternating magnetic field~\cite{becker2022fine,le2022double}, single layer system with quadratic band crossing point (QBCP) under periodic strain field~\cite{wan2023topological}, twisted bilayer Fe-based superlattices~\cite{eugenio2022twisted}, and TBG with strong second harmonic tunneling~\cite{becker2023degenerate}. Curiously, even though the origin of the exact FBs in all these systems are same, there are some tangible differences between the exact FBs reported in these articles. Firstly, in some cases the number of FBs are doubled~\cite{becker2022fine,le2022double,wan2023topological,becker2023degenerate}. Second, in case of TBCL, the Chern number is $C=\pm 2$ rather than $\pm 1$. High number of FBs as well as high Chern number can potentially give rise to exotic fractional and correlated states~\cite{neupert2011fractional,qi2011generic,wang2011nearly,wang2012fractional,yang2012topological,wu2013bloch,andrews2020fractional,liu2021gate,andrews2021stability,zhang2019nearly}. 

However, the different number of FBs and different Chern numbers lead to puzzles. To get a better understanding, let us describe the origin of the exact FBs in these systems. The general form of the $\mathbf{k}\cdot\mathbf{p}$ Hamiltonian with chiral symmetry is
\begin{equation}
\label{eq:chiralH}
    \mathcal{H}(\mathbf{r}) = \begin{pmatrix}
        \mathbf{0} & \mathcal{D}^\dagger(\mathbf{r})\\
        \mathcal{D}(\mathbf{r}) & \mathbf{0}
    \end{pmatrix}
\end{equation}
near a high symmetry momentum (HSM) $\mathbf{k}_0$ in the Brillouin zone (BZ) of the constituent material having a DP or QBCP, at $\mathbf{k}_0$, protected by $n$-fold rotation symmetry $\mathcal{C}_{nz}$ ($n=3$ for DP in TBG, $n=3$ or $4$ for QBCP) and an antiunitary symmetry $\mathcal{A}$ ($=\mathcal{T}$ for QBCP and $=\mathcal{C}_{2z}\mathcal{T}$ for DP, where $\mathcal{T}$ is time reversal symmetry; $\mathcal{A}\mathbf{r} = \pm \mathbf{r}$ ($\mathcal{A}\mathbf{k} = \mp \mathbf{k}$) when $\mathcal{A} = \mathcal{T}$ and $\mathcal{C}_{2z} \mathcal{T}$, respectively; also $\mathcal{A}\mathbf{k}_0\equiv \mathbf{k}_0$). Note that $\mathcal{D}(\mathbf{r})$ is $l\times l$ matrix, where $l$ is the number of layers. The operator $\mathcal{D}(\mathbf{r})$ has kinetic part $\mathcal{D}_k(\mathbf{r})$ consisting of spatial derivatives and potential term $D_U(\mathbf{r};\boldsymbol{\alpha})$ which arises from moir\'e periodic interlayer tunneling~\cite{tarnopolsky2019origin,li2022magic,becker2022fine,le2022double,eugenio2022twisted,becker2023degenerate} or externally applied field~\cite{le2022double,wan2023topological}, and $\boldsymbol{\alpha}$ is the vector of tuning parameters of the potential. Common to all of these systems is the fact that the kinetic part is antiholomorphic: $\mathcal{D}_k(\mathbf{r}) = (-2i\overline{\partial_{z}})^m \mathds{1}$, where $z = x+ i y$, $\mathds{1}$ is the identity matrix, $^*$ stands for complex conjugation, and $m =1$ and $2$ for DP and QBCP, respectively. 
Furthermore, under $\mathcal{C}_{nz}$ rotation, $\mathcal{D}(\mathbf{r}) = \mathcal{D}_k(\mathbf{r})+\mathcal{D}_U(\mathbf{r};\boldsymbol{\alpha})$ transforms as $\mathcal{D}(\mathcal{C}_{nz}\mathbf{r}) =(\omega^*)^2 \mathcal{D}(\mathbf{r})$, where $\omega = e^{2\pi i (m+3)/n}$. Consequently, the Hamiltonian satisfies 
$\text{Diag}\{\omega \mathds{1},\omega^* \mathds{1}\} \mathcal{H}(\mathbf{r})\text{Diag}\{\omega^* \mathds{1},\omega \mathds{1}\} = \mathcal{H}(\mathcal{C}_{nz}\mathbf{r})$.
The antiunitary operator $\mathcal{A}$ acts as $\sigma_x\otimes \mathds{1} \mathcal{H}^*(\mathbf{r})\sigma_x\otimes \mathds{1} = \mathcal{H}(\mathcal{A}\mathbf{r})$ ({\color{red}see Supplemental Material (SM)~\cite{SM2022} Sec.~S-I for more details}). 
An exact FB of $\mathcal{H}(\mathbf{r})$ at energy $E= 0$ with WF $\Psi_\mathbf{k}(\mathbf{r})$ satisfies $\mathcal{H}(\mathbf{r})\Psi_\mathbf{k}(\mathbf{r}) = \mathbf{0}$ for all $\mathbf{k}$. The construction of such a $\Psi_\mathbf{k}(\mathbf{r})$ is as follows. Note that due to $\mathcal{C}_{nz}$ and chiral symmetry $\mathcal{S} = \sigma_z\otimes \mathds{1}$, the two fold degeneracy of DP or QBCP at HSM $\mathbf{k}_0^m$ ($\mathbf{k}_0$ gets mapped to $\mathbf{k}_0^m$ in the moir\'e Brillouin zone (mBZ)) remains at $E= 0$ for any $\mathcal{D}_U (\mathbf{r};\boldsymbol{\alpha})$ that keeps $\mathcal{C}_{nz}$ symmetry. This means that there are always two sublattice polarized WFs $\Psi_{\mathbf{k}_0^m,1}(\mathbf{r}) = \{\psi_{\mathbf{k}_0^m}(\mathbf{r}),\mathbf{0}\}$ and $\Psi_{\mathbf{k}_0^m,2}(\mathbf{r}) = \{\mathbf{0},\psi_{\mathbf{k}_0^m}^*(\mathcal{A}\mathbf{r})\}$ satisfying $\mathcal{H}(\mathbf{r})\Psi_{\mathbf{k}_0^m,i}(\mathbf{r})=\mathbf{0}$, or equivalently $\mathcal{D}(\mathbf{r})\psi_{\mathbf{k}_0^m}(\mathbf{r})=\mathbf{0}$. If there are exact FBs, the WFs can be written as $\{\psi_\mathbf{k}(\mathbf{r}),\mathbf{0}\}$ and $\{\mathbf{0},\psi_{\mathcal{A}\mathbf{k}}^*(\mathcal{A}\mathbf{r})\}$. Since the kinetic part $\mathcal{D}_k(\mathbf{r})$ is antiholomorphic, the trial WF is naturally $\psi_{\mathbf{k}+\mathbf{k}_0^m}(\mathbf{r}) = f_\mathbf{k}(z)\psi_{\mathbf{k}_0^m}(\mathbf{r})$, where $f_\mathbf{k}(z)$ is holomorphic function satisfying $\overline{\partial_{z}}f_\mathbf{k}(z) = 0$. The function $f_\mathbf{k}(z)$ needs to satisfy Bloch periodicity (translation by moir\'e lattice vector $\mathbf{a}^{m}$ gives phase shift $e^{i\mathbf{k}\cdot\mathbf{a}^{m}}$). However, from Louiville's theorem, such a holomorphic function must have poles, making $\psi_\mathbf{k}(\mathbf{r})$ divergent, unless $\psi_{\mathbf{k}_0^m}(\mathbf{r})$ has a zero that cancels the pole. Conversely, if $\psi_{\mathbf{k}_0^m}(\mathbf{r})$ has zero at $\mathbf{r}_0$, a Bloch periodic holomorphic function
\begin{equation}
\label{eq:fkz}
\begin{split}
    &f_\mathbf{k}(z;\mathbf{r}_0)  = e^{i (\mathbf{k}\cdot\mathbf{a}^{m}_1) z/a_1}\frac{\vartheta\left(\frac{z-z_0}{a_1^m}-\frac{k}{b_2^m},\tau\right)}{\vartheta\left(\frac{z-z_0}{a_1^m},\tau\right)}\\
    &\hspace{1.3cm}= e^{i\mathbf{k}\cdot\mathbf{r}} \tilde{f}_\mathbf{k}(\mathbf{r};\mathbf{r}_0),\\
    & \tilde{f}_\mathbf{k}(\mathbf{r};\mathbf{r}_0) = e^{-i(\mathbf{b}_2^m\cdot\mathbf{r})k/b_2^m}\frac{\vartheta\left(\frac{z-z_0}{a_1^m}-\frac{k}{b_2^m},\tau\right)}{\vartheta\left(\frac{z-z_0}{a_1^m},\tau\right)}
\end{split}
\end{equation}
with a pole at $\mathbf{r}_0$ can be constructed. Here $\vartheta(z,\tau)$ is the Jacobi theta function of the first type~\cite{ledwith2020fractional}, $\mathbf{a}^{m}_i$ are lattice vectors, $\mathbf{b}^{m}_i$ are the corresponding reciprocal lattice vectors ($\mathbf{a}^{m}_i\cdot\mathbf{b}^{m}_j = 2\pi \delta_{ij}$), $a_i^m = (\mathbf{a}^{m}_i)_x+ i (\mathbf{a}^{m}_i)_y$, $b_i^m = (\mathbf{b}^{m}_i)_x+ i (\mathbf{b}^{m}_i)_y$, $z_0 = (\mathbf{r}_0)_x+ i (\mathbf{r}_0)_y$, $k = k_x + i k_y$, and $\tau = a_2^m/a_1^m$. Remarkably, at ``magic'' values of $\boldsymbol{\alpha}$, the WF $\psi_{\mathbf{k}_0^m}(\mathbf{r})$ has a zero, which allows for such $f_\mathbf{k}(z;\mathbf{r}_0)$, and in turn gives rise to two exact FBs. Furthermore, the periodic part $\tilde{f}_\mathbf{k}(\mathbf{r};\mathbf{r}_0)$ is a holomorphic function of $k$~\cite{kharchev2015theta,ledwith2020fractional}; this property along with the presence of the zero in the WF can be used to prove that  wave functions of this form carry Chern number $C = \pm 1$ (see~\cite{wang2021exact}, also {\color{red}SM~\cite{SM2022} S-IV for a proof}).

It was found in~\cite{wan2023topological}, that in some cases $\psi_{\mathbf{k}_0^m}(\mathbf{r})$ can have two zeros at $\mathbf{r}_0^{(1)}$ and $\mathbf{r}_0^{(2)}$ in the unit cell allowing construction of two holomorphic functions $f_\mathbf{k}(z;\mathbf{r}_0^{(1)})$ and $f_\mathbf{k}(z;\mathbf{r}_0^{(2)})$ having poles at $\mathbf{r}_0^{(1)}$ and $\mathbf{r}_0^{(2)}$, respectively, giving rise to 4 FBs. Naturally, the following questions arise: (i) how many zeros (equivalently, how many FBs) can $\psi_{\mathbf{k}_0^m}(\mathbf{r})$ have? (ii) do all zeros of $\psi_{\mathbf{k}_0^m}(\mathbf{r})$ of allow for construction of exact FB WFs? Related to this was the observation in~\cite{le2022double}; four FBs were found even though $\psi_{\mathbf{k}_0^m}(\mathbf{r})$ had a single quadratic zero, and it was not clear how to construct the FB WFs fully analytically.
\begin{figure}[t]
     \centering
\includegraphics[scale=1]{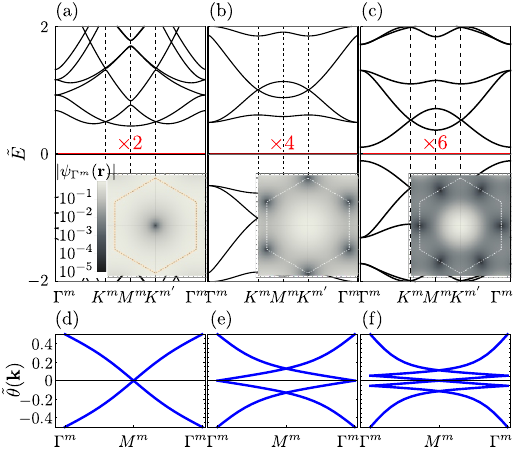}
     \caption{Different number of exact flat bands in single layer system with QBCP under different periodic strain fields $\tilde{A}(\mathbf{r})$ all having $p6mm$ symmetry. For (a-b), $\tilde{A}(\mathbf{r})= -\frac{\alpha}{2}\sum_{n=1}^{3}e^{i(4-n)\phi} \cos\left(\mathbf{b}_n^m \cdot \mathbf{r} \right)$, whereas for (c), $\tilde{A}(\mathbf{r})= -\frac{\alpha}{2} \sum_{n=1}^{3}(-e^{i(1-n)\phi} \cos\left(2\mathbf{b}_n^m \cdot \mathbf{r}\right)+ 2 e^{i(2-n)\phi} \cos\left((\mathbf{b}_n^m-\mathbf{b}_{n+1}^m) \cdot \mathbf{r}\right) + 2 e^{i(4-n)\phi} \cos\left(\mathbf{b}_n^m \cdot \mathbf{r} \right))$, where $\phi=2\pi /3$, $\mathbf{b}^m_{1} = \frac{4\pi}{\sqrt{3}a^m}(0,1)$ and $\mathbf{b}^m_{2,3} = \frac{4\pi}{\sqrt{3}a^m}(\mp\sqrt{3}/2,-1/2)$ are the reciprocal lattice vectors, and $a^m$ is the lattice constant of the superlattice.  The vertical axes in (a-c) are normalized eigen-energies $\tilde{E}=\frac{E}{|\mathbf{b}^m|^2}$. The ``magic'' values of $\alpha$ at which exact FBs appear are $\tilde{\alpha} =\frac{\alpha}{|\mathbf{b}^m| ^{2}} =0.62, -2.88$ and $2.93$ for (a), (b) and (c), respectively. The number of FBs in (a), (b) and (c) are 2, 4 and 6, respectively. Density plots of 
     $|\psi_{\Gamma^{m}}(\mathbf{r})|$ (normalized by its maximum) are shown at the lower right corner of (a-c). The dark spots in the density plots denote zeros, the numbers of which in the unit cell are 1, 2 and 3 for (a), (b) and (c), respectively. The dashed lines in the density plots mark the boundary of the unit cell. (d-f) Wilson loop spectrum $\tilde{\theta}({\mathbf{k}})=\frac{\theta({\mathbf{k}})}{2\pi}$ of the flat bands of (a-c), respectively.
     }
     \label{fig:p6mm}
\end{figure}

In this article, we show that not all zeros of $\psi_{\mathbf{k}_0^m}(\mathbf{r})$ allow for construction of exact FB WF, only the ones at high symmetry points (HSPs) having $\mathcal{C}_{n'z}$ ($n'\geq 2$) symmetry allow such construction. Using this result, we systematically classify systems with different PG symmetries, and enumerate the number of FBs they can possess depending on the multiplicity of the HSPs, where the zeros appear, in the unit cell. We show that all known examples of exact FBs fall under this classification in Fig.~\ref{fig:flowchart}(b). Furthermore, we find that a maximum of 6-FBs is possible in a system with $\mathcal{C}_{6}$ or $\mathcal{C}_{6v}$ PG symmetry (unless the system is fine-tuned), we show an example of $\mathcal{C}_{6v}$ PG symmetry in a single layer system with QBCP under periodic strain field. Additionally, we clarify the topology of these FBs, and explain the observation of $C=\pm 2$ bands in~\cite{li2022magic}.

\textit{Only the zeros of $\psi_{\mathbf{k}_0^m}(\mathbf{r})$ at HSPs give FBs.}--For a single layer system, $\psi_{\mathbf{k}_0^m}(\mathbf{r})$ is a one-component function. To the lowest order, $\psi_{\mathbf{k}_0^m}(\mathbf{r}=\mathbf{r}_0+\delta (x,y)) = \delta(c_1 x + c_2 y) +\mathcal{O}(\delta^2)$ near a generic zero at $\mathbf{r}_0$ of $\psi_{\mathbf{k}_0^m}$, whereas $f_\mathbf{k}(\mathbf{r}=\mathbf{r}_0+\delta (x,y);\mathbf{r}_0)\sim\frac{c}{\delta(x+iy)}$ since $f_\mathbf{k}$ is holomorphic. Hence, $\psi_{\mathbf{k}+\mathbf{k}_{0}}(\mathbf{r}=\mathbf{r}_0+\delta (x,y))\sim \frac{c_1 x + c_2 y}{x+ i y}$; such a function is singular at $\mathbf{r} =\mathbf{r}_0$ for generic values of $c_1$ and $c_2$, hence $\psi_{\mathbf{k}+\mathbf{k}_{0}}(\mathbf{r})$ is not a good WF. At high symmetry points, however, the relative value of $c_1$ and $c_2$ is constrained. For example, around $\mathcal{C}_{n'z}$ ($n' \geq 2$) symmetric HSP, 
if $\psi_{\mathbf{k}_0^m}(\mathbf{r}_0+\delta \mathcal{C}_{n'z} (x,y)) = \psi_{\mathbf{k}_0^m}(\mathbf{r}_0+\delta (x,y))$ (see {\color{red}SM~\cite{SM2022} S-II} for the transformation properties of the WF), then $\psi_{\mathbf{k}_0^m}(\mathbf{r}=\mathbf{r}_0+\delta (x,y)) = \mathcal{O}(\delta^2)$ implying $c_1=c_2 = 0$ which makes the WF $\psi_\mathbf{k}(\mathbf{r})$ regular (non-singular). 
In the bilayer case, where the WF $\psi_{\mathbf{k}_0^m}(\mathbf{r})$ is a two-component function, more care is required to prove similar result; this is done in the {\color{red}SM~\cite{SM2022} S-II}.

\textit{Multiplicity of flat-bands in systems with different PG symmetries.}-- We established that among all zeros of $\psi_{\mathbf{k}_0^m}(\mathbf{r})$ only the ones at the HSPs allow construction of exact flat-band WF. However, notice that there can $m>1$ symmetry related HSPs in one unit cell.
Let the positions of these HSPs be $\mathbf{r}_0^{(1)},\dots, \mathbf{r}_0^{(m)}$. Then $m$ zeros would appear in the unit cell together, and hence $m$ complex analytic Bloch-periodic functions $f_\mathbf{k}(z;\mathbf{r}_0^{(1)}),\dots,f_\mathbf{k}(z;\mathbf{r}_0^{(m)})$ having poles at $\mathbf{r}_0^{(1)},\dots, \mathbf{r}_0^{(m)}$, respectively, can be constructed using Eq.~\ref{eq:fkz}, and WFs $\psi_{\mathbf{k}+\mathbf{k}_0^m}^{(i)}(\mathbf{r})=f_\mathbf{k}(z;\mathbf{r}_0^{(i)})\psi_{\mathbf{k}_0^m}(\mathbf{r})$ ($i=1,\dots,m$) form $m$ flat-bands. Moreover, using the antiunitary symmetry $\mathcal{A}$, we can construct another $m$ flat-band WFs. This implies that if the HSP, where the zero of $\psi_{\mathbf{k}_0^m}(\mathbf{r})$ appears, has multiplicity $m$, there are $2m$ flat-bands. We show examples of this for space group $p6mm$ in Fig.~\ref{fig:p6mm}, where we consider a single layer system with QBCP at $\mathbf{k}_0^m = \Gamma$ point under periodic strain field~\cite{wan2023topological} with $\mathcal{D}_U(\mathbf{r};\boldsymbol{\alpha}) = \tilde{A}(\mathbf{r}) = A_x(\mathbf{r})+iA_y(\mathbf{r})$ where $A_x=u_{xx}-u_{yy}$ and $A_y=u_{xy}$ are moir\'e periodic shear strain fields. In Fig.~\ref{fig:p6mm}(a), (b) and (c), we show that for specific forms of $\mathcal{D}_U(\mathbf{r};\boldsymbol{\alpha})$ (see the caption of Fig.~\ref{fig:p6mm} for details), at a critical value of $\alpha$ (the strength of the strain field), 2, 4 and 6 FBs appear, respectively. Correspondingly, the plots of $\psi_\Gamma(\mathbf{r})$ show zeros at HSPs with multiplicity 1, 2 (related by $\mathcal{C}_{2z}$) and 3 (related by $\mathcal{C}_{3z}$), respectively, in the unit cell, in agreement with the argument for the number of FBs presented above.
\begin{figure}[t]
     \centering
\includegraphics[scale=1]{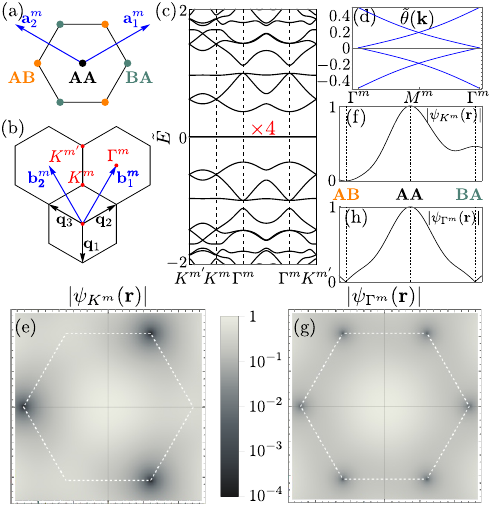}
     \caption{Four flat bands in TBG Hamiltonian under alternating magnetic field~\cite{le2022double}.
     (a), (b) Moir\'e unit cell and Brillouin zone, respectively.
     (c) Band structure for $\alpha=1.379e^{i 0.254 \pi}$ showing 4 exact FBs. 
     (d) Wilson loop spectrum $\tilde{\theta}({\mathbf{k}})=\frac{\theta({\mathbf{k}})}{2\pi}$ of the 4 flat bands.
     (e) Density plot of $|\psi_{K^m}(\mathbf{r})|$ (normalized by its maximum), which has a single quadratic zero at the AB position, as shown in (f). 
     (g) Density plot of $|\psi_{\Gamma^{m}}(\mathbf{r})|$ (normalized by its maximum), which has two linear zeros at the AB and BA positions, as shown in (h).
     The dashed lines in the density plots mark the boundary of the unit cell.
    }
     \label{fig:tbg4}
\end{figure}
\begin{figure*}[t]
     \centering
\includegraphics[scale=1]{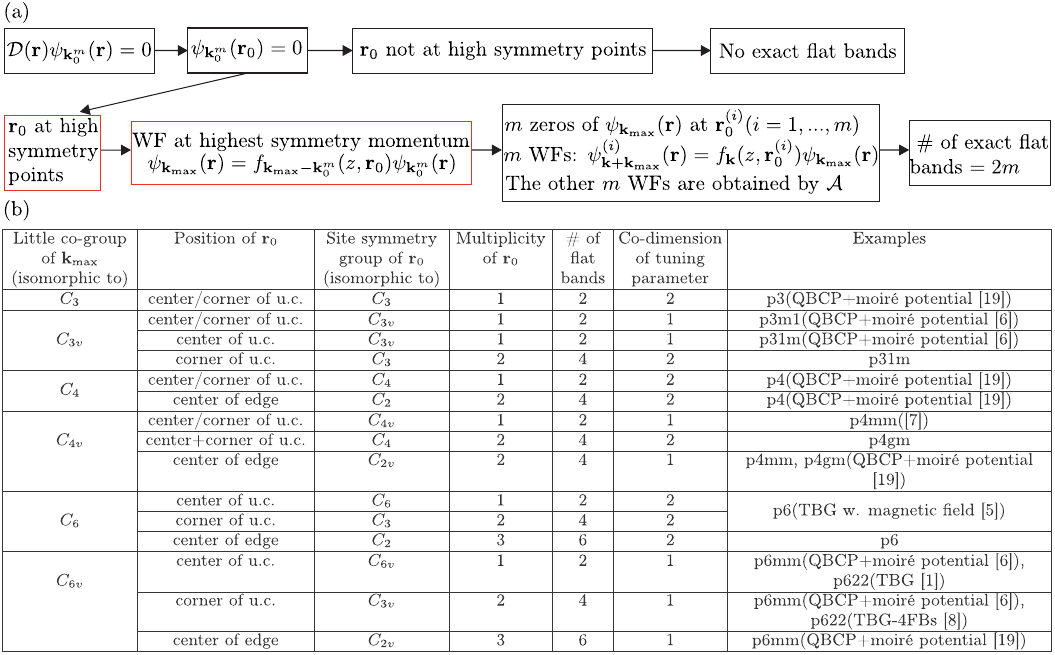}
     \caption{
     (a) Flowchart of the construction of exact flat band wave-functions starting from the wave-function at $\mathbf{k}_0^m$.
     (b) List of all possible point groups of $\mathbf{k}_\text{max}$ for space groups where DP or QBCP is possible, possible HSPs $\mathbf{r}_0$ (having $C_2$ or higher symmetry), their multiplicity in the unit cell, number of flat bands if zeros appear at these HSPs, the number of parameters need to be varied (co-dimension) to achieve the flat bands. The last column shows examples of different cases.
     }
     \label{fig:flowchart}
\end{figure*}

It is important to note that the momentum $\mathbf{k}_0^m$, where the DP or QBCP is, may not be the momentum ($\mathbf{k}_\text{max}$) with highest symmetry in the BZ. An example of this is chiral TBG: the $K$ point (where the DP is) has magnetic little co-group $6'$ (generated by $\mathcal{C}_{3z}$, $\mathcal{C}_{2x}$ and $\mathcal{C}_{2z}\mathcal{T}$). But, due to a unitary particle-hole symmetry $P = \mathds{1}\otimes i\sigma_y$ ($P\mathbf{r} =-\mathbf{r}$, $P\mathbf{k} =-\mathbf{k}$ and $P \mathcal{H}(\mathbf{r}) = -\mathcal{H}(-\mathbf{r})P$)~\cite{song2019all,hejazi2019multiple} and chiral symmetry $\mathcal{S} =\sigma_z\otimes \mathds{1}$ ($\mathcal{S}\mathbf{r} =\mathbf{r}$, $\mathcal{S}\mathbf{k} =\mathbf{k}$ and $\mathcal{S} \mathcal{H}(\mathbf{r}) = -\mathcal{H}(\mathbf{r})\mathcal{S}$), an effective $\mathcal{C}_{2z}$ symmetry can be obtained: $\mathcal{C}_{2z} = P\mathcal{S} = \sigma_z\otimes i\sigma_y$ (this symmetry was found previously in~\cite{wang2021chiral}, and was called intra-valley inversion symmetry). As a consequence, in chiral TBG, $\mathbf{k}_\text{max} = \Gamma$ with magnetic little co-group $6221'$ (generated by $\mathcal{C}_{3z}$, $\mathcal{C}_{2z}$, $\mathcal{C}_{2x}$ and $\mathcal{T}$; see {\color{red}SM Sec.~S-I} for a review of symmetries of chiral TBG). In these cases, where $\mathbf{k}_0^m \neq \mathbf{k}_\text{max}$, if there is a zero at some HSP $\mathbf{r}_0$ of the WF $\psi_{\mathbf{k}_0^m}(\mathbf{r})$, one should first construct $\psi_{\mathbf{k}_\text{max}}(\mathbf{r}) = f_{\mathbf{k}_\text{max}-\mathbf{k}_0^m}(z,\mathbf{r}_0)\psi_{\mathbf{k}_0^m}(\mathbf{r})$ using the zero, and then check the number of zeros of $\psi_{\mathbf{k}_\text{max}}(\mathbf{r})$ in the moir\'e unit cell (since $\psi_{\mathbf{k}_\text{max}}(\mathbf{r})$ has more symmetry than $\psi_{\mathbf{k}_0^m}(\mathbf{r})$) to construct the flat-band WFs. 
Interestingly, this resolves the ambiguity in the construction of 4 flat-band WFs in TBG under alternating magnetic field in~\cite{le2022double} as is described below. The $\mathbf{k}\cdot\mathbf{p}$ Hamiltonian around $\mathbf{k}_0 = K$ considered in this paper is the same as in Eq.~\ref{eq:chiralH} with 
\begin{equation}
\begin{split}
    \mathcal{D}(\mathbf{r}) &= \begin{pmatrix}
        -2i\overline{\partial_{z}} & \alpha U(\mathbf{r})\\
        \alpha U(-\mathbf{r}) & -2i\overline{\partial_{z}}
    \end{pmatrix},\\
    U(\mathbf{r}) &= e^{-i\mathbf{q}_1\cdot\mathbf{r}}+e^{i \phi} e^{-i\mathbf{q}_2\cdot\mathbf{r}} +e^{-i \phi} e^{-i\mathbf{q}_3\cdot\mathbf{r}},
\end{split}
\end{equation}
where $\phi=\frac{2\pi}{3}$, $\mathbf{q}_1 =\frac{4\pi}{3a^{m}}(0,-1)$, $\mathbf{q}_{2,3} = \frac{2\pi}{3a^{m}}(\pm \sqrt{3},1)$, $a^{m}$ is the moir\'e lattice constant (see Fig.~\ref{fig:tbg4}(a-b) for details of the moir\'e lattice vectors and reciprocal lattice vectors). This is exactly the same as chiral TBG model considered in~\cite{tarnopolsky2019origin} except that here, unlike~\cite{tarnopolsky2019origin}, $\alpha$ is allowed to take complex values. It was found in~\cite{le2022double} that, at certain critical complex values of $\alpha$, there are 4 exact FBs (see Fig.~\ref{fig:tbg4}(c-d)). At these critical $\alpha$, the WF $\psi_{\mathbf{k}_0^m = K}(\mathbf{r})$ has a single quadratic zero at the AB position $\mathbf{r}_0^{(1)} = a^{m}/\sqrt{3}(-1,0)$ as shown in Fig.~\ref{fig:tbg4}(e-f). Therefore, it was pointed out in~\cite{le2022double} how to construct the FB WFs using complex analytic function Eq.~\ref{eq:fkz} fully. However, the WF $\psi_{\mathbf{k}_\text{max} = \Gamma}(\mathbf{r}) = f_{\mathbf{k}_\text{max}-\mathbf{k}_0^m}(z,\mathbf{r}_0^{(1)})\psi_{\mathbf{k}_0^m}(\mathbf{r}) = f_{\mathbf{q}_3}(z,\mathbf{r}_0^{(1)})\psi_{K}(\mathbf{r})$ has two linear zeros at AB ($\mathbf{r}_0^{(1)} = a^{m}/\sqrt{3}(-1,0)$) and BA  ($\mathbf{r}_0^{(2)} = a^{m}/\sqrt{3}(1,0)$) positions, as shown in Fig.~\ref{fig:tbg4}(g-h). This is due to the fact that this Hamiltonian has the effective $\mathcal{C}_{2z} = P\mathcal{S}$ as discussed earlier (note that when $\alpha$ is complex, $\mathcal{C}_{2x}$ is broken, hence this system has $\mathcal{C}_6$ point group and $p6$ space group symmetry), hence $\psi_\Gamma(\mathbf{r})$ is $\mathcal{C}_{2z}$ symmetric, and AB and BA positions are related by $\mathcal{C}_{2z}$. Using these two zeros, 4 FB WFs can be constructed as follows:
\begin{equation}
\begin{split}
    \Psi_\mathbf{k}^{(1)}(\mathbf{r}) &= f_\mathbf{k}(z;\mathbf{r}_0^{(1)})f_{\mathbf{q}_3}(z,\mathbf{r}_0^{(1)})\begin{bmatrix}\psi_K(\mathbf{r})\\\mathbf{0}\end{bmatrix},\\
    \Psi_\mathbf{k}^{(2)}(\mathbf{r}) &= f_\mathbf{k}(z;\mathbf{r}_0^{(2)})f_{\mathbf{q}_3}(z,\mathbf{r}_0^{(1)})\begin{bmatrix}\psi_K(\mathbf{r})\\\mathbf{0}\end{bmatrix},\\
    \Psi_\mathbf{k}^{(3)}(\mathbf{r}) &= \sigma_x\otimes \mathds{1}{\Psi_{\mathcal{A}\mathbf{k}}^{(1)}}^*(\mathcal{A}\mathbf{r}),\\
    \Psi_\mathbf{k}^{(4)}(\mathbf{r}) &= \sigma_x\otimes \mathds{1}{\Psi_{\mathcal{A}\mathbf{k}}^{(2)}}^*(\mathcal{A}\mathbf{r}).
\end{split}
\end{equation}

We summarize the procedure to find the number of exact FBs from the number of zeros of $\psi_{\mathbf{k}_\text{max}}(\mathbf{r})$ as well as the construction of the exact FB WFs in the flow-chart in Fig.~\ref{fig:flowchart}(a). This insight also allows us to predict the number of possible FBs in systems with all different point group symmetries that protect DP or QBCP. This is done via the following steps: (i) we list all possible point group symmetries (little co-group~\cite{bradley2010mathematical}) of $\mathbf{k}_\text{max}$ for space groups that allow DP or QBCP, (ii) then we list the HSPs (maximal wyckoff positions) of these space groups, their site-symmetry groups and multiplicity in the unit cell, (iii) the number of possible symmetry-dictated FBs is then twice the multiplicity of the HSPs. These results are tabulated in Fig.~\ref{fig:flowchart}(b). The table also contains all examples of exact flat bands (in single and bilayer systems) found in the literature~\cite{tarnopolsky2019origin,le2022double,becker2022fine,becker2023degenerate,wan2023topological,eugenio2022twisted} as well as some new ones: 2 FBs in systems with QBCP under periodic strain having $p3$ or $p4$ space group symmetries, 4 FBs in systems with QBCP under periodic strain having $p4$, $p4mm$ and $p4gm$ space group symmetry (details of these are given in {\color{red}SM~\cite{SM2022} S-VI}), and 6 flat bands in systems with QBCP under periodic strain having $p6mm$ space group symmetry (Fig.~\ref{fig:p6mm}(c)). Furthermore, since in 2D space groups, the maximum multiplicity of a HSP is 3, the maximum number of FBs can be 6 (not counting fine tuned cases where zeros can accidentally appear at HSPs not related by symmetries).

\textit{Construction of WFs at} $\mathbf{k}_\text{max}$.--Notice that, in the construction of $2m$ ($m>1$) FB WFs from zeros of $\psi_{\mathbf{k}_\text{max}}$ at $\mathbf{r}_0^{(i)}$, the functions $f_\mathbf{k}(z;\mathbf{r}_0^{(i)})$ are independent of each other for all $\mathbf{k}$ except $\mathbf{k} = \mathbf{0}$ since $f_{\mathbf{k}= \mathbf{0}}(z;\mathbf{r}_0^{(i)}) = 1$ for all $i = 1, \dots, m$. Therefore, the construction above gives $2m$ independent WFs for $\mathbf{k}\neq \mathbf{k}_\text{max}$, but only 2 independent WFs at $\mathbf{k}_\text{max}$; hence the other $2(m-1)$ WFs at $\mathbf{k}_\text{max}$ need to be obtained separately. To this end, note that the WFs $\psi_{\mathbf{k}+\mathbf{k}_\text{max}}^{(i)}(\mathbf{r})=f_{\mathbf{k}}(z;\mathbf{r}_0^{(i)})\psi_{\mathbf{k}_\text{max}}(\mathbf{r})$ are independent, but not mutually orthogonal for $\mathbf{k}\neq \mathbf{k}_\text{max}$. We can modify the later $(m-1)$ WFs to be orthogonal to $\psi_\mathbf{k}^{(1)}(\mathbf{r})$, and define new WFs:
\begin{equation}
\label{eq:PartialOrthogonalization}
    \tilde{\psi}_{\mathbf{k}}^{(i)}(\mathbf{r}) = \begin{cases}
        \psi_\mathbf{k}^{(1)}(\mathbf{r}) &\text{ if } i =1,\\
        \psi_\mathbf{k}^{(i)}(\mathbf{r}) -\frac{\langle \psi_\mathbf{k}^{(1)}|\psi_\mathbf{k}^{(i)}\rangle}{\langle \psi_\mathbf{k}^{(1)}|\psi_\mathbf{k}^{(1)}\rangle}\psi_\mathbf{k}^{(1)}(\mathbf{r}) &\text{ if } i \neq 1,\\
    \end{cases}
\end{equation}
where $\langle f | g \rangle = \int_\text{unit cell} d^2\mathbf{r} f^*(\mathbf{r})g(\mathbf{r})$. Remembering $f_\mathbf{k}(z;\mathbf{r}_0)$ is product of a holomorphic function \xw{$\tilde{f}_\mathbf{k}(\mathbf{r};\mathbf{r}_0)$} 
of $k = k_x +i k_y$ and a phase factor $e^{i\mathbf{k}\cdot\mathbf{r}}$, and $\tilde{f}_\mathbf{0}(\mathbf{r};\mathbf{r}_0) = 1$, we can expand these functions as $f_\mathbf{k}(z,\mathbf{r}_0) = 1 + \tilde{f}_\mathbf{0}'(\mathbf{r};\mathbf{r}_0) k +i(k z^*+k^* z)/2 +\mathcal{O}(k^2)$, where $\tilde{f}_\mathbf{0}'(\mathbf{r};\mathbf{r}_0) \equiv \frac{1}{2}[(\partial_{k_x}-i\partial_{k_y})\tilde{f}_\mathbf{k}(\mathbf{r};\mathbf{r}_0)]|_{\mathbf{k} = \mathbf{0}}$. Plugging this back in Eq.~\ref{eq:PartialOrthogonalization}, we find close to $\mathbf{k}_\text{max}$
\begin{widetext}
\begin{equation}
\label{eq:PartialOrthogonalizationTE}
    \tilde{\psi}_{\mathbf{k}_\text{max}+\mathbf{k}}^{(i)}(\mathbf{r}) \approx\begin{cases}
        (1+k \tilde{f}_\mathbf{0}'(\mathbf{r};\mathbf{r}_0^{(1)}) +(k z^*+k^* z)/2 )\psi_{\mathbf{k}_\text{max}}(\mathbf{r})  &\text{ if } i =1,\\
        k(\tilde{f}_\mathbf{0}'(\mathbf{r};\mathbf{r}_0^{(i)})-\tilde{f}_\mathbf{0}'(\mathbf{r};\mathbf{r}_0^{(1)})-\langle\psi_{\mathbf{k}_\text{max}}|(\tilde{f}_\mathbf{0}'(\mathbf{r};\mathbf{r}_0^{(i)})-\tilde{f}_\mathbf{0}'(\mathbf{r};\mathbf{r}_0^{(1)}))\psi_{\mathbf{k}_\text{max}}\rangle)\psi_{\mathbf{k}_\text{max}}(\mathbf{r})&\text{ if } i \neq 1.\\
    \end{cases}
\end{equation}
Continuing these WFs to $\mathbf{k}_\text{max}$, we get
\begin{equation}
\label{eq:WFKmax}
    \tilde{\psi}_{\mathbf{k}_\text{max}}^{(i)}(\mathbf{r}) =\begin{cases}
        \psi_{\mathbf{k}_\text{max}}(\mathbf{r})  &\text{ if } i =1,\\
        (\tilde{f}_\mathbf{0}'(\mathbf{r};\mathbf{r}_0^{(i)})-\tilde{f}_\mathbf{0}'(\mathbf{r};\mathbf{r}_0^{(1)})-\langle\psi_{\mathbf{k}_\text{max}}|(\tilde{f}_\mathbf{0}'(\mathbf{r};\mathbf{r}_0^{(i)})-\tilde{f}_\mathbf{0}'(\mathbf{r};\mathbf{r}_0^{(1)}))\psi_{\mathbf{k}_\text{max}}\rangle)\psi_{\mathbf{k}_\text{max}}(\mathbf{r})&\text{ if } i \neq 1.\\
    \end{cases}
\end{equation}
\end{widetext}
Note that $\tilde{f}_\mathbf{0}'(\mathbf{r};\mathbf{r}_0^{(i)})-\tilde{f}_\mathbf{0}'(\mathbf{r};\mathbf{r}_0^{(1)})$ is a holomorphic function of $z = x+iy$ even though \xw{$\tilde{f}_\mathbf{k}(\mathbf{r};\mathbf{r}_0)$}
is not holomorphic, hence $ \tilde{\psi}_{\mathbf{k}_\text{max}}^{(i)}(\mathbf{r})$ satisfy $\mathcal{D}(\mathbf{r}) \tilde{\psi}_{\mathbf{k}_\text{max}}^{(i)}(\mathbf{r}) = \mathbf{0}$. Furthermore, the last $(m-1)$ WFs at $\mathbf{k}_\text{max}$ are automatically orthogonal to $\psi_{\mathbf{k}_\text{max}}(\mathbf{r})$. We can further mutually orogonalize them using Gram-Schmidt orthogonalization procedure. The $m$ WFs at $\mathbf{k}_\text{max}$ polarized on the other sublattice are, as usual, $(\tilde{\psi}_{\mathbf{k}_\text{max}}^{(i)}(\mathcal{A}\mathbf{r}))^*$. We compare WFs obtained in this way with numerically obtained WFs at $\mathbf{k}_\text{max}=\Gamma$ for the 4 and 6 FBs systems 
({\color{red}see SM~\cite{SM2022} S-III for details}).

\textit{Co-dimension of the tuning parameters.}--Since $\psi_{\mathbf{k}_\text{max}}(\mathbf{r})$ is complex $l$ component vector ($l$ is the number of layers), the vector equation $\psi_{\mathbf{k}_\text{max}}(\mathbf{r}_0) = \mathbf{0}$ is actually a set of $2l$ equations (real and imaginary part of each component need to be zero). Therefore, in principle, $2l$ parameters has to be varied to obtain FBs; hence the co-dimension is generically $2l$. In case of TBG, since one of the components of $\psi_{\mathbf{k}_\text{max}}(\mathbf{r})$ is automatically zero at the corners of the moir\'e unit cell (AB or BA stacking positions) due to $\mathcal{C}_{3z}$ symmetry (see~\cite{tarnopolsky2019origin}, also {\color{red}SM~\cite{SM2022} S-II} for details), in all cases, that we are considering, the co-dimension is 2, unless there are extra symmetries. If the HSP $\mathbf{r}_0$ has mirror-type symmetry $\mathcal{M}$, such that the representation of the composition $\mathcal{M}\mathcal{A}$ is diagonal with entries $\pm 1$ (by mirror-type we mean: (i) the symmetry does not commute with $\mathcal{C}_{nz}$ ($n\geq 3$), (ii) it commutes with $\mathcal{A}$ whose representation is $\sigma_x\otimes \mathds{1}\mathcal{K}$ and (iii) $\mathcal{M}^2 = \mathds{1}$. Note that $C_{2x}$ is also mirror-type according to this.), then the components of $\psi_{\mathbf{k}_\text{max}}(\mathbf{r})$ are purely real or imaginary at $\mathbf{r} = \mathbf{r}_0$, which reduces the co-dimension to 1. In case of single layer QBCP, mirror symmetry necessarily has representation $\rho(\mathcal{M}) =\sigma_x$. In case of chiral TBG, one can construct an $\mathcal{MA}$ like symmetry $P\mathcal{S}\mathcal{C}_{2x} \mathcal{C}_{2z}\mathcal{T}$ (under which $P\mathcal{S}\mathcal{C}_{2x} \mathcal{C}_{2z}\mathcal{T}(x,y) = (x,-y)$ and $P\mathcal{S}\mathcal{C}_{2x} \mathcal{C}_{2z}\mathcal{T}(k_x,k_y) = (-k_x,k_y)$, hence it is a symmetry of the AB and BA stacking point in real space, as well as a symmetry of the $K^m$ point), that has representation $\rho(P\mathcal{S}\mathcal{C}_{2x} \mathcal{C}_{2z}\mathcal{T}) 
= \sigma_z\otimes\sigma_z \mathcal{K}$. Therefore, in QBCP systems with mirror-type symmetry as well as in case of chiral TBG, the co-dimension is 1. Furthermore, in case of chiral TBG with alternating magnetic field~\cite{le2022double}, $\mathcal{C}_{2x}$ is broken, hence the co-dimension is 2. The co-dimension of tuning parameters for systems with different symmetries are tabulated in table in Fig.~\ref{fig:flowchart}(b). Our results for co-dimensions are in agreement with the ones given in~\cite{sheffer2023symmetries}.

\textit{Topology and quantum geometry of the FBs.}--As was mentioned earlier, it can be proven that the sublattice polarized exact FB WFs have Chern number $C = \pm 1$ using the fact that the WFs are (anti)holomorphic in $k$ (the WFs on one sublattice are holomorphic, the WFs on the other sublattice are antiholomorphic). This is why the Wilson loop spectra 
({\color{red}see SM~\cite{SM2022} S-IV for details}), for the case where there are only two FBs, have $\pm 1$ winding
 as shown in Fig.~\ref{fig:p6mm}(d). Surprisingly, for the cases of 4 or 6 FBs, the Wilson loop spectrum still has winding $\pm 1$ as can be seen from Figs.~\ref{fig:p6mm}(e-f) and~\ref{fig:tbg4}(d). This implies that the $m$ WFs polarized to the same sublattice together actually have Chern number $|C| = 1$, instead of $|C| = m$ as we would have na\"ively guessed from the fact that these WFs are holomorphic functions. We explain this in the following. As mentioned earlier, the sublattice polarized WFs $\psi_{\mathbf{k}+\mathbf{k}_\text{max}}^{(i)}(\mathbf{r}) = f_\mathbf{k}(z;\mathbf{r}_0^{(i)})\psi_{\mathbf{k}_\text{max}}(\mathbf{r})$ are not mutually orthogonal. When we mutually orthogonalize them (similar to Eq.~\ref{eq:PartialOrthogonalization}), $m-1$ of the WFs have a singularity; they are proportional to $k$ near $\mathbf{k} = 0$ (Eq.~\ref{eq:PartialOrthogonalizationTE}). Therefore, they have a Berry phase winding around $\mathbf{k} = 0$, this Berry phase winding cancels the Berry phase winding along the edge of the BZ to give Chern number $C = 0$ ({\color{red}see SM~\cite{SM2022} S-IV for details}). As a consequence, only one of the sublattice polarized WFs carry $|C|=1$, the others are topologically trivial. This result is in agreement with~\cite{TBGIIBernevig}, where it was shown that for systems with an antiunitary particle-hole symmetry $\mathcal{P}$ that satisfies $\mathcal{P}^2 = -\mathds{1}$, if a set of bands symmetric around zero energy has $4p+2$ ($p\in \mathds{N}$) number of Dirac points between them, then they are $\mathds{Z}_2$ nontrivial having Chern number $C = \pm 1$. Indeed, the systems considered here have $\mathcal{P}$ symmetry; for QBCP systems it can be constructed as $\mathcal{P} = \mathcal{S}\mathcal{T}$, whereas in TBG it is $\mathcal{P} = P\mathcal{C}_{2z}\mathcal{T}$. Interestingly, TBCL~\cite{li2022magic} has 2 FBs with Chern numbers $\pm2$ which seemingly is in disagreement with our result. However, it can be shown that the TBCL Hamiltonian is just two copies of a single layer system with QBCP under a periodic moir\'e potential with $p4mm$ space group symmetry unfolded to a BZ twice the size of single layer moir\'e BZ, and this BZ doubling gives results in the doubling of the Chern number ({\color{red}see SM~\cite{SM2022} S-VII for details}). Also, if one replaces the $p4mm$ symmetric tunneling potential with a $p4gm$ potential in TBCL (this can be achieved by simply making the amplitude purely imaginary), one finds 4 FBs with $C =\pm 2$ instead of 2 FBs at some magic angle ({\color{red}see SM~\cite{SM2022} S-VII}); this doubling of number of FBs in $p4gm$ case is due to ``band sticking'' inherent in nonsymmorphic space groups~\cite{dresselhaus2008applications}. We further show in the {\color{red} SM~\cite{SM2022} S-V}, that the FBs with higher degeneracy satisfy ideal non-Abelian trace condition: $\text{tr}(g^{mn}_{\alpha \beta}(\mathbf{k})) = \sum_{n}'\sum_{\alpha \in \{x,y\}} g^{nn}_{\alpha\alpha}(\mathbf{k}) = |\text{tr}(F_{xy}^{mn}(\mathbf{k}))| = |\sum_{n}'F_{xy}^{nn}(\mathbf{k})|$, where $g_{ij}^{mn}(\mathbf{k})$ is the non-Abelian Fubini-Study metric and $F_{xy}^{mn}(\mathbf{k})$ is the non-Abelian Berry curvature, and the prime on the sum over $n$ means that the sum is restricted to FBs polarized on one sublattice.
\begin{figure}[t]
     \centering
\includegraphics[scale=1]{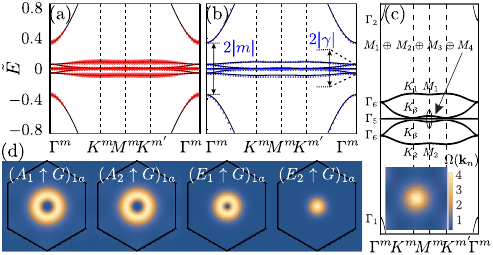}
     \caption{Topological heavy fermion model for the system with 6 flat bands in Fig.~\ref{fig:p6mm}(c).
     (a) The band structure at $\tilde{\alpha}=2.53$, which is a little deviated from the ``magic'' value $\tilde{\alpha}=2.93$, at which we have 6 exact FBs.  The overlaps between the Bloch states and the trial Wannier functions are represented by the red circles.
     (b) Band structure given by the topological heavy fermion model (blue crosses) compared to that of QBCP 6 nearly FBs full Hamiltonian (black lines).
     (c) The band structure of QBCP 6 nearly FBs model plotted within $\tilde{E}\in (-0.4,0.4)$. The irreps at HSMs are shown. Total Berry curvature distribution $\Omega(\mathbf{k}_n)$ (normalized by average Berry curvature) of the 3 FBs polarized on one sublattice is shown at the bottom left corner of (c).}
     \label{fig:hf}
\end{figure}

\textit{Topological heavy fermion (THF) model.}--Due to the antiunitary particle-hole symmetry $\mathcal{P}$ discussed above, any set of bands symmetric about the charge neutrality point is topological~\cite{TBGIIBernevig}. As a consequence, a tight binding description of these bands is never possible. However, in the case of TBG, due to the fact that the Berry curvature distribution is peaked at $\Gamma$ point in the mBZ, it was shown in~\cite{song2022magic} that hybridization of 2 atomic limit HF bands with 4 topological conduction bands (having nontrivial winding) at $\Gamma$ point can describe the 2 topological FBs of TBG. This THF model keeps all the relevant symmetries of TBG and captures the correct topology of the bands. We find that similar THF description of the high number of FBs discussed in this article is possible as long as the Berry curvature distribution has a pronounced peak around some point in the BZ. We show an example of this in Fig.~\ref{fig:hf} for the system (space group $G = p6mm$) with 6 FBs in Fig.~\ref{fig:p6mm}(c). The irreps of the 6 FBs at the HSMs are $\Gamma_5 \oplus 2\Gamma_6 - 2M_1 \oplus 2M_2 \oplus M_3 \oplus M_4 - K_1\oplus K_2 \oplus 2K_3$, which is not a linear combination of elementary band representations (EBR)~\cite{bradlyn2017topological}. On the other hand, the two lowest higher energy bands have representations $\Gamma_1$ and $\Gamma_2$ at $\Gamma$. Furthermore, replacement of one $\Gamma_6$ with $\Gamma_1\oplus\Gamma_2$ allows for band representation $BR = (A_1\uparrow G)_{1a}\oplus (A_2\uparrow G)_{1a} \oplus (E_1\uparrow G)_{1a} \oplus (E_2\uparrow G)_{1a}$~\cite{aroyo2011crystallography,aroyo2006bilbaoI,aroyo2006bilbaoII}. This and the fact that the Berry curvature distribution of the 6 FBs being peaked at $\Gamma$ (Fig~\ref{fig:hf}(c)) suggests a THF model composed of local orbitals having band representation $BR$ and topological conduction bands (we will refer to them as $c$-bands) with representation $\Gamma_6$. Thus we construct 6 Gaussian-type trial Wannier function that transform as $A_1$, $A_2$, $E_1$ and $E_2$ representations of PG $C_{6v}$, and calculate the overlaps of them with the 6 FB WFs. As shown in Fig.~\ref{fig:hf}(a), the Wannier functions are fully supported by the 6 FBs everywhere in the BZ except near the $\Gamma$ point, where the $A_1$ and $A_2$ type Wannier functions are supported by the lowest higher energy bands. We obtain the maximally localized Wannier functions (we will refer to them as $f$-orbitals) using the program Wannier90~\cite{marzari1997maximally,souza2001maximally,pizzi2020wannier90} ({\color{red}see SM~\cite{SM2022} S-VIII for details}), these MLWFs are extremely localized as can be seen in Fig.~\ref{fig:hf}(d). Projecting the Hamiltonian into the space spanned by $f$-orbitals and the remaining two $c$-bands, we obtain the following single-particle Hamiltonian:
\begin{align}
     &\hat{\mathcal{H}} = \sum_{|\mathbf{k}|< \Lambda_c} H^{(c)}_{ab}(\mathbf{k}) c^\dagger_a(\mathbf{k}) c_b(\mathbf{k}) + \sum_{\mathbf{R}} H^{(f)}_{\alpha\beta} f^\dagger_\alpha(\mathbf{R}) f_\beta(\mathbf{R})+\nonumber\\
     &\phantom{\hat{\mathcal{H}}}\sum_{|\mathbf{k}|< \Lambda_c,\mathbf{R}} \left(H^{(cf)}_{a\alpha}(\mathbf{k}) e^{-i\mathbf{k}\cdot\mathbf{R}-|\mathbf{k}|^2\lambda^2/2}c^\dagger_a(\mathbf{k})f_\alpha(\mathbf{R})+\text{h.c.}\right),\nonumber\\
     &H^{(c)}(\mathbf{k}) \approx c_1 (k_x^2-k_y^2)\sigma_x - 2c_1 k_xk_y\sigma_y,\nonumber\\
     &H^{(f)} \approx \begin{pmatrix}
         m\sigma_z & \mathbf{0}_{2\times 4}\\
         \mathbf{0}_{4\times 2} & \mathbf{0}_{4\times 4}
     \end{pmatrix},\nonumber\\
     &H^{(cf)}(\mathbf{k}) \approx \begin{pmatrix}
         -ic_2(k_x-ik_y) & c_2(k_x-ik_y) & 0 & \gamma & 0 & 0\\
         -ic_2(k_x+ik_y) & -c_2(k_x+ik_y) & \gamma & 0 & 0 & 0\\
     \end{pmatrix}
\end{align}
where $\alpha \in \{A_1,A_2, E_1^1,E_1^2,E_2^1,E_2^2\}$ goes over the $f$-orbitals, $m$ and $\gamma$ set the gap between $\Gamma_1$-$\Gamma_2$ and $\Gamma_6$-$\Gamma_6$ irreps, respectively (see Fig.~\ref{fig:hf}(b)), $c_1$ and $c_2$ are real constants. The spectrum of $\hat{\mathcal{H}}$ matches very well with that of the full Hamiltonian (Fig.~\ref{fig:hf}(b)).

\textit{Conclusion.}--In this letter, we systematically studied single and bilayer continuum Hamiltonians under moir\'e periodic modulations, and classified all possible number of exact flat bands that can appear in systems with different point group symmetries, and examined their topology. We not only showed that all single and bilayer systems in literature (to our knowledge) fall into this classification, but also found several new examples of model systems that can host high number of flat bands as well as high Chern number. In the future, it would be interesting to study the interplay between interactions and high number of flat bands that can enrich the correlated states in these systems.

\noindent \textit{Acknowledgements}.--This work was supported in part by Air Force Office of Scientific Research MURI FA9550-23-1-0334 and the Office of Naval Research MURI N00014-20-1-2479 (XW, SS and KS) and Award N00014-21-1-2770 (XW and KS), and by the Gordon and Betty Moore Foundation Award N031710 (KS). The work at LANL (SZL) was carried out under the auspices of the U.S. DOE NNSA under contract No. 89233218CNA000001 through the LDRD Program, and was supported by the Center for Nonlinear Studies at LANL, and was performed, in part, at the Center for Integrated Nanotechnologies, an Office of Science User Facility operated for the U.S. DOE Office of Science, under user proposals $\#2018BU0010$ and $\#2018BU0083$.

\let\oldaddcontentsline\addcontentsline
\renewcommand{\addcontentsline}[3]{}
\bibliographystyle{apsrev4-1}
\bibliography{ref}
\pagebreak
\let\addcontentsline\oldaddcontentsline
\onecolumngrid

\makeatletter
\renewcommand \thesection{S-\@arabic\c@section}
\renewcommand\thetable{S\@arabic\c@table}
\renewcommand \thefigure{S\@arabic\c@figure}
\renewcommand \theequation{S\@arabic\c@equation}
\makeatother
\setcounter{equation}{0}  
\setcounter{figure}{0}  
\setcounter{section}{0}  

{
    \center \bf \large 
    Supplemental Material\vspace*{0.1cm}\\ 
    \vspace*{0.0cm}
}
\maketitle
\tableofcontents
\let\oldsection\section
\renewcommand\section{\clearpage\oldsection}
\section{Continuum chiral symmetric moir\'e Hamiltonian and symmetries} 
\subsection{$\mathbf{k}\cdot\mathbf{p}$ Hamiltonian near a Dirac or quadratic band crossing point}\label{sec:k.p} 
A Dirac point (DP) or quadratic band crossing point (QBCP) (at a high symmetry momentum (HSM) $\mathbf{k}_0$) is a two fold degeneracy in the band structure. Let the corresponding eigenstates be $|\mathbf{k}_0,A\rangle$ and $|\mathbf{k}_0,B\rangle$. In case of graphene, $A$ and $B$ are the sublattices. However, in general case, they denote the two eigenvectors. These degeneracies are generally protected by $\mathcal{C}_{nz}$ ($n \in \{3,4\}$) and an antiunitary symmetry $\mathcal{A}$, which in case of QBCP is time reversal symmetry $\mathcal{T}$ and in case of DP is $\mathcal{C}_{2z}\mathcal{T}$ ({\color{red} note that here and elsewhere we will use the letter $n$ for $n$-fold rotation symmetry $\mathcal{C}_{nz}$ that protects the DP or QBCP. If there are other rotation symmetries in a particular system, we will use $n'$ or write it explicitly}). The eigenstates $|\mathbf{k}_0,A\rangle$ and $|\mathbf{k}_0,B\rangle$ transform as 2-dimensional (co-)irreducible representation (irrep) under these two symmetries. Furthermore, we will restrict ourselves to chiral ($\mathcal{S}$) symmetric limit. Here and in the rest of the article, we use the following gauge of the two eigenstates: 
\begin{equation} 
\begin{split} 
\mathcal{C}_{nz}\{|\mathbf{k}_0,A\rangle,|\mathbf{k}_0,B\rangle\} &= \{|\mathbf{k}_0,A\rangle,|\mathbf{k}_0,B\rangle\} \rho(\mathcal{C}_{nz}),\\ 
\mathcal{A}\{|\mathbf{k}_0,A\rangle,|\mathbf{k}_0,B\rangle\} &= \{|\mathbf{k}_0,A\rangle,|\mathbf{k}_0,B\rangle\} \rho(\mathcal{A}),\\ 
\mathcal{S}\{|\mathbf{k}_0,A\rangle,|\mathbf{k}_0,B\rangle\} &= \{|\mathbf{k}_0,A\rangle,|\mathbf{k}_0,B\rangle\} \rho(\mathcal{S}),\\ 
\end{split} 
\end{equation} 
where 
\begin{equation} 
\rho(\mathcal{C}_{nz}) = \begin{pmatrix} \omega & 0\\ 0 & \omega^*\end{pmatrix}, \rho(\mathcal{A}) = \sigma_x, \rho(\mathcal{S}) = \sigma_z, 
\end{equation} 
where $\omega = \exp(i2\pi (m+3)/n)$ and $m$ is an integer which takes values $m=1$ for DP and $m=2$ for QBCP and $n \in \{3,4\}$. Near the HSM, we choose similar gauge: 
\begin{equation} 
\begin{split} 
\mathcal{C}_{nz}\{|\mathbf{k}_0+\mathbf{k},A\rangle,|\mathbf{k}_0+\mathbf{k},B\rangle\} &= \{|\mathbf{k}_0+ \mathcal{C}_{nz} \mathbf{k},A\rangle,|\mathbf{k}_0+\mathcal{C}_{nz} \mathbf{k},B\rangle\} \rho(\mathcal{C}_{nz}),\\ 
\mathcal{A}\{|\mathbf{k}_0+ \mathbf{k},A\rangle,|\mathbf{k}_0+ \mathbf{k},B\rangle\} &= \{|\mathbf{k}_0+\mathcal{A}\mathbf{k},A\rangle,|\mathbf{k}_0+\mathcal{A} \mathbf{k},B\rangle\} \rho(\mathcal{A}),\\ 
\mathcal{S}\{|\mathbf{k}_0+ \mathbf{k},A\rangle,|\mathbf{k}_0+ \mathbf{k},B\rangle\} &= \{|\mathbf{k}_0+\mathbf{k},A\rangle,|\mathbf{k}_0+\mathbf{k},B\rangle\} \rho(\mathcal{S}),\\ 
\end{split} 
\end{equation} 
where $\mathcal{A}\mathbf{k} = \pm \mathbf{k}$ for $\mathcal{A} = \mathcal{C}_{2z}\mathcal{T}$ and $\mathcal{A} = \mathcal{T}$ respectively. Note that the HSM $\mathbf{k}_0$ may have some extra symmetries such as mirror reflection $\mathcal{M}$ and/or two-fold rotation symmetry $\mathcal{C}_{2z}$. For mirror symmetry, since $[\mathcal{M}, \mathcal{C}_{nz}]\neq 0$ ($n\in \{3,4\}$), $[\mathcal{M}, \mathcal{A}] = 0$ and $\mathcal{M}^2 = \mathds{1}$, the only representation it can have is $\rho(\mathcal{M}) = \sigma_x$. On the other hand, $\mathcal{C}_{2z}$ satisfies $[\mathcal{C}_{2z}, \mathcal{C}_{nz}] = 0$ ($n\in \{3,4\}$), $[\mathcal{C}_{2z}, \mathcal{A}] = 0$ and $\mathcal{C}_{2z}^2 = \mathds{1}$. Hence, representation of $\mathcal{C}_{2z}$ can only be $\pm \mathds{1}$. In four-fold rotation symmetric case it is necessarily $\rho(\mathcal{C}_{2z}) = \rho^2(\mathcal{C}_{4z}) = -\mathds{1}$. In $6$-fold rotation symmetric case there are two choices: $\mathcal{C}_{2z}= \mathds{1}$ (example: $\Gamma_5$ representation in $p6mm$ space group; here we are following the notation given in Bilbao Crystallography Server~\citesupp{aroyo2011crystallographys,aroyo2006bilbaoIs,aroyo2006bilbaoIIs}) and $\mathcal{C}_{2z}= -\mathds{1}$ (example: $\Gamma_6$ representation in $p6mm$ space group).

Then, the $\mathbf{k}\cdot\mathbf{p}$ Hamiltonian near HSM $\mathbf{k}_0$ is given by: 
\begin{equation} 
\hat{H} = \Omega\sum_\mathbf{k} \{|\mathbf{k}_0+\mathbf{k},A\rangle,|\mathbf{k}_0+\mathbf{k},B\rangle\} \mathcal{H}(\mathbf{k}) \{\langle\mathbf{k}_0+\mathbf{k},A|,\langle\mathbf{k}_0+\mathbf{k},B|\}^T. 
\end{equation} 
where $\Omega$ is the volume of the system. Since the Hamiltonian invariant under $\mathcal{C}_{nz}$ and $\mathcal{A}$, we have 
\begin{equation} 
\begin{split} 
&\mathcal{C}_{nz} \hat{H}\mathcal{C}_{nz}^{-1} = \Omega\sum_\mathbf{k} \mathcal{C}_{nz}\{|\mathbf{k}_0+\mathbf{k},A\rangle,|\mathbf{k}_0+\mathbf{k},B\rangle\} \mathcal{H}(\mathbf{k}) \{\langle\mathbf{k}_0+\mathbf{k},A|,\langle\mathbf{k}_0+\mathbf{k},B|\}^T\mathcal{C}_{nz}^{-1}\\ 
&\phantom{\mathcal{C}_{nz} \hat{H}\mathcal{C}_{nz}^{-1} }= \Omega\sum_\mathbf{k} \{|\mathbf{k}_0+\mathcal{C}_{nz}\mathbf{k},A\rangle,|\mathbf{k}_0+\mathcal{C}_{nz}\mathbf{k},B\rangle\} \rho(\mathcal{C}_{nz})\mathcal{H}(\mathbf{k}) \rho(\mathcal{C}_{nz})^\dagger \{\langle\mathbf{k}_0+\mathcal{C}_{nz}\mathbf{k},A|,\langle\mathbf{k}_0+\mathcal{C}_{nz}\mathbf{k},B|\}^T\\ 
&\phantom{\mathcal{C}_{nz} \hat{H}\mathcal{C}_{nz}^{-1} }= \Omega\sum_\mathbf{k} \{|\mathbf{k}_0+\mathbf{k},A\rangle,|\mathbf{k}_0+\mathbf{k},B\rangle\} \rho(\mathcal{C}_{nz})\mathcal{H}(\mathcal{C}_{nz}^{-1}\mathbf{k}) \rho(\mathcal{C}_{nz})^\dagger \{\langle\mathbf{k}_0+\mathbf{k},A|,\langle\mathbf{k}_0+\mathbf{k},B|\}^T\\ 
&\phantom{\mathcal{C}_{nz} \hat{H}\mathcal{C}_{nz}^{-1} }=\hat{H},\\ 
\Rightarrow & \rho(\mathcal{C}_{nz})\mathcal{H}(\mathbf{k}) \rho(\mathcal{C}_{nz})^\dagger= \mathcal{H}(\mathcal{C}_{nz}\mathbf{k}). 
\end{split} 
\end{equation} 
Similarly, we also have 
\begin{equation} 
\begin{split} 
\rho(\mathcal{A})\mathcal{H}^*(\mathbf{k}) \rho(\mathcal{A})^\dagger= \mathcal{H}(\mathcal{A}\mathbf{k}),\\ 
\end{split} 
\end{equation} 
where the complex conjugation $^*$ is due to antiunitarity of $\mathcal{A}$. Furthermore, since $\hat{H}$ anticommutes with chiral symmetry $\mathcal{S}$, we have 
\begin{equation} 
\begin{split} 
\rho(\mathcal{S})\mathcal{H}(\mathbf{k}) \rho(\mathcal{S})^\dagger= -\mathcal{H}(\mathbf{k}).\\ 
\end{split} 
\end{equation} 
These symmetries together imply that to the lowest order the Hamiltonian matrix has the following form: 
\begin{equation} 
\begin{split} 
\mathcal{H}_\text{DP}(\mathbf{k}) &= \begin{pmatrix}0 & c^* k^*\\ ck & 0\end{pmatrix},\\ 
\mathcal{H}_\text{QBCP}(\mathbf{k}) &= \begin{pmatrix}0 & c_1^* (k_x^2-k_y^2)-2ic_2^*k_xk_y\\ c_1 (k_x^2-k_y^2)+2ic_2k_xk_y& 0\end{pmatrix}, 
\end{split} 
\end{equation} 
where $k = k_x+ik_y$, and $c$, $c_1$ and $c_2$ are complex constants. Note that for QBCPs protected by $\mathcal{C}_{3z}$ or $\mathcal{C}_{6z}$, $c_1 = c_2$ making the Hamiltonian isotropic. However, for QBCP protected by $\mathcal{C}_{4z}$, generically $c_1 \neq c_2$. But, in this article, we will consider assume isotropic limit for all cases. Then, the complex phase in $c = |c|e^{i\varphi}$ and $c_1=|c_1|e^{i\varphi}$ can be absorbed by rotating the coordinate system $k \rightarrow k e^{-i\varphi}$ and $k \rightarrow k e^{-i\varphi/2}$ in case of DP and QBCP, respectively. Then dropping the multiplicative constant $|c|$ and $|c_1|$, we obtain the following form of the Hamiltonian: 
\begin{equation} 
\mathcal{H}_m(\mathbf{k}) = \begin{pmatrix}0 & (k^m)^*\\ k^m & 0\end{pmatrix}, 
\end{equation} 
with $m=1$ and $m=2$ for DP and QBCP, respectively. 

Next, we perform Fourier transform on the basis states $|\mathbf{k}_0+\mathbf{k},A\rangle$ and $|\mathbf{k}_0+\mathbf{k},B\rangle$: 
\begin{equation} 
\begin{split} 
|\mathbf{k}_0,\mathbf{r},\alpha\rangle &= \sum_\mathbf{k} e^{-i\mathbf{k}\cdot\mathbf{r}} |\mathbf{k}_0+\mathbf{k},\alpha\rangle,\\ 
|\mathbf{k}_0+\mathbf{k},\alpha\rangle &=\frac{1}{\Omega} \int d^2\mathbf{r} e^{i\mathbf{k}\cdot\mathbf{r}} |\mathbf{k}_0,\mathbf{r},\alpha\rangle\text{ }\alpha \in \{A,B\}. 
\end{split} 
\end{equation} 
Note that under lattice translations $\mathbf{a}_1$ and $\mathbf{a}_2$ the state $|\mathbf{k}_0+\mathbf{k},\alpha\rangle$ transforms as $T_{\mathbf{a}_i}|\mathbf{k}_0+\mathbf{k},\alpha\rangle = e^{-i(\mathbf{k}_0+\mathbf{k})\cdot\mathbf{a}_i}|\mathbf{k}_0+\mathbf{k},\alpha\rangle$. This implies 
\begin{equation} 
\begin{split} 
T_{\mathbf{a}_i}|\mathbf{k}_0,\mathbf{r},\alpha\rangle &= \sum_\mathbf{k} e^{-i\mathbf{k}\cdot\mathbf{r}} T_{\mathbf{a}_i}|\mathbf{k}_0+\mathbf{k},\alpha\rangle\\ 
&= \sum_\mathbf{k} e^{-i\mathbf{k}\cdot\mathbf{r}}e^{-i(\mathbf{k}_0+\mathbf{k})\cdot\mathbf{a}_i} |\mathbf{k}_0+\mathbf{k},\alpha\rangle\\ 
&=e^{-i\mathbf{k}_0\cdot\mathbf{a}} |\mathbf{k}_0,\mathbf{r}+\mathbf{a},\alpha\rangle 
\end{split} 
\end{equation} 
In this basis, the Hamiltonian becomes 
\begin{equation} 
\begin{split} 
\hat{H} &=\Omega\sum_\mathbf{k} \{|\mathbf{k}_0+\mathbf{k},A\rangle,|\mathbf{k}_0+\mathbf{k},B\rangle\} \mathcal{H}_m(\mathbf{k}) \{\langle\mathbf{k}_0+\mathbf{k},A|,\langle\mathbf{k}_0+\mathbf{k},B|\}^T\\ 
&= \frac{1}{\Omega} \sum_\mathbf{k} \int d^2\mathbf{r}\int d^2\mathbf{r}' \{|\mathbf{k}_0,\mathbf{r},A\rangle,|\mathbf{k}_0,\mathbf{r},B\rangle\} e^{i\mathbf{k}\cdot\mathbf{r}}\mathcal{H}_m(\mathbf{k}) e^{-i\mathbf{k}\cdot\mathbf{r}'} \{\langle\mathbf{k}_0,\mathbf{r}',A|,\langle\mathbf{k}_0,\mathbf{r}',B|\}^T\\ 
&= \frac{1}{\Omega} \sum_\mathbf{k} \int d^2\mathbf{r}\int d^2\mathbf{r}' \{|\mathbf{k}_0,\mathbf{r},A\rangle,|\mathbf{k}_0,\mathbf{r},B\rangle\} e^{i\mathbf{k}\cdot(\mathbf{r}-\mathbf{r}')}\mathcal{H}_m(\mathbf{r}') \{\langle\mathbf{k}_0,\mathbf{r}',A|,\langle\mathbf{k}_0,\mathbf{r}',B|\}^T\\ 
&= \frac{1}{\Omega}\int d^2\mathbf{r}\int d^2\mathbf{r}' \{|\mathbf{k}_0,\mathbf{r},A\rangle,|\mathbf{k}_0,\mathbf{r},B\rangle\} \Omega \delta^{(2)}(\mathbf{r}-\mathbf{r}')\mathcal{H}_m(\mathbf{r}') \{\langle\mathbf{k}_0,\mathbf{r}',A|,\langle\mathbf{k}_0,\mathbf{r}',B|\}^T\\ 
&= \int d^2\mathbf{r} \{|\mathbf{k}_0,\mathbf{r},A\rangle,|\mathbf{k}_0,\mathbf{r},B\rangle\} \mathcal{H}_m(\mathbf{r}) \{\langle\mathbf{k}_0,\mathbf{r},A|,\langle\mathbf{k}_0,\mathbf{r},B|\}^T, 
\end{split} 
\end{equation} 
where $\delta^{(2)}(\mathbf{r})$ is the 2-dimensional Dirac delta function and 
\begin{equation} 
\mathcal{H}_m(\mathbf{r}) = \mathcal{H}_m(\mathbf{k} \rightarrow -i\boldsymbol{\nabla}) = \begin{pmatrix}0 & (-2i\partial_{z})^m\\ (-2i\overline{\partial_{z}})^m & 0\end{pmatrix}, 
\end{equation} 
where $\partial_z = (\partial_x -i\partial_y)/2$. Note that in this real space representation the Hamiltonian $\mathcal{H}_m(\mathbf{r})$ satisfies the following symmetry relations: 
\begin{equation} 
\label{eq:HRealSym1} 
\begin{split} 
\rho(\mathcal{C}_{nz})\mathcal{H}_m(\mathbf{r}) \rho(\mathcal{C}_{nz})^\dagger &= \mathcal{H}_m(\mathcal{C}_{nz}\mathbf{r}),\\ 
\rho(\mathcal{A})\mathcal{H}_m^*(\mathbf{r}) \rho(\mathcal{A})^\dagger &= \mathcal{H}_m(\mathcal{A}\mathbf{r}),\\ 
\rho(\mathcal{S})\mathcal{H}_m(\mathbf{r}) \rho(\mathcal{S})^\dagger &= -\mathcal{H}_m(\mathbf{r}),\\ 
\end{split} 
\end{equation} 
where $\mathcal{A}\mathbf{r} = \pm \mathbf{r}$ for $\mathcal{A} =\mathcal{T}$ and $\mathcal{A} =\mathcal{C}_{2z}\mathcal{T}$, respectively. If there are mirror or two fold rotation symmetries: 
\begin{equation} 
\label{eq:HRealSym2} 
\begin{split} 
\rho(\mathcal{C}_{2z})\mathcal{H}_m(\mathbf{r}) \rho(\mathcal{C}_{2z})^\dagger &= \mathcal{H}_m(\mathcal{C}_{2z}\mathbf{r}),\\ 
\rho(\mathcal{M})\mathcal{H}_m(\mathbf{r}) \rho(\mathcal{M})^\dagger &= \mathcal{H}_m(\mathcal{M}\mathbf{r}). 
\end{split} 
\end{equation} 
\subsection{Adding moir\'e potential and Brillouin zone folding} 
Adding a moir\'e potential whose periodicity (let the moir\'e lattice vectors be $\mathbf{a}_1^m$ and $\mathbf{a}_2^m$) is much larger than the periodicity of the lattice (let the lattice vectors be $\mathbf{a}_1$ and $\mathbf{a}_2$) makes the moir\'e Brillouin zone (mBZ) (with moir\'e reciprocal lattice vectors $\mathbf{b}_1^m$ and $\mathbf{b}_2^m$) much smaller than the BZ (with reciprocal lattice vectors $\mathbf{b}_1$ and $\mathbf{b}_2$) of the lattice. For simplicity, here we are going to consider the moir\'e periodicity to be commensurate with the lattice periodicity. The point $\mathbf{k}_0$ in the lattice BZ, gets mapped to some point $\mathbf{k}_0^m=\mathbf{k}_0-m_1\mathbf{b}_1^m-m_2\mathbf{b}_2^m$. We know that under translation by lattice vectors $n_1\mathbf{a}_1+n_2\mathbf{a}_2$, the basis states transform as $T_{n_1\mathbf{a}_1+n_2\mathbf{a}_2}|\mathbf{k}_0,\mathbf{r},\alpha\rangle = e^{-i\mathbf{k}_0\cdot(n_1\mathbf{a}_1+n_2\mathbf{a}_2)}|\mathbf{k}_0,\mathbf{r},\alpha\rangle$. Under translations by moir\'e lattice vectors $n_1\mathbf{a}_1^m+n_2\mathbf{a}_2^m$, they transform as 
\begin{equation} 
\label{eq:k0k0m} 
\begin{split} 
T_{n_1\mathbf{a}_1^m+n_2\mathbf{a}_2^m}|\mathbf{k}_0,\mathbf{r},\alpha\rangle &= e^{-i\mathbf{k}_0\cdot(n_1\mathbf{a}_1^m+n_2\mathbf{a}_2^m)}|\mathbf{k}_0,\mathbf{r},\alpha\rangle\\ 
&= e^{-i(\mathbf{k}_0-m_1\mathbf{b}_1^m-m_2\mathbf{b}_2^m)\cdot(n_1\mathbf{a}_1^m+n_2\mathbf{a}_2^m)}|\mathbf{k}_0,\mathbf{r},\alpha\rangle e^{-i(m_1\mathbf{b}_1^m+m_2\mathbf{b}_2^m)\cdot(n_1\mathbf{a}_1^m+n_2\mathbf{a}_2^m)}\\ 
&=e^{-i\mathbf{k}_0^m\cdot(n_1\mathbf{a}_1^m+n_2\mathbf{a}_2^m)}|\mathbf{k}_0,\mathbf{r},\alpha\rangle, 
\end{split} 
\end{equation} 
where in the first equality we used the fact that moir\'e lattice vectors are integer multiples of lattice vectors, and in last equality we used $\mathbf{b}_i^{m}.\mathbf{a}_j^m = 2\pi\delta_{ij}$. Therefore, under moir\'e translations they transform as basis states at momentum $\mathbf{k}_0^m$ in the mBZ. 

\subsection{Single layer system with QBCP under periodic potential} 
If we add a moir\'e potential to single layer QBCP Hamiltonian that does not break $\mathcal{C}_{nz}$, $\mathcal{A} =\mathcal{T}$ and $\mathcal{S}$, the general form of the Hamiltonian becomes: 
\begin{equation} 
\label{eq:QBCPS}
\mathcal{H}_S(\mathbf{r}) = \begin{pmatrix}0 &\mathcal{D}_k^\dagger(\mathbf{r})+\mathcal{D}_U^*(\mathbf{r};\boldsymbol{\alpha})\\ \mathcal{D}_k(\mathbf{r})+\mathcal{D}_U(\mathbf{r};\boldsymbol{\alpha}) & 0\end{pmatrix} =\begin{pmatrix}0 & (-2i\partial_{z})^m+\mathcal{D}_U^*(\mathbf{r};\boldsymbol{\alpha})\\ (-2i\overline{\partial_{z}})^m+\mathcal{D}_U(\mathbf{r};\boldsymbol{\alpha}) & 0\end{pmatrix}. 
\end{equation} 
Under moir\'e periodicities $\mathbf{a}_1^m$ and $\mathbf{a}_2^m$, the basis states $|\mathbf{k}_0,\mathbf{r},\alpha\rangle$ transform as $T_{\mathbf{a}_i^m}|\mathbf{k}_0,\mathbf{r},\alpha\rangle = e^{-i\mathbf{k}_0^m\cdot\mathbf{a}_i^m}|\mathbf{k}_0,\mathbf{r}+\mathbf{a}_i^m,\alpha\rangle $. Hence, the representation of the moir\'e translations are $\rho(\sum_{i=1}^2n_i\mathbf{a}_i^m) = e^{-i\mathbf{k}_0^m\cdot(\sum_{i=1}^2n_i\mathbf{a}_i^m)}\mathds{1}$, and $\mathcal{D}_U(\mathbf{r} + \mathbf{a}_i^m;\boldsymbol{\alpha}) = \mathcal{D}_U(\mathbf{r} ;\boldsymbol{\alpha}) $. Due to Eq.~\eqref{eq:HRealSym1}, $\mathcal{D}_U(\mathbf{r};\boldsymbol{\alpha})$ satisfies 
\begin{equation} 
\begin{split} 
\mathcal{D}_U(\mathcal{C}_{nz}\mathbf{r};\boldsymbol{\alpha}) = (\omega^*)^2 \mathcal{D}_U(\mathbf{r};\boldsymbol{\alpha}),\\ 
\end{split} 
\end{equation} 
and if additionally there are mirror and rotation symmetries in Eq.~\eqref{eq:HRealSym2}, then 
\begin{equation} 
\begin{split} 
\mathcal{D}_U(\mathcal{C}_{2z}\mathbf{r};\boldsymbol{\alpha}) = \mathcal{D}_U(\mathbf{r};\boldsymbol{\alpha}),\\ 
\mathcal{D}_U(\mathcal{M}\mathbf{r};\boldsymbol{\alpha}) = \mathcal{D}_U^*(\mathbf{r};\boldsymbol{\alpha}).\\ 
\end{split} 
\end{equation} 
A composition of chiral and time reversal symmetry in these system gives a anti-unitary particle-hole symmetry $\mathcal{P} = \mathcal{S}\mathcal{T}$, its representation is $\rho(\mathcal{P}) = \rho(\mathcal{S})\rho(\mathcal{T}) = \sigma_y$, hence $\mathcal{P}^2 =-\mathds{1}$ since $\rho(\mathcal{P}) \rho(\mathcal{P}) ^*=\sigma_y\sigma_y^* = -\mathds{1}$. 
\subsection{Twisted bilayer systems with DP or QBCP} 
In twisted bilayer systems, two layers are twisted from each other by small angle $\theta$. That creates an offset between the HSMs of the two layers $\mathbf{k}_0^{(1)} = \mathbf{R}(-\theta/2)\mathbf{k}_0$ and $\mathbf{k}_0^{(2)}= \mathbf{R}(\theta/2)\mathbf{k}_0$: $\mathbf{k}_0^{(1)} - \mathbf{k}_0^{(2)} = \mathbf{q}_1 \approx \mathbf{k}_0\times\theta\hat{z}$ (here the numbers in the superscripts denote the layer index). In case of TBG, $\mathbf{q}_1 =-(\mathbf{b}_1^m+\mathbf{b}_2^m)/3$ in the notation specified in {\color{red}Fig.~2} of the main text. In case of twisted bilayer checkerboard lattice, $\mathbf{q}_1 =(\mathbf{b}_1^m-\mathbf{b}_2^m)/2$ in the notation specified in Fig.~\ref{fig:TBCL}. The general form of the twisted bilayer system in the chiral limit is 
\begin{equation} 
\hat{H}_{TB} = \int d^2\mathbf{r}\{|\mathbf{k}_0^{(1)},\mathbf{r},A\rangle, |\mathbf{k}_0^{(1)},\mathbf{r},B\rangle,|\mathbf{k}_0^{(2)},\mathbf{r},A\rangle,|\mathbf{k}_0^{(2)},\mathbf{r},B\rangle \}\mathcal{H}_{TB}(\mathbf{r}) \{\langle\mathbf{k}_0^{(1)},\mathbf{r},A|, \langle\mathbf{k}_0^{(1)},\mathbf{r},B|,\langle\mathbf{k}_0^{(2)},\mathbf{r},A|,\langle\mathbf{k}_0^{(2)},\mathbf{r},B| \}^T, 
\end{equation} 
with 
\begin{equation} 
\mathcal{H}_{TB}(\mathbf{r}) = \left(\begin{array}{c|c} \mathcal{H}_m(\mathbf{R}(-\theta/2)\mathbf{r}) & \begin{array}{cc}0 & U_2(\mathbf{r})\\U_1(\mathbf{r}) & 0\end{array}\\ \hline \begin{array}{cc}0 & U_1^*(\mathbf{r})\\U_2^*(\mathbf{r}) & 0 \end{array} & \mathcal{H}_m(\mathbf{R}(\theta/2)\mathbf{r})\end{array}\right), 
\end{equation} 
where the zeros in the off-diagonal are due to chiral symmetry, which has representation $\rho(\mathcal{S}) = \mathds{1}\otimes \sigma_{z}$ in this basis. Note that $ \mathcal{H}_m(\mathbf{R}(-\theta/2)\mathbf{r}) = e^{i m\theta\sigma_z/4} \mathcal{H}_m(\mathbf{r}) e^{-i m\theta\sigma_z/4}$. Then transforming $\splitatcommas{H_{TB}(\mathbf{r}) \rightarrow \text{Diag}\{e^{-i m\theta\sigma_z/4},e^{i m\theta\sigma_z/4}\} H_{TB}(\mathbf{r}) \text{Diag}\{e^{i m\theta\sigma_z/4},e^{-i m\theta\sigma_z/4}\}}$, we obtain 
\begin{equation} 
\mathcal{H}_{TB}(\mathbf{r}) = \left(\begin{array}{c|c} \mathcal{H}_m(\mathbf{r}) & \begin{array}{cc}0 & U_2(\mathbf{r})\\U_1(\mathbf{r}) & 0\end{array}\\ \hline \begin{array}{cc}0 & U_1^*(\mathbf{r})\\U_2^*(\mathbf{r}) & 0 \end{array} & \mathcal{H}_m(\mathbf{r})\end{array}\right). 
\end{equation} 
Note that the last transformation does not change the off-diagonal part because $\{\sigma_z,\sigma_{x/y}\} = 0$. Furthermore, the last transformation changes the basis; however, it can be seen straightforwardly that the representations of the relevant symmetries remain the same. It is customary in the literature to change the order of the basis states to write the Hamiltonian in the following form: 
\begin{equation} 
\hat{H}_{TB} = \int d^2\mathbf{r}\{|\mathbf{k}_0^{(1)},\mathbf{r},A\rangle, |\mathbf{k}_0^{(2)},\mathbf{r},A\rangle,|\mathbf{k}_0^{(1)},\mathbf{r},B\rangle,|\mathbf{k}_0^{(2)},\mathbf{r},B\rangle \}\mathcal{H}_{TB}(\mathbf{r}) \{\langle\mathbf{k}_0^{(1)},\mathbf{r},A|, \langle\mathbf{k}_0^{(2)},\mathbf{r},A|,\langle\mathbf{k}_0^{(1)},\mathbf{r},B|,\langle\mathbf{k}_0^{(2)},\mathbf{r},B| \}^T, 
\end{equation} 
with 
\begin{equation} 
\label{eq:HTB}
\mathcal{H}_{TB}(\mathbf{r}) = \begin{pmatrix} 
0 & 0 & (-2i\partial_z)^m & U_2(\mathbf{r})\\ 
0 & 0 & U_1^*(\mathbf{r}) & (-2i\partial_z)^m\\ 
(-2i\overline{\partial_{z}})^m & U_1(\mathbf{r}) & 0 & 0\\ 
U_2^*(\mathbf{r}) & (-2i\overline{\partial_{z}})^m\ & 0 & 0\ 
\end{pmatrix} 
\end{equation} 
In this basis the representation of the symmetries are $\rho(\mathcal{C}_{nz}) = \text{Diag}\{\omega,\omega,\omega^*,\omega^*\}$, $\rho(\mathcal{A}) =\sigma_x\otimes\mathds{1}$, $\rho(\mathcal{S}) = \sigma_z\otimes\mathds{1}$. The symmetries $\mathcal{C}_{nz}$ and $\mathcal{A}$ put constraints on $U_1(\mathbf{r})$ and $U_2(\mathbf{r})$: 
\begin{equation} 
\label{eq:Urotation} 
\begin{split} 
U_1(\mathcal{A}\mathbf{r}) &= U_2^*(\mathbf{r})\\ 
U_1(\mathcal{C}_{nz}\mathbf{r}) &= (\omega^*)^2 U_1(\mathbf{r}) 
\end{split} 
\end{equation} 
There are further constraints on $U_i(\mathbf{r})$ from moir\'e translations. The basis states for the $\mathbf{k}\cdot\mathbf{p}$ theory of each layer satisfy: $T_{\mathbf{a}_i} |\mathbf{k}_0^{(l)},\mathbf{r},\alpha\rangle = e^{-i\mathbf{k}_0^{(l)}\cdot\mathbf{a}_i}|\mathbf{k}_0^{(l)},\mathbf{r}+\mathbf{a}_i,\alpha\rangle$. Then, under a moir\'e transaltion $\mathbf{t}^m =\sum_{i=1}^{2} n_i\mathbf{a}_i^m = n_1\mathbf{a}_1^m+n_2\mathbf{a}_2^m$, the basis states transform as 
\begin{equation} 
\begin{split} 
&T_{\sum_{i=1}^{2} n_i\mathbf{a}_i^m}\{|\mathbf{k}_0^{(1)},\mathbf{r},A\rangle, |\mathbf{k}_0^{(2)},\mathbf{r},A\rangle, |\mathbf{k}_0^{(1)},\mathbf{r},B\rangle,|\mathbf{k}_0^{(2)},\mathbf{r},B\rangle\}\\ 
=& \{|\mathbf{k}_0^{(1)},\mathbf{r}+\sum_{i=1}^{2} n_i\mathbf{a}_i^m,A\rangle, |\mathbf{k}_0^{(2)},\mathbf{r}+\sum_{i=1}^{2} n_i\mathbf{a}_i^m,A\rangle, |\mathbf{k}_0^{(1)},\mathbf{r}+\sum_{i=1}^{2} n_i\mathbf{a}_i^m,B\rangle,|\mathbf{k}_0^{(2)},\mathbf{r}+\sum_{i=1}^{2} n_i\mathbf{a}_i^m,B\rangle\}\rho(T_{\sum_{i=1}^{2} n_i\mathbf{a}_i^m}) 
\end{split} 
\end{equation} 
where 
\begin{equation} 
\begin{split} 
\rho(T_{\sum_{i=1}^{2} n_i\mathbf{a}_i^m}) &= \text{Diag}\{e^{-i\mathbf{k}_0^{(1)}\cdot(\sum_{i=1}^{2} n_i\mathbf{a}_i^m)},e^{-i\mathbf{k}_0^{(2)}\cdot(\sum_{i=1}^{2} n_i\mathbf{a}_i^m)},e^{-i\mathbf{k}_0^{(1)}\cdot(\sum_{i=1}^{2} n_i\mathbf{a}_i^m)},e^{-i\mathbf{k}_0^{(2)}\cdot(\sum_{i=1}^{2} n_i\mathbf{a}_i^m)}\}\\ 
&= e^{-i\mathbf{k}_0^{(1)}\cdot(\sum_{i=1}^{2} n_i\mathbf{a}_i^m)} \text{Diag}\{1,e^{i(\mathbf{k}_0^{(1)}-\mathbf{k}_0^{(2)})\cdot(\sum_{i=1}^{2} n_i\mathbf{a}_i^m)},1,e^{i(\mathbf{k}_0^{(1)}-\mathbf{k}_0^{(2)})\cdot(\sum_{i=1}^{2} n_i\mathbf{a}_i^m)}\}\\ 
&= e^{-i\mathbf{k}_0^{(1)}\cdot(\sum_{i=1}^{2} n_i\mathbf{a}_i^m)} \text{Diag}\{1,e^{i\mathbf{q}_1\cdot(\sum_{i=1}^{2} n_i\mathbf{a}_i^m)},1,e^{i\mathbf{q}_1\cdot(\sum_{i=1}^{2} n_i\mathbf{a}_i^m)}\}\\
&= e^{-i\mathbf{k}_0^m\cdot(\sum_{i=1}^{2} n_i\mathbf{a}_i^m)} \text{Diag}\{1,e^{i\mathbf{q}_1\cdot(\sum_{i=1}^{2} n_i\mathbf{a}_i^m)},1,e^{i\mathbf{q}_1\cdot(\sum_{i=1}^{2} n_i\mathbf{a}_i^m)}\}
\end{split} 
\end{equation} 
where we define $\mathbf{k}_0^m$ as the mapping of $\mathbf{k}_0^{(1)}$ in the mBZ after the folding of the BZ. Then, the Hamiltonian of the twisted bilayer transforms as 
\begin{equation} 
\begin{split} 
&T_{\sum_{i=1}^{2} n_i\mathbf{a}_i^m}\hat{H}_{TB}T_{\sum_{i=1}^{2} n_i\mathbf{a}_i^m}^{-1} \\ 
=& \int d^2\mathbf{r}\,T_{\sum_{i=1}^{2} n_i\mathbf{a}_i^m}\{|\mathbf{k}_0^{(1)},\mathbf{r},A\rangle, |\mathbf{k}_0^{(2)},\mathbf{r},A\rangle, |\mathbf{k}_0^{(1)},\mathbf{r},B\rangle,|\mathbf{k}_0^{(2)},\mathbf{r},B\rangle\} \mathcal{H}_{TB}(\mathbf{r})\times \\ 
&\hspace{3cm}\{\langle\mathbf{k}_0^{(1)},\mathbf{r},A|, \langle\mathbf{k}_0^{(1)},\mathbf{r},B|,\langle\mathbf{k}_0^{(2)},\mathbf{r},A|,\langle\mathbf{k}_0^{(2)},\mathbf{r},B| \}^T T_{\sum_{i=1}^{2} n_i\mathbf{a}_i^m}^{-1}\\ 
=& \int d^2\mathbf{r}\,\{|\mathbf{k}_0^{(1)},\mathbf{r}+\sum_{i=1}^{2} n_i\mathbf{a}_i^m,A\rangle, |\mathbf{k}_0^{(2)},\mathbf{r}+\sum_{i=1}^{2} n_i\mathbf{a}_i^m,A\rangle, |\mathbf{k}_0^{(1)},\mathbf{r}+\sum_{i=1}^{2} n_i\mathbf{a}_i^m,B\rangle,|\mathbf{k}_0^{(2)},\mathbf{r}+\sum_{i=1}^{2} n_i\mathbf{a}_i^m,B\rangle\}\times\\ 
&\hspace{3cm}\rho(T_{\sum_{i=1}^{2} n_i\mathbf{a}_i^m})\mathcal{H}_{TB}(\mathbf{r}) \rho^\dagger(T_{\sum_{i=1}^{2} n_i\mathbf{a}_i^m})\times\\ 
&\hspace{3cm}\{\langle\mathbf{k}_0^{(1)},\mathbf{r}+\sum_{i=1}^{2} n_i\mathbf{a}_i^m,A|, \langle\mathbf{k}_0^{(1)},\mathbf{r}+\sum_{i=1}^{2} n_i\mathbf{a}_i^m,B|,\langle\mathbf{k}_0^{(2)},\mathbf{r}+\sum_{i=1}^{2} n_i\mathbf{a}_i^m,A|,\langle\mathbf{k}_0^{(2)},\mathbf{r}+\sum_{i=1}^{2} n_i\mathbf{a}_i^m,B| \}^T \\ 
=& \int d^2\mathbf{r}\,\{|\mathbf{k}_0^{(1)},\mathbf{r},A\rangle, |\mathbf{k}_0^{(2)},\mathbf{r},A\rangle, |\mathbf{k}_0^{(1)},\mathbf{r},B\rangle,|\mathbf{k}_0^{(2)},\mathbf{r},B\rangle\}\times\\ 
&\hspace{3cm}\rho(T_{\sum_{i=1}^{2} n_i\mathbf{a}_i^m})\mathcal{H}_{TB}(\mathbf{r}-\sum_{i=1}^{2} n_i\mathbf{a}_i^m) \rho^\dagger(T_{\sum_{i=1}^{2} n_i\mathbf{a}_i^m})\{\langle\mathbf{k}_0^{(1)},\mathbf{r},A|, \langle\mathbf{k}_0^{(1)},\mathbf{r},B|,\langle\mathbf{k}_0^{(2)},\mathbf{r},A|,\langle\mathbf{k}_0^{(2)},\mathbf{r},B| \}^T \\ 
&=\hat{H}_{TB}\\ 
\Rightarrow & \rho(T_{\sum_{i=1}^{2} n_i\mathbf{a}_i^m})\mathcal{H}_{TB}(\mathbf{r}) \rho^\dagger(T_{\sum_{i=1}^{2} n_i\mathbf{a}_i^m}) = \mathcal{H}_{TB}(\mathbf{r}+\sum_{i=1}^{2} n_i\mathbf{a}_i^m). 
\end{split} 
\end{equation} 
This puts the following constraints on the moir\'e potentials 
\begin{equation} 
\label{eq:Utranslation} 
\begin{split} 
U_1(\mathbf{r}+\sum_{i=1}^{2} n_i\mathbf{a}_i^m)= e^{-i\mathbf{q}_1\cdot(\sum_{i=1}^{2} n_i\mathbf{a}_i^m)} U_1(\mathbf{r}),\\ 
U_2(\mathbf{r}+\sum_{i=1}^{2} n_i\mathbf{a}_i^m) = e^{-i\mathbf{q}_1\cdot(\sum_{i=1}^{2} n_i\mathbf{a}_i^m)} U_2(\mathbf{r}), 
\end{split} 
\end{equation} 

\subsection{Symmetries of chiral TBG} 
In this section we specialize to TBG and summarize what we obtained in the previous subsection as well as point out some extra symmetries of TBG. The fact $U_2^*(\mathbf{r}) = U_1(-\mathbf{r})$ from Eq.~\eqref{eq:Urotation} (since for TBG $\mathcal{A} = \mathcal{C}_{2z}\mathcal{T}$, we have $\mathcal{A}\mathbf{r} = -\mathbf{r}$) gives 
\begin{equation} 
\label{eq:TBGH} 
\mathcal{H}_{TBG}(\mathbf{r}) = \begin{pmatrix} 
0 & 0 & (-2i\partial_z)^m & U_1^*(-\mathbf{r})\\ 
0 & 0 & U_1^*(\mathbf{r}) & (-2i\partial_z)^m\\ 
(-2i\overline{\partial_{z}})^m & U_1(\mathbf{r}) & 0 & 0\\ 
U_1(-\mathbf{r}) & (-2i\overline{\partial_{z}})^m\ & 0 & 0\ 
\end{pmatrix}. 
\end{equation} 
From Eq.~\eqref{eq:Utranslation} we obtain 
\begin{equation}\label{eq:TBGUFourier} 
U_1(\mathbf{r}) = \sum_{\mathbf{b}^m} a_{\mathbf{b}^m}e^{-i(\mathbf{q}_1+\mathbf{b}^m)\cdot \mathbf{r}}, 
\end{equation} 
where the sum goes over all moir\'e reciprocal lattice vectors, and $a_{\mathbf{b}^m}$ are Fourier amplitudes. Also from Eq.~\eqref{eq:Urotation}, noting that in case of TBG $\omega = e^{2\pi i/3}$, we get 
\begin{equation} 
\begin{split}\label{eq:TBGUFourierC3} 
U_1(\mathcal{C}_{3z}\mathbf{r}) = e^{2\pi i/3} U_1(\mathbf{r}) &\Rightarrow \sum_{\mathbf{b}^m} a_{\mathbf{b}^m}e^{-i(\mathbf{q}_1+\mathbf{b}^m)\cdot \mathcal{C}_{3z}\mathbf{r}} = e^{2\pi i/3} \sum_{\mathbf{b}^m} a_{\mathbf{b}^m}e^{-i(\mathbf{q}_1+\mathbf{b}^m)\cdot \mathbf{r}}\\ 
&\Rightarrow \sum_{\mathbf{b}^m} a_{\mathbf{b}^m}e^{-i( \mathcal{C}_{3z}^{-1}\mathbf{q}_1+ \mathcal{C}_{3z}^{-1} \mathbf{b}^m)\cdot\mathbf{r}} = e^{2\pi i/3} \sum_{\mathbf{b}^m} a_{\mathbf{b}^m}e^{-i(\mathbf{q}_1+\mathbf{b}^m)\cdot \mathbf{r}}\\ 
&\Rightarrow \sum_{\mathbf{b}^m} a_{\mathbf{b}^m}e^{-i(\mathbf{q}_1+\mathbf{b}_2^m+ \mathcal{C}_{3z}^{-1} \mathbf{b}^m)\cdot\mathbf{r}} = e^{2\pi i/3} \sum_{\mathbf{b}^m} a_{\mathbf{b}^m}e^{-i(\mathbf{q}_1+\mathbf{b}^m)\cdot \mathbf{r}}\\ 
&\Rightarrow \sum_{\mathbf{b}^m} a_{\mathcal{C}_{3z}(\mathbf{b}^m-\mathbf{b}^m_2)}e^{-i(\mathbf{q}_1 +\mathbf{b}^m)\cdot\mathbf{r}} = e^{2\pi i/3} \sum_{\mathbf{b}^m} a_{\mathbf{b}^m}e^{-i(\mathbf{q}_1+\mathbf{b}^m)\cdot \mathbf{r}},\\ 
&\Rightarrow a_{\mathcal{C}_{3z}(\mathbf{b}^m-\mathbf{b}^m_2)} = e^{2\pi i/3}a_{\mathbf{b}^m} \text{ for all reciprocal lattice vector }\mathbf{b}^m. 
\end{split} 
\end{equation} 
Eq.~\eqref{eq:TBGH} along with Eqs.~\eqref{eq:TBGUFourier} and~\eqref{eq:TBGUFourierC3} is the most general form of the TBG Hamiltonian. Using Eq.~\eqref{eq:TBGUFourierC3}, starting from $a_\mathbf{0}\equiv\alpha$, we get 
\begin{equation} 
\begin{split} 
&a_{-\mathcal{C}_{3z}\mathbf{b}^m_2} = a_{\mathbf{b}^m_1} = e^{2\pi i/3}a_{\mathbf{0}} = e^{2\pi i/3}\alpha, \\ 
&a_{\mathcal{C}_{3z}(\mathbf{b}^m_1-\mathbf{b}^m_2)} = a_{\mathbf{b}^m_2} = e^{2\pi i/3}a_{\mathbf{b}^m_1} = e^{4\pi/3}a_{\mathbf{0}} = e^{4\pi i/3}\alpha. 
\end{split} 
\end{equation} 
Hence, if we keep only lowest harmonics the expression for $U_1(\mathbf{r})$ becomes 
\begin{equation} 
\begin{split} 
\label{eq:UTBGTarnopolsky} 
U_1(\mathbf{r}) &= a_\mathbf{0} (e^{-i\mathbf{q}_1\cdot \mathbf{r}}+e^{2\pi i/3} e^{-i(\mathbf{q}_1+\mathbf{b}_1^m)\cdot \mathbf{r}}+e^{4\pi i/3} e^{-i(\mathbf{q}_1-\mathbf{b}_2^m)\cdot \mathbf{r}})\\ 
&= \alpha(e^{-i\mathbf{q}_1\cdot \mathbf{r}}+e^{2\pi i/3} e^{-i\mathbf{q}_2\cdot \mathbf{r}}+e^{4\pi i/3} e^{-i\mathbf{q}_3\cdot \mathbf{r}}). 
\end{split} 
\end{equation} 
This form of $U_1(\mathbf{r})$ was used in~\citesupp{tarnopolsky2019origins} (and in {\color{red}Eq.~3} in the main text). Similarly, if one starts from $a_\mathbf{\mathbf{b}_1^m+\mathbf{b}_2^m}$, one gets relations to $a_\mathbf{\mathbf{b}_2^m-\mathbf{b}_1^m}$ and $a_\mathbf{\mathbf{b}_1^m-\mathbf{b}_2^m}$. Furthermore, if one sets $a_\mathbf{\mathbf{b}_1^m+\mathbf{b}_2^m}= -a_{\mathbf{0}} = -\alpha$, in that fine-tuned case, $U_1(\mathbf{r})$ becomes 
\begin{equation} 
\label{eq:UTBGZworski} 
U_1(\mathbf{r}) = \alpha \left[(e^{-i\mathbf{q}_1\cdot \mathbf{r}}+e^{2\pi i/3} e^{-i\mathbf{q}_2\cdot \mathbf{r}}+e^{4\pi i/3} e^{-i\mathbf{q}_3\cdot \mathbf{r}})-(e^{2i\mathbf{q}_1\cdot \mathbf{r}}+e^{2\pi i/3} e^{2i\mathbf{q}_2\cdot \mathbf{r}}+e^{4\pi i/3} e^{2i\mathbf{q}_3\cdot \mathbf{r}})\right], 
\end{equation} 
this form of $U_1(\mathbf{r})$ was used in~\citesupp{becker2023degenerates} to obtain 4 exact flat bands for some ``magic'' values of $\alpha$. 

Chiral TBG also has the following symmetries: 
\begin{itemize} 
\item $\mathcal{C}_{2x}$ symmetry that swaps $|\mathbf{k}_0^{(1)},\mathbf{r},A\rangle \leftrightarrow |\mathbf{k}_0^{(2)},\mathcal{M}_y\mathbf{r},B\rangle$ and $|\mathbf{k}_0^{(1)},\mathbf{r},B\rangle \leftrightarrow |\mathbf{k}_0^{(2)},\mathcal{M}_y\mathbf{r},A\rangle$, where $\mathcal{M}_y\mathbf{r} = \mathcal{M}_y\{x,y\}=\{x,-y\}$. This means the following 
\begin{equation} 
\mathcal{C}_{2x}\{|\mathbf{k}_0^{(1)},\mathbf{r},A\rangle, |\mathbf{k}_0^{(2)},\mathbf{r},A\rangle, |\mathbf{k}_0^{(1)},\mathbf{r},B\rangle,|\mathbf{k}_0^{(2)},\mathbf{r},B\rangle\} = \{|\mathbf{k}_0^{(1)}, \mathcal{M}_y\mathbf{r},A\rangle, |\mathbf{k}_0^{(2)}, \mathcal{M}_y\mathbf{r},A\rangle, |\mathbf{k}_0^{(1)}, \mathcal{M}_y\mathbf{r},B\rangle,|\mathbf{k}_0^{(2)}, \mathcal{M}_y\mathbf{r},B\rangle\}\rho(\mathcal{C}_{2x})\\ 
\end{equation} 
where $\rho(\mathcal{C}_{2x}) = \sigma_x\otimes\sigma_x$. This also implies 
\begin{equation} 
\rho(\mathcal{C}_{2x}) \mathcal{H}_{TBG}(\mathbf{r}) \rho^\dagger(\mathcal{C}_{2x}) = \mathcal{H}_{TBG}(\mathcal{M}_y\mathbf{r}) \Rightarrow U_1^*(\mathbf{r}) = U_1(\mathcal{M}_y\mathbf{r}). 
\end{equation} 
This and the facts that $\mathbf{q}\cdot \mathcal{M}_y\mathbf{r} = (\mathcal{M}_y\mathbf{q})\cdot \mathbf{r}$, $\mathcal{M}_y\mathbf{q}_1 = -\mathbf{q}_1$, $\mathcal{M}_y\mathbf{q}_2 = -\mathbf{q}_3$ and $\mathbf{M}_y\mathbf{q}_3 = -\mathbf{q}_2$ imply $\alpha^* = \alpha$ in Eqs.~\eqref{eq:UTBGTarnopolsky} and~\eqref{eq:UTBGZworski}. Applying a magnetic field normal to TBG plane breaks $\mathcal{C}_{2x}$, this is why in~\citesupp{le2022doubles} the parameter $\alpha$ was a complex number under alternating magnetic field. 
\item Unitary particle-hole symmetry $P$ whose representation is $\rho(P) = \mathds{1}\otimes i\sigma_y$ with $P\mathbf{r} = -\mathbf{r}$ and 
\begin{equation} 
P\{|\mathbf{k}_0^{(1)},\mathbf{r},A\rangle, |\mathbf{k}_0^{(2)},\mathbf{r},A\rangle, |\mathbf{k}_0^{(1)},\mathbf{r},B\rangle,|\mathbf{k}_0^{(2)},\mathbf{r},B\rangle\} = \{|\mathbf{k}_0^{(1)}, -\mathbf{r},A\rangle, |\mathbf{k}_0^{(2)}, -\mathbf{r},A\rangle, |\mathbf{k}_0^{(1)}, -\mathbf{r},B\rangle,|\mathbf{k}_0^{(2)}, -\mathbf{r},B\rangle\}\rho(P)\\ 
\end{equation} 
such that $\rho(P) \mathcal{H}_{TBG}(\mathbf{r}) = - \mathcal{H}_{TBG}(-\mathbf{r})\rho(P)$. Note that $P\mathbf{k} = -\mathbf{k}$, but it does not change the valley. 
\item Composition of $P$ and $\mathcal{S}$ acts like a $\mathcal{C}_{2z}$ symmetry: 
\begin{enumerate} 
\item since both $P$ and $\mathcal{S}$ anticommute with $\hat{H}_{TBG}$, $P\mathcal{S}$ commutes with $\hat{H}_{TBG}$, 
\item since $P\mathbf{r} = -\mathbf{r}$, $P\mathbf{k} = -\mathbf{k}$, and $\mathcal{S}$ is local in both position and momentum space, $P\mathcal{S}\mathbf{r} = -\mathbf{r}$ and $P\mathcal{S}\mathbf{k} = -\mathbf{k}$, 
\item $\rho(P\mathcal{S})$ commutes with $\rho(\mathcal{C}_{3z})$, $\rho(\mathcal{C}_{2x})$ and $\rho(\mathcal{C}_{2z}\mathcal{T})$. 
\end{enumerate} 
This has representation $\rho(P\mathcal{S}) = \sigma_z \otimes i\sigma_y$.
\item A composition of $P\mathcal{S}$ with $\mathcal{C}_{2x}$ gives a $\mathcal{C}_{2y}$ type symmetry with representation $\rho(P\mathcal{SC}_{2x})= i\sigma_y\otimes \sigma_z $. 
\item A composition of $P$ with $\mathcal{A} = \mathcal{C}_{2z}\mathcal{T}$ gives an antiunitary particle-hole symmetry $\mathcal{P} = P\mathcal{C}_{2z}\mathcal{T}$, which squares to $\rho(P\mathcal{C}_{2z}\mathcal{T})\rho^*(P\mathcal{C}_{2z}\mathcal{T}) = (\sigma_x\otimes i\sigma_y)^2 = -\mathds{1}$. 
\item Since $P\mathcal{S}$ is a $\mathcal{C}_{2z}$ type symmetry, a composition of $P\mathcal{S}$ and $\mathcal{C}_{2z}\mathcal{T}$ gives a time-reversal like symmetry with representation $\rho(P\mathcal{S}\mathcal{C}_{2z}\mathcal{T}) = i\sigma_y\otimes\sigma_z$. 
\item The composition $P\mathcal{S}\mathcal{C}_{2x}\mathcal{C}_{2z}\mathcal{T}$ commutes with the Hamiltonian, its representation is diagonal $\rho(P\mathcal{S}\mathcal{C}_{2x}\mathcal{C}_{2z}\mathcal{T}) = \sigma_z\otimes\sigma_z$, and it acts on position and momentum as $P\mathcal{S}\mathcal{C}_{2x}\mathcal{C}_{2z}\mathcal{T}\mathbf{r} = P\mathcal{S}\mathcal{C}_{2x}\mathcal{C}_{2z}\mathcal{T}\{x,y\}=\{x,-y\} = \mathcal{M}_y\mathbf{r}$ whereas $P\mathcal{S}\mathcal{C}_{2x}\mathcal{C}_{2z}\mathcal{T}\mathbf{k} = P\mathcal{S}\mathcal{C}_{2x}\mathcal{C}_{2z}\mathcal{T}\{x,y\}=\{-k_x,k_y\} = \mathcal{M}_x\mathbf{k}$. Therefore, it has an effect on position and momentum like $\mathcal{C}_{2x}\mathcal{T}$. 
\end{itemize} 
Because of the $\mathcal{C}_{2z}$ type symmetry $P\mathcal{S}$, the space group is enhanced in single valley chiral TBG to $p622$ from $p6'2'2$ for single-valley non-chiral TBG. 

\section{Properties of zero energy eigen wave-function $\psi_{\mathbf{k}_0^m}(\mathbf{r})$} 
In the previous section we discussed the moir\'e Hamiltonians for single and bilayer DP or QBCP and their symmetry properties. In this section we discuss the zero energy sublattice polarized eigen wave-functions $\psi_{\mathbf{k}_0^m}(\mathbf{r})$ of the moir\'e Hamiltonian, how they transform under the symmetries and how symmetries dictate appearance of zeros of $\psi_{\mathbf{k}_0^m}(\mathbf{r})$. 
\subsection{Zero energy eigen wave-function $\psi_{\mathbf{k}_0^m}(\mathbf{r})$ and its transformation under symmetries} \label{sec:psiSymmetries}
First, to treat single and bilayer system together, we introduce shorthand $\{|\mathbf{k}_0,\mathbf{r}\rangle\}_{1\times 2l}$ for the basis states, where $\{|\mathbf{k}_0,\mathbf{r}\rangle\}_{1\times 2} = \{|\mathbf{k}_0,\mathbf{r},A\rangle,|\mathbf{k}_0,\mathbf{r},B\rangle\}$ and $\{|\mathbf{k}_0,\mathbf{r}\rangle\}_{1\times 4} = \{|\mathbf{k}_0^{(1)},\mathbf{r},A\rangle,|\mathbf{k}_0^{(2)},\mathbf{r},A\rangle,|\mathbf{k}_0^{(1)},\mathbf{r},B\rangle,|\mathbf{k}_0^{(2)},\mathbf{r},B\rangle\}$ for single and bilayer systems respectively. In this notation, the Hamiltonian is written as 
\begin{equation} 
\hat{H}_{S/TB} = \int d^2\mathbf{r} \{|\mathbf{k}_0,\mathbf{r}\rangle\}_{1\times 2l} \mathcal{H}_{S/TB}(\mathbf{r}) \{\langle\mathbf{k}_0,\mathbf{r}|\}^T_{2l\times 1}. 
\end{equation} 
Before adding any moir\'e potential, there are two zero energy eigenmodes at $\mathbf{k}_0$ protected by $\mathcal{C}_{nz}$ and $\mathcal{A}$ for both DP and QBCP. They have $\mathcal{C}_{nz}$ eigenvalues $\omega$ and $\omega^*$. Now, if the moir\'e potential does not break $\mathcal{C}_{nz}$ and $\mathcal{A}$, the degeneracy of the DP and QBCP remain. If the moir\'e potential does not break $\mathcal{S}$, then the DP or QBCP has to remain at zero energy since $[\mathcal{C}_{nz},\mathcal{S}] =0$. This means there are two states $|\Psi_{\mathbf{k}_0^m,1}\rangle$ and $|\Psi_{\mathbf{k}_0^m,2}\rangle$ such that $\hat{H}_{S/TB}|\Psi_{\mathbf{k}_0^m,i}\rangle =0 $ for any moir\'e potential that keeps $\mathcal{C}_{nz}$, $\mathcal{A}$ and $\mathcal{S}$ intact (recall that $\mathbf{k}_0^m$ is the momentum in the mBZ where $\mathbf{k}_0$ gets mapped to after BZ folding). Furthermore, the irrep of the degeneracy at $\mathbf{k}_0^m$ at zero energy before and after adding the moir\'e potential is the same. These two states can be chosen to be sublattice polarized with following properties: 
\begin{equation} 
\begin{split} 
|\Psi_{\mathbf{k}_0^m,1}\rangle &= \int d^2\mathbf{r} \{|\mathbf{k}_0,\mathbf{r}\rangle\}_{1\times 2l} \begin{Bmatrix}\psi_{\mathbf{k}_0^m}(\mathbf{r})_{l\times 1}\\\mathbf{0}_{l\times 1}\end{Bmatrix},\\ 
|\Psi_{\mathbf{k}_0^m,2}\rangle &=\mathcal{A}|\Psi_{\mathbf{k}_0,1}\rangle= \int d^2\mathbf{r} \{|\mathbf{k}_0,\mathbf{r}\rangle\}_{1\times 2l} \begin{Bmatrix}\mathbf{0}_{l\times 1}\\\psi_{\mathbf{k}_0^m}^*(\mathcal{A}\mathbf{r})_{l\times 1}\end{Bmatrix}.
\end{split} 
\end{equation} 
Below we discuss the effect of symmetries on the wave-function {\color{black}$\psi_{\mathbf{k}_0^m}(\mathbf{r})$}. 
\begin{enumerate}
\item Due to DP or QBCP {\color{black}protected by} $\mathcal{C}_{nz}$, we have
\begin{equation}
\begin{split}
\mathcal{C}_{nz}|\Psi_{\mathbf{k}_0^m,1}\rangle &= \omega |\Psi_{\mathbf{k}_0^m,1}\rangle\\ 
\mathcal{C}_{nz}|\Psi_{\mathbf{k}_0^m,2}\rangle &= \omega^* |\Psi_{\mathbf{k}_0^m,2}\rangle. 
\end{split}
\end{equation}
The last two relations mean 
\begin{equation} 
\begin{split} 
\omega |\Psi_{\mathbf{k}_0^m,1}\rangle =\mathcal{C}_{nz}|\Psi_{\mathbf{k}_0^m,1}\rangle &= \int d^2\mathbf{r} \mathcal{C}_{nz}\{|\mathbf{k}_0,\mathbf{r}\rangle\}_{1\times 2l} \begin{Bmatrix}\psi_{\mathbf{k}_0^m}(\mathbf{r})_{l\times 1}\\\mathbf{0}_{l\times 1}\end{Bmatrix}\\ 
&= \int d^2\mathbf{r} \{|\mathbf{k}_0,\mathcal{C}_{nz}\mathbf{r}\rangle\}_{1\times 2l}\rho(\mathcal{C}_{nz}) \begin{Bmatrix}\psi_{\mathbf{k}_0^m}(\mathbf{r})_{l\times 1}\\\mathbf{0}_{l\times 1}\end{Bmatrix}\\ 
&= \int d^2\mathbf{r} \{|\mathbf{k}_0,\mathcal{C}_{nz}\mathbf{r}\rangle\}_{1\times 2l}{\color{black}\text{Diag}\{\omega,\omega^*\}\otimes \mathds{1}_{l \times l}} \begin{Bmatrix}\psi_{\mathbf{k}_0^m}(\mathbf{r})_{l\times 1}\\\mathbf{0}_{l\times 1}\end{Bmatrix}\\ 
&= \int d^2\mathbf{r} \{|\mathbf{k}_0,\mathcal{C}_{nz}\mathbf{r}\rangle\}_{1\times 2l}\omega\begin{Bmatrix}\psi_{\mathbf{k}_0^m}(\mathbf{r})_{l\times 1}\\\mathbf{0}_{l\times 1}\end{Bmatrix}\\
&= \int d^2\mathbf{r} \{|\mathbf{k}_0,\mathbf{r}\rangle\}_{1\times 2l}\omega \begin{Bmatrix}\psi_{\mathbf{k}_0^m}(\mathcal{C}_{nz}^{-1}\mathbf{r})_{l\times 1}\\\mathbf{0}_{l\times 1}\end{Bmatrix}\\ 
\Rightarrow \omega \psi_{\mathbf{k}_0^m}(\mathcal{C}_{nz}^{-1}\mathbf{r})_{l\times 1} &= \omega \psi_{\mathbf{k}_0^m}(\mathbf{r})_{l\times 1} \Rightarrow \psi_{\mathbf{k}_0^m}(\mathcal{C}_{nz}\mathbf{r})= \psi_{\mathbf{k}_0^m}(\mathbf{r}).\\ 
\end{split} 
\end{equation} 
\item If there is a mirror $\mathcal{M}$ (or $\mathcal{C}_{2y}$ like symmetry $P\mathcal{S}\mathcal{C}_{2x}$ in case of chiral TBG) at $\mathbf{k}_0$, we can choose its sewing matrix $[B_{\mathbf{k}_0^m}(\mathcal{M})]_{ij} = \langle \Psi_{\mathbf{k}_0^m,i}|\rho(\mathcal{M})|\Psi_{\mathbf{k}_0^m,j} \rangle=\sigma_x$ (see, for example, Bilbao Crystallography server~\citesupp{aroyo2011crystallographys,aroyo2006bilbaoIs,aroyo2006bilbaoIIs}). This implies 
\begin{equation} 
\mathcal{M}|\Psi_{\mathbf{k}_0^m,1}\rangle = |\Psi_{\mathbf{k}_0^m,2}\rangle \Rightarrow \psi_{\mathbf{k}_0^m}(\mathbf{r})_{l\times 1} = \psi_{\mathbf{k}_0^m}^*(\mathcal{MA}\mathbf{r})_{l\times 1}. 
\end{equation} 
This implies that along $\mathcal{MA}$ invariant line (given by $\mathcal{MA}\mathbf{r} = \mathbf{r}$), the wave-function $\psi_{\mathbf{k}_0^m}(\mathbf{r})_{l\times 1}$ is a real function. 
\item If there is $\mathcal{C}_{2z}$ symmetry such that $\mathcal{C}_{2z}\mathbf{k}_0 = \mathbf{k}_0$ (this is true for QBCP in $4$ and $6$-fold rotation symmetry), its representation $\rho(\mathcal{C}_{2z})$ is $\zeta \mathds{1}$ with $\zeta = \pm 1$ (see Sec.~\ref{sec:k.p}), and the sewing matrix $[B_{\mathbf{k}_0^m}(\mathcal{C}_{2z})]_{ij} = \langle \Psi_{\mathbf{k}_0^m,i}|\rho(\mathcal{C}_{2z})|\Psi_{\mathbf{k}_0^m,j} \rangle = \zeta \delta_{ij}$ since the irrep of the two degenerate zero energy modes remains the same before and after adding moir\'e potential as discussed earlier. From this we obtain: 
\begin{equation} 
\begin{split}
    &\mathcal{C}_{2z}|\Psi_{\mathbf{k}_0^m,1}\rangle = \zeta |\Psi_{\mathbf{k}_0^m,1}\rangle\\
    \Rightarrow &\psi_{\mathbf{k}_0^m}(\mathcal{C}_{2z}\mathbf{r})_{l\times 1} = \psi_{\mathbf{k}_0^m}(\mathbf{r})_{l\times 1}.
\end{split}
\end{equation} 
\item Under moir\'e lattice translation $n_1\mathbf{a}_1^m+n_2\mathbf{a}_2^m$, the states $|\Psi_{\mathbf{k}_0^m,i}\rangle$ should transform as a Bloch states at $\mathbf{k}_0^m$, hence
\begin{equation}
\label{eq:WFtranslation}
\begin{split}
	&T_{\sum_{i=1}^2n_i\mathbf{a}_i^m} |\Psi_{\mathbf{k}_0^m,i}\rangle = e^{-i\mathbf{k}_0^m\cdot(\sum_{i=1}^2n_i\mathbf{a}_i^m)} |\Psi_{\mathbf{k}_0^m,i}\rangle\\
	\Rightarrow& \rho(T_{\sum_{i=1}^2n_i\mathbf{a}_i^m})\begin{Bmatrix}\psi_{\mathbf{k}_0^m}(\mathbf{r})_{l\times 1}\\\mathbf{0}_{l\times 1}\end{Bmatrix} = e^{-i\mathbf{k}_0^m\cdot(\sum_{i=1}^2n_i\mathbf{a}_i^m)}\begin{Bmatrix}\psi_{\mathbf{k}_0^m}(\mathbf{r}+\sum_{i=1}^2n_i\mathbf{a}_i^m)_{l\times 1}\\\mathbf{0}_{l\times 1}\end{Bmatrix}.
\end{split}
\end{equation}
The consequences of this for single and twisted bilayer cases are as follows:
\begin{itemize}
\item For single layer QBCP systems, as discussed earlier, $\rho(T_{\sum_{i=1}^2n_i\mathbf{a}_i^m}) = e^{-i\mathbf{k}_0^m\cdot(\sum_{i=1}^2n_i\mathbf{a}_i^m)}\mathds{1}$. This implies $\psi_{\mathbf{k}_0^m}(\mathbf{r}+\sum_{i=1}^2n_i\mathbf{a}_i^m) =\psi_{\mathbf{k}_0^m}(\mathbf{r})$, or in other words $\psi_{\mathbf{k}_0^m}(\mathbf{r})$ is periodic function with periodicity of moir\'e lattice vector.
\item For twisted bilayer systems, as discussed earlier, $\splitatcommas{\rho(T_{\sum_{i=1}^2n_i\mathbf{a}_i^m}) = e^{-i\mathbf{k}_0^m\cdot(\sum_{i=1}^2n_i\mathbf{a}_i^m)}\text{Diag}\{1,e^{i\mathbf{q}_1\cdot(\sum_{i=1}^{2} n_i\mathbf{a}_i^m)},1,e^{i\mathbf{q}_1\cdot(\sum_{i=1}^{2} n_i\mathbf{a}_i^m)}\}}$. This implies that the two components of $\psi_{\mathbf{k}_0^m}(\mathbf{r})$ transform differently: The first component satisfies $\psi_{\mathbf{k}_0^m,1}(\mathbf{r}+\sum_{i=1}^2n_i\mathbf{a}_i^m) =\psi_{\mathbf{k}_0^m,1}(\mathbf{r})$, hence it is a periodic function with moir\'e periodicity. The second component satisfies $\psi_{\mathbf{k}_0^m,2}(\mathbf{r}+\sum_{i=1}^2n_i\mathbf{a}_i^m) =e^{i\mathbf{q}_1\cdot(\sum_{i=1}^{2} n_i\mathbf{a}_i^m)}\psi_{\mathbf{k}_0^m,2}(\mathbf{r})$. In case of TBG, in the notation of {\color{red}Fig.~2}, the phase factor is $e^{-i(\mathbf{b}_1+\mathbf{b}_2)/3\cdot(\sum_{i=1}^{2} n_i\mathbf{a}_i^m)} = e^{-2\pi i (n_1+n_2)/3}$. In case of TBCL, in the notation of {\color{red}Fig.~\ref{fig:TBCL}}, the phase factor is ${\color{black}e^{i(\mathbf{b}_1-\mathbf{b}_2)/2\cdot(\sum_{i=1}^{2} n_i\mathbf{a}_i^m)}= e^{\pi i(n_1-n_2)}}$.
\end{itemize}
\end{enumerate}

\subsection{Zeros of $\psi_{\mathbf{k}_0^m}(\mathbf{r})$, and constraints from symmetries}
As discussed in the main text, the general construction of exact flat-band (FB) wave-function (WF) requires a zero in $\psi_{\mathbf{k}_0^m}(\mathbf{r})$ somewhere in the unit cell since that can cancel the pole of the holomorphic function $f_\mathbf{k}(z;\mathbf{r}_0)$ such the FB WF is finite everywhere in real spalce. However, not any zero of $\psi_{\mathbf{k}_0^m}(\mathbf{r})$ is suitable for construction of exact flat band wave-function, only the ones at high symmetry points (HSPs) are suitable. Below, we present a more detailed discussion on this for single and bilayer systems.
\subsubsection{Single layer system with QBCP}
For single layer systems, $\psi_{\mathbf{k}_0^m}(\mathbf{r})$ is a one component function. Near a zero at position $\mathbf{r}_0$ of  $\psi_{\mathbf{k}_0^m}(\mathbf{r})$ ($\psi_{\mathbf{k}_0^m}(\mathbf{r}_0) = 0$), generically the function behaves as $\psi_{\mathbf{k}_0^m}(\mathbf{r}_0+\delta(x,y)) = \delta(a x + b y)+\mathcal{O}(\delta^2)$ for small $\delta$, where $a$ and $b$ are complex constants. Since the holomorphic function $f_\mathbf{k}(z;\mathbf{r}_0)$ has a pole at $\mathbf{r}_0$, near $\mathbf{r}_0$ it behave as $f_{\mathbf{k}}(\mathbf{r}_0+\delta(x,y);\mathbf{r}_0) = \frac{c_0}{\delta(x + i y)}+\mathcal{O}(\delta^0)$. Hence the function $\psi_{\mathbf{k}+\mathbf{k}_0^m}(\mathbf{r}) = f_{\mathbf{k}}(z;\mathbf{r}_0)\psi_{\mathbf{k}_0^m}(\mathbf{r})$ behaves as $\psi_{\mathbf{k}+\mathbf{k}_0^m}(\mathbf{r}_0+\delta(x,y)) = \frac{c_0(ax+by)}{x+iy}+\mathcal{O}(\delta)$. However, this function is singular at $\mathbf{r}_0$ for generic values of $a$ and $b$ since the function oscillates around the point $\mathbf{r}_0$. However, if $\mathbf{r}_0$ is a HSP meaning that there exists a rotation symmetry $\mathcal{C}_{n'z}$ ($n'\geq 2$) and some moir\'e lattice vector $\mathbf{a}^m$ such that $\mathcal{C}_{n'z}\mathbf{r}_0 +\mathbf{a}^m = \mathbf{r}_0$, then we know from Sec.~\ref{sec:psiSymmetries} that $\psi_{\mathbf{k}_0^m}(\mathbf{r}_0+\delta(x,y))=\psi_{\mathbf{k}_0^m}(\mathcal{C}_{n'z}(\mathbf{r}_0+\delta(x,y))) = \psi_{\mathbf{k}_0^m}(\mathbf{r}_0-\mathbf{a}^m+\mathcal{C}_{n'z}\delta(x,y)) = \psi_{\mathbf{k}_0^m}(\mathbf{r}_0+\mathcal{C}_{n'z}\delta(x,y)))$. This implies $a=b=0$. This would make the function $\psi_{\mathbf{k}+\mathbf{k}_0}(\mathbf{r})$ regular.
\subsubsection{Twisted bilayer graphene}
In twisted bilayer graphene, $\psi_{\mathbf{k}_0^m}(\mathbf{r})$ is a two component function. At a zero of both real and imaginary parts of each component of $\psi_{\mathbf{k}_0^m}(\mathbf{r})$ have to be zero. This means that four equations need to be satisfied at once for a zero. However, it turns out that due to the twisting between the two layers, one of the components is automatically zero at some HSPs for any value of the tuning parameter $\boldsymbol{\alpha}$. Then, only the other component of $\psi_{\mathbf{k}_0^m}(\mathbf{r})$ need to be set to zero by tuning the parameters $\boldsymbol{\alpha}$. We show this below. 

As was shown in~\ref{sec:psiSymmetries}, under $\mathcal{C}_{3z}$, each of the two components satisfies $\psi_{\mathbf{k}_0^m,i}(\mathcal{C}_{3z}\mathbf{r})= \psi_{\mathbf{k}_0^m,i}(\mathbf{r})$, whereas under moir\'e lattice translations they satisfy $\psi_{\mathbf{k}_0^m,1}(\mathbf{r}+\sum_{i=1}^2n_i\mathbf{a}_i^m) =\psi_{\mathbf{k}_0^m,1}(\mathbf{r})$ and $\psi_{\mathbf{k}_0^m,2}(\mathbf{r}+\sum_{i=1}^2n_i\mathbf{a}_i^m) =e^{-2\pi i (n_1+n_2)/3}\psi_{\mathbf{k}_0^m,2}(\mathbf{r})$. Near the corners (BA and AB positions ($\mathbf{r}_0=\mp (\mathbf{a}_2^m-\mathbf{a}_1^m)/3$ respectively, see {\color{red}Fig.~2 of main text}) of the moir\'e unit cell, we have
\begin{equation}
\begin{split}
	\psi_{\mathbf{k}_0^m,1}(\mathbf{r}_0+\mathbf{r}) &=\psi_{\mathbf{k}_0^m,1}(\mathcal{C}_{3z}\mathbf{r}_0+\mathcal{C}_{3z}\mathbf{r}) =\psi_{\mathbf{k}_0^m,1}(\mathbf{r}_0\pm \mathbf{a}_2^m+\mathcal{C}_{3z}\mathbf{r}) = \psi_{\mathbf{k}_0^m,1}(\mathbf{r}_0+\mathcal{C}_{3z}\mathbf{r}),\\
	\Rightarrow \psi_{\mathbf{k}_0^m,1}(\mathbf{r}_0+\mathcal{C}_{3z}\mathbf{r}) &= \psi_{\mathbf{k}_0^m,1}(\mathbf{r}_0+\mathbf{r}),\\
	\psi_{\mathbf{k}_0^m,2}(\mathbf{r}_0+\mathbf{r}) &=\psi_{\mathbf{k}_0^m,2}(\mathcal{C}_{3z}\mathbf{r}_0+\mathcal{C}_{3z}\mathbf{r}) =\psi_{\mathbf{k}_0^m,2}(\mathbf{r}_0\pm \mathbf{a}_2^m+\mathcal{C}_{3z}\mathbf{r}) = e^{\mp2\pi i/3}\psi_{\mathbf{k}_0^m,2}(\mathbf{r}_0+\mathcal{C}_{3z}\mathbf{r}),\\
	 \Rightarrow \psi_{\mathbf{k}_0^m,2}(\mathbf{r}_0+\mathcal{C}_{3z}\mathbf{r}) &= e^{\pm2\pi i/3}\psi_{\mathbf{k}_0^m,2}(\mathbf{r}_0+\mathbf{r}),
\end{split}
\end{equation} 
for BA and AB positions, respectively. The latter of the two equations implies that $\psi_{\mathbf{k}_0^m,2}(\mathbf{r}_0) = 0$ at BA and AB positions for any $\boldsymbol{\alpha}$. The above equation also means the following:
\begin{itemize}
\item\underline{BA:} If for a ``magic'' value of $\boldsymbol{\alpha}$,  $\psi_{\mathbf{k}_0^m}(\mathbf{r}_0) =\mathbf{0}$,  near the zero we have 
\begin{equation}
\begin{split}
	\psi_{\mathbf{k}_0^m,1}(\mathbf{r}_0+\mathcal{C}_{3z}(\delta x,\delta y)) &= \psi_{\mathbf{k}_0^m,1}(\mathbf{r}_0+(\delta x,\delta y)),\\
	\Rightarrow \psi_{\mathbf{k}_0^m,1}(\mathbf{r}_0+(\delta x,\delta y)) &= c_1 \delta^2 (x^2+y^2) +\mathcal{O}(\delta^3),\\
	\psi_{\mathbf{k}_0^m,2}(\mathbf{r}_0+\mathcal{C}_{3z}(\delta x,\delta y)) &= e^{2\pi i/3}\psi_{\mathbf{k}_0^m,2}(\mathbf{r}_0+(\delta x,\delta y)),\\
	\Rightarrow \psi_{\mathbf{k}_0^m,2}(\mathbf{r}_0+(\delta x,\delta y)) &= c_2 \delta (x+iy) +\mathcal{O}(\delta^2).
\end{split}
\end{equation}
These two together imply that $\psi_{\mathbf{k}_0^m}(\mathbf{r}_0+(\delta x,\delta y))\sim \delta(x+iy)\times \text{(regular function)}$. Then $f_\mathbf{k}(z;\mathbf{r}) \psi_{\mathbf{k}_0^m}(\mathbf{r})$ is a regular function near $\mathbf{r}_0$, and a FB WF construction is possible.
\item\underline{AB:} The situation is more complicated here. If for a ``magic'' value of $\boldsymbol{\alpha}$,  $\psi_{\mathbf{k}_0^m}(\mathbf{r}_0) =\mathbf{0}$,  near the zero we have generically
\begin{equation}
\begin{split}
	\psi_{\mathbf{k}_0^m,1}(\mathbf{r}_0+\mathcal{C}_{3z}(\delta x,\delta y)) &= \psi_{\mathbf{k}_0^m,1}(\mathbf{r}_0+(\delta x,\delta y)),\\
	\Rightarrow \psi_{\mathbf{k}_0^m,1}(\mathbf{r}_0+(\delta x,\delta y)) &= c_1 \delta^2 (x^2+y^2) +\mathcal{O}(\delta^3),\\
	\psi_{\mathbf{k}_0^m,2}(\mathbf{r}_0+\mathcal{C}_{3z}(\delta x,\delta y)) &= e^{-2\pi i/3}\psi_{\mathbf{k}_0^m,2}(\mathbf{r}_0+(\delta x,\delta y)),\\
	\Rightarrow \psi_{\mathbf{k}_0^m,2}(\mathbf{r}_0+(\delta x,\delta y)) &= c_2 \delta (x-iy) +\mathcal{O}(\delta^2).
\end{split}
\end{equation}
If this were the case, the function $f_\mathbf{k}(z;\mathbf{r}_0) \psi_{\mathbf{k}_0^m}(\mathbf{r})~\sim (x-iy)/(x+iy)$ would be singular, and a FB WF construction would not be possible. However, the following result (taken from~\citesupp{becker2023degenerates}) comes to rescue. \\
\underline{Thoerem:} If $\mathcal{D}(\mathbf{r})\psi_{\mathbf{k}_0^m}(\mathbf{r}) =\mathbf{0}$ ($\mathcal{D}(\mathbf{r})$ is defined in {\color{red}Eq.~3 of main text}), and $\psi_{\mathbf{k}_0^m}(\mathbf{r}_0) = \mathbf{0}$, then $\psi_{\mathbf{k}_0^m}(\mathbf{r}_0+(x,y)) = (x+iy)g(\mathbf{r})$ where $g(\mathbf{r})$ is a regular two component function. In other words, if $\mathcal{D}(\mathbf{r})\psi_{\mathbf{k}_0^m}(\mathbf{r}) =\mathbf{0}$, and $\psi_{\mathbf{k}_0^m}(\mathbf{r}_0) = \mathbf{0}$, then $\overline{\partial_{z}}^p \psi_{\mathbf{k}_0^m}(\mathbf{r})|_{\mathbf{r}=\mathbf{r}_0} = \mathbf{0}$ for any $p \in \mathds{N}$.\\
\underline{Proof:} Notice that 
\begin{equation}
\mathcal{D}(\mathbf{r})\psi_{\mathbf{k}_0^m}(\mathbf{r}) =\mathbf{0} \Rightarrow (2i\overline{\partial_{z}})\psi_{\mathbf{k}_0^m}(\mathbf{r}) = -\begin{pmatrix}
0 & U_1(\mathbf{r})\\ U_1(-\mathbf{r}) & 0
\end{pmatrix} \psi_{\mathbf{k}_0^m}(\mathbf{r}) \Rightarrow \overline{\partial_{z}} \psi_{\mathbf{k}_0^m}(\mathbf{r}) = \frac{i}{2}\tilde{U}(\mathbf{r})  \psi_{\mathbf{k}_0^m}(\mathbf{r}).
\end{equation}
Then, clearly we have $\overline{\partial_{z}}\psi_{\mathbf{k}_0^m}(\mathbf{r})|_{\mathbf{r}=\mathbf{r}_0} = \frac{i}{2}\tilde{U}(\mathbf{r})  \psi_{\mathbf{k}_0^m}(\mathbf{r}))|_{\mathbf{r}=\mathbf{r}_0} = \mathbf{0}$ since $\psi_{\mathbf{k}_0^m}(\mathbf{r}_0) = \mathbf{0}$. Similarly,
\begin{equation}
\begin{split}
	\left.\overline{\partial_{z}}^2\psi_{\mathbf{k}_0^m}(\mathbf{r})\right|_{\mathbf{r}=\mathbf{r}_0} &=  \left.\frac{i}{2}\overline{\partial_{z}}(\tilde{U}(\mathbf{r})  \psi_{\mathbf{k}_0^m}(\mathbf{r}))\right|_{\mathbf{r}=\mathbf{r}_0}\\
	&=  \left.\frac{i}{2}((\overline{\partial_{z}}\tilde{U}(\mathbf{r}) ) \psi_{\mathbf{k}_0^m}(\mathbf{r})+\tilde{U}(\mathbf{r})  \overline{\partial_{z}}\psi_{\mathbf{k}_0^m}(\mathbf{r}))\right|_{\mathbf{r}=\mathbf{r}_0}\\
	&= \left. \frac{i}{2}((\overline{\partial_{z}}\tilde{U}(\mathbf{r}) ) \psi_{\mathbf{k}_0^m}(\mathbf{r})+\frac{i}{2}\tilde{U}^2(\mathbf{r})  \psi_{\mathbf{k}_0^m}(\mathbf{r}))\right|_{\mathbf{r}=\mathbf{r}_0}\\
	&= \mathbf{0}.
\end{split}
\end{equation}
Iterating this procedure of replacing $\overline{\partial_{z}} \psi_{\mathbf{k}_0^m}(\mathbf{r})$ with $\frac{i}{2}\tilde{U}(\mathbf{r})  \psi_{\mathbf{k}_0^m}(\mathbf{r})$ and using $\psi_{\mathbf{k}_0^m}(\mathbf{r}_0) = \mathbf{0}$, we see that $\overline{\partial_{z}}^p \psi_{\mathbf{k}_0^m}(\mathbf{r})|_{\mathbf{r}=\mathbf{r}_0} = \mathbf{0}$ for any $p \in \mathds{N}$.

In fact this theorem implies that, near a zero of $\psi_{\mathbf{k}_0^m,2}(\mathbf{r})$ is locally a holomorphic function.

This theorem together with transformation under $\mathcal{C}_{3z}$ implies that if there is a zero at AB, the function $\psi_{\mathbf{k}_0^m}(\mathbf{r})$ near AB behaves as
\begin{equation}
\begin{split}
	\psi_{\mathbf{k}_0^m,1}(\mathbf{r}_0+(\delta x,\delta y)) &= c_1 \delta^3 (x+iy)^3 +\mathcal{O}(\delta^4),\\
	\psi_{\mathbf{k}_0^m,2}(\mathbf{r}_0+(\delta x,\delta y)) &= c_2 \delta^2 (x+iy)^2 +\mathcal{O}(\delta^3).
\end{split}
\end{equation}
This makes $f_\mathbf{k}(z;\mathbf{r}_0)\psi_{\mathbf{k}_0^m}(\mathbf{r})$ a regular function near AB and, consequently, construction of FB WF possible.
\end{itemize}
\section{Comparing analytically obtained wave-functions $\tilde{\psi}_{\mathbf{k}_\text{max}}^{(i)}$ in Eq.~7 of main text with numerical ones for single layer QBCP under periodic strain field with $p6mm$ symmetry}

\begin{figure}[h!]
     \centering
\includegraphics[scale=1]{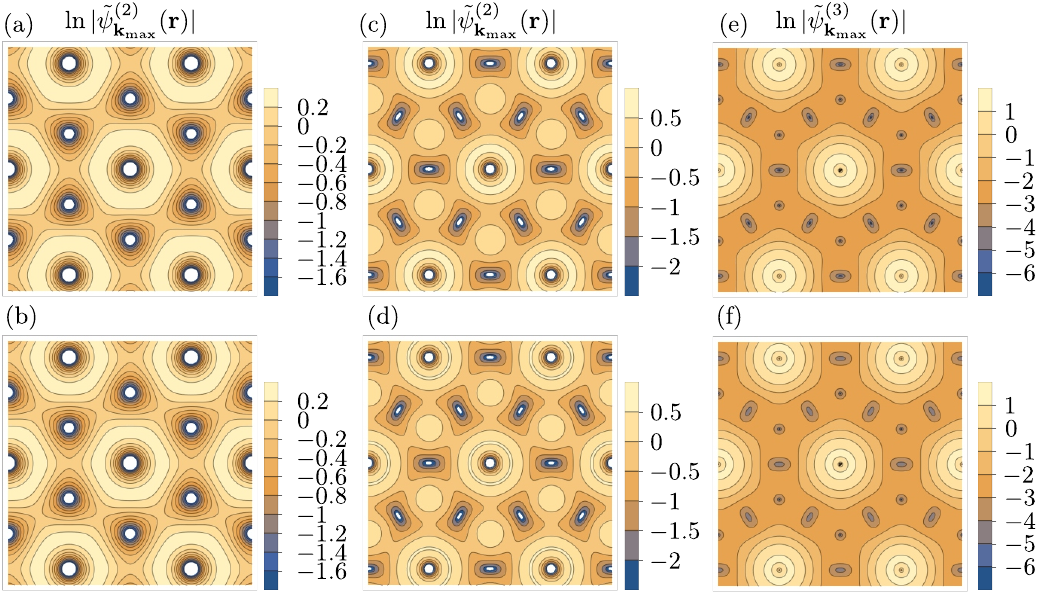}
     \caption{Comparison between analytically obtained wave-functions $\tilde{\psi}_{\mathbf{k}_\text{max}}^{(i)}$ in {\color{red}Eq,~7 of main text} (bottom row) with numerical ones (top row) for single layer QBCP under periodic strain field with $p6mm$ symmetry. (a-b) correspond to the 4 flat band case in {\color{red}Fig.~1(b) of the main text}, (c-f) correspond to 6 flat bands case in  {\color{red}Fig.~1(c) of the main text}. Note that in the 6 flat band case, $\tilde{\psi}_{\mathbf{k}_\text{max}}^{(2)}$ and $\tilde{\psi}_{\mathbf{k}_\text{max}}^{(3)}$ from {\color{red}Eq,~7 of main text} are not mutually orthogonal. Therefore, we orthogonalize them by redefining $\tilde{\psi}_{\mathbf{k}_\text{max}}^{(3)} \rightarrow \tilde{\psi}_{\mathbf{k}_\text{max}}^{(3)} - \frac{\langle\tilde{\psi}_{\mathbf{k}_\text{max}}^{(2)}|\tilde{\psi}_{\mathbf{k}_\text{max}}^{(3)}\rangle}{\langle\tilde{\psi}_{\mathbf{k}_\text{max}}^{(2)}|\tilde{\psi}_{\mathbf{k}_\text{max}}^{(2)}\rangle} \tilde{\psi}_{\mathbf{k}_\text{max}}^{(2)}$. To ensure same gauge between numerics and analytical expression, we further change the basis such that $\tilde{\psi}_{\mathbf{k}_\text{max}}^{(2)}(\mathcal{C}_{3z}\mathbf{r}) = \tilde{\psi}_{\mathbf{k}_\text{max}}^{(2)}(\mathbf{r})$ and $\tilde{\psi}_{\mathbf{k}_\text{max}}^{(3)}(\mathcal{C}_{3z}\mathbf{r}) = e^{2\pi i/3}\tilde{\psi}_{\mathbf{k}_\text{max}}^{(3)}(\mathbf{r})$.} 
     \label{fig:WFcomp}
\end{figure}
\section{Topology of the flat-bands}
In the main text, we claim that among all the FBs polarized to the same sub-lattice, only one carries Chern number $|C| = 1$, the others are trivial. Here we prove this statement. We start by re-deriving the well known result for when there is a single sub-lattice polarized flat band, and then tackle the cases when there are more than one FBs polarized to one sub-lattice.
\subsection{Topology of two flatbands -- single flat band on each sub-lattice}
In the main text, we found that the FB WF can be written as
\begin{equation}
	\psi_{\mathbf{k}+\mathbf{k}_0^m}(\mathbf{r}) = f_\mathbf{k}(z;\mathbf{r}_0)\psi_{\mathbf{k}_0^m}(\mathbf{r})  =  e^{i (\mathbf{k}\cdot\mathbf{a}_1^m) z/a_1^m}\frac{\vartheta\left(\frac{z-z_0}{a_1^m}-\frac{k}{b_2^m},\tau\right)}{\vartheta\left(\frac{z-z_0}{a_1^m},\tau\right)}\psi_{\mathbf{k}_0^m}(\mathbf{r}),
\end{equation}
and the corresponding Bloch periodic part of the WF 
\begin{equation}
	u_{\mathbf{k}+\mathbf{k}_0^m}(\mathbf{r}) =e^{-i\mathbf{k}\cdot\mathbf{r}}\psi_{\mathbf{k}+\mathbf{k}_0^m}(\mathbf{r})  = e^{-i\mathbf{k}\cdot\mathbf{r}}e^{i (\mathbf{k}\cdot\mathbf{a}_1^m) z/a_1^m}\frac{\vartheta\left(\frac{z-z_0}{a_1^m}-\frac{k}{b_2^m},\tau\right)}{\vartheta\left(\frac{z-z_0}{a_1^m},\tau\right)}\psi_{\mathbf{k}_0^m}(\mathbf{r})= e^{-i(\mathbf{b}_2^m\cdot\mathbf{r})k/b_2^m}\frac{\vartheta\left(\frac{z-z_0}{a_1^m}-\frac{k}{b_2^m},\tau\right)}{\vartheta\left(\frac{z-z_0}{a_1^m},\tau\right)}\psi_{\mathbf{k}_0^m}(\mathbf{r}),
\end{equation}
where $\vartheta (z,\tau) = -i\sum_{n=-\infty}^\infty(-1)^n e^{\pi i \tau(n+1/2)^2+\pi i(2n+1)z}$ is the Jacobi theta function of first kind, $\mathbf{a}_i^m$ are Moir\'e lattice vectors, $\mathbf{b}_i^m$ are the corresponding reciprocal lattice vectors ($\mathbf{a}_i^m\cdot\mathbf{b}_j^m= 2\pi \delta_{ij}$), $a_i^m = (\mathbf{a}_i^m)_x+ i (\mathbf{a}_i^m)_y$, $b_i^m = (\mathbf{b}_i)_x+ i (\mathbf{b}_i)_y$, $z_0 = (\mathbf{r}_0)_x+ i (\mathbf{r}_0)_y$ is complexification of the position of the zero of $\psi_{\mathbf{k}_0^m}(\mathbf{r})$, $k = k_x + i k_y$, and $\tau = a_2^m/a_1^m$. Clearly, the function $u_{\mathbf{k}+\mathbf{k}_0^m}(\mathbf{r})$ is holomorphic function of $k$ ($\partial_{k^*}u_{\mathbf{k}+\mathbf{k}_0^m}(\mathbf{r}) = 0$). Furthermore, the function $u_{\mathbf{k}+\mathbf{k}_0^m}(\mathbf{r})$ is smooth throughout the mBZ but not periodic. The Chern number can be calculated by taking line integral of Berry connection around the Brillouin zone. To this end, first notice that the norm of the function $N_\mathbf{k}^{-2} = \langle u_{\mathbf{k}+\mathbf{k}_0^m}(\mathbf{r})|u_{\mathbf{k}+\mathbf{k}_0^m}(\mathbf{r})\rangle \equiv \int_\text{moir\'e unit cell} d^2\mathbf{r}\,u_{\mathbf{k}+\mathbf{k}_0^m}^*(\mathbf{r})u_{\mathbf{k}+\mathbf{k}_0^m}(\mathbf{r}) \neq 1$. Define normalized function $u_{N,\mathbf{k}+\mathbf{k}_0^m}(\mathbf{r}) = N_\mathbf{k}u_{\mathbf{k}+\mathbf{k}_0^m}(\mathbf{r})$. Then, the Berry connection is defined as $\mathbf{A}_{\mathbf{k}} \equiv i\langle u_{N,\mathbf{k}+\mathbf{k}_0}|\boldsymbol{\nabla}_\mathbf{k} u_{N,\mathbf{k}+\mathbf{k}_0}\rangle = i\int_\text{moir\'e unit cell} d^2\mathbf{r}\,u_{N,\mathbf{k}+\mathbf{k}_0^m}^*(\mathbf{r})\boldsymbol{\nabla}_\mathbf{k} u_{N,\mathbf{k}+\mathbf{k}_0^m}(\mathbf{r})$. Hence the Chern number is
\begin{equation}
\label{eq:ChernBZBoundary}
\begin{split}
	C &= \frac{1}{2\pi} (\int_{3}-\int_{1}+\int_4-\int_2)d\mathbf{k}\cdot\mathbf{A}_\mathbf{k}\\
	&=\frac{1}{2\pi}\int_{\mathbf{0}\rightarrow\mathbf{b}_1^m}d\mathbf{k}\cdot (\mathbf{A}_\mathbf{k}-\mathbf{A}_{\mathbf{k}+\mathbf{b}_2^m})+\frac{1}{2\pi}\int_{\mathbf{0}\rightarrow\mathbf{b}_2^m}d\mathbf{k}\cdot (\mathbf{A}_{\mathbf{k}+\mathbf{b}_1^m}-\mathbf{A}_\mathbf{k})
\end{split}
\end{equation}
where paths 1, 2, 3 and 4 are shown in {\color{red}Fig.~\ref{fig:path}}. However, paths $i$ and $i+2$ are related by a moir\'e recirprocal lattice vector $\mathbf{b}_1^m$, hence we can try to find the relations between $\mathbf{A}_{\mathbf{k}+\mathbf{b}_i^m}$ and $\mathbf{A}_{\mathbf{k}}$. Using the expression for $u_{\mathbf{k}+\mathbf{k}_0}(\mathbf{r})$, we find
\begin{equation}
\label{eq:ub2periodicity}
\begin{split}
	u_{\mathbf{k}+\mathbf{k}_0^m+\mathbf{b}_2^m}(\mathbf{r}) &= e^{-i(\mathbf{b}_2^m\cdot\mathbf{r})(k+b_2^m)/b_2^m}\frac{\vartheta\left(\frac{z-z_0}{a_1^m}-\frac{k+b_2^m}{b_2^m},\tau\right)}{\vartheta\left(\frac{z-z_0}{a_1^m},\tau\right)}\psi_{\mathbf{k}_0^m}(\mathbf{r})\\
	&= -e^{-i\mathbf{b}_2^m\cdot\mathbf{r}} u_{\mathbf{k}+\mathbf{k}_0^m}(\mathbf{r}) \equiv e^{i\theta_{\mathbf{k},\mathbf{b}_2^{\color{black}{m}}}} e^{-i\mathbf{b}_2^m\cdot\mathbf{r}} u_{\mathbf{k}+\mathbf{k}_0^m}(\mathbf{r}),
\end{split}
\end{equation}
where we used the fact that $\vartheta(z+1,\tau) = -\vartheta(z,\tau)$. Also,
\begin{equation}
\begin{split}
	&u_{\mathbf{k}+\mathbf{k}_0^m+\mathbf{b}_1^m}(\mathbf{r}) \\
	&= e^{-i(\mathbf{b}_2^m\cdot\mathbf{r})(k+b_1^m)/b_2^m}\frac{\vartheta\left(\frac{z-z_0}{a_1^m}-\frac{k+b_1^m}{b_2^m},\tau\right)}{\vartheta\left(\frac{z-z_0}{a_1^m},\tau\right)}\psi_{\mathbf{k}_0^m}(\mathbf{r})\\
	&= e^{-i\mathbf{b}_1^m\cdot\mathbf{r}} e^{i(\mathbf{b}_1^m - \frac{b_1^m}{b_2^m}\mathbf{b}_2^m)\cdot\mathbf{r}} (-i)\sum_{n=-\infty}^\infty(-1)^n e^{\pi i \frac{a_2^m}{a_1^m}(n+1/2)^2+\pi i(2n+1)(\frac{z-z_0}{a_1^m}-\frac{k+b_1^m}{b_2^m})}\frac{1}{\vartheta\left(\frac{z-z_0}{a_1^m},\tau\right)}\psi_{\mathbf{k}_0^m}(\mathbf{r})\\
	&= e^{-i\mathbf{b}_1^m\cdot\mathbf{r}} e^{i(\mathbf{b}_1^m - \frac{b_1^m}{b_2^m}\mathbf{b}_2^m)\cdot\mathbf{r}} (-i)\sum_{n=-\infty}^\infty(-1)^n e^{\pi i \frac{a_2^m}{a_1^m}[(n+1/2)^2+(2n+1)]+\pi i(2n+1)(\frac{z-z_0}{a_1^m}-\frac{k}{b_2^m})}\frac{1}{\vartheta\left(\frac{z-z_0}{a_1^m},\tau\right)}\psi_{\mathbf{k}_0^m}(\mathbf{r})\\
	&= e^{-i\mathbf{b}_1^m\cdot\mathbf{r}} e^{i(\mathbf{b}_1^m - \frac{b_1^m}{b_2^m}\mathbf{b}_2^m)\cdot\mathbf{r}} (-i)\sum_{n=-\infty}^\infty(-1)^n e^{\pi i \frac{a_2^m}{a_1^m}[(n+3/2)^2-1]+\pi i(2n+1)(\frac{z-z_0}{a_1^m}-\frac{k}{b_2^m})}\frac{1}{\vartheta\left(\frac{z-z_0}{a_1^m},\tau\right)}\psi_{\mathbf{k}_0^m}(\mathbf{r})\\
	&= e^{-i\mathbf{b}_1^m\cdot\mathbf{r}} e^{i(\mathbf{b}_1^m - \frac{b_1^m}{b_2^m}\mathbf{b}_2^m)\cdot\mathbf{r}}  e^{-\pi i \frac{a_2^m}{a_1^m}}(-1)e^{-2\pi i (\frac{z-z_0}{a_1^m}-\frac{k}{b_2^m})}(-i)\sum_{n=-\infty}^\infty(-1)^{n+1} e^{\pi i \frac{a_2^m}{a_1^m}[(n+3/2)^2]+\pi i(2n+3)(\frac{z-z_0}{a_1^m}-\frac{k}{b_2^m})}\frac{\psi_{\mathbf{k}_0^m}(\mathbf{r})}{\vartheta\left(\frac{z-z_0}{a_1^m},\tau\right)}\\
	&= e^{i(\mathbf{b}_1^m - \frac{b_1^m}{b_2^m}\mathbf{b}_2^m)\cdot\mathbf{r}}  e^{-\pi i \frac{a_2^m}{a_1^m}}(-1)e^{-2\pi i (\frac{z-z_0}{a_1^m}-\frac{k}{b_2^m})}e^{-i\mathbf{b}_1^m\cdot\mathbf{r}} u_{\mathbf{k}+\mathbf{k}_0^m}(\mathbf{r}),
\end{split}
\end{equation}
where we used the relations that if $a_1^m = a^m e^{i\phi_1}$ and $a_2^m = a^m e^{i\phi_2}$ ($a^m$ is length of the vector), we have $b_1^m = \frac{2\pi a^m}{|\mathbf{a}_1^m\times \mathbf{a}_2^m|}e^{i\phi_2-i\pi/2}$ and $b_2^m = \frac{2\pi a^m}{|\mathbf{a}_1^m\times \mathbf{a}_2^m|}e^{i\phi_1+i\pi/2}$, and hence $b_2^m/b_1^m = -a_1^m/a_2^m$. To further simplify the expression, we notice that
\begin{equation}
\begin{split}
	\left(\mathbf{b}_1^m - \frac{b_1^m}{b_2^m}\mathbf{b}_2^m\right)\cdot\mathbf{r} &= \frac{1}{2}\left(b_1^m z^*+(b_1^m)^*z - \frac{b_1^m}{b_2^m}(b_2^m z^*+(b_2^m)^*z)\right)\\
	&=\frac{1}{2}\left((b_1^m)^* - \frac{b_1^m}{b_2^m}(b_2^m)^*\right)z\\
	&=i\frac{(\mathbf{b}_1^m\times\mathbf{b}_2^m)\cdot \hat{z}}{b_2^m}z\\
	&= i (-\frac{2\pi i}{a_1^m})z = \frac{2\pi z}{a_1^m},
\end{split}
\end{equation}
where we used $b_2^m (b_1^m)^* -b_1^m (b_2^m)^* = 2i (\mathbf{b}_1^m\times\mathbf{b}_2^m)\cdot \hat{z}$ and $(\mathbf{b}_1^m\times\mathbf{b}_2^m)\cdot \hat{z} = \frac{4\pi^2}{\hat{z}\cdot(\mathbf{a}_1^m\times\mathbf{a}_2^m)}$. These together imply that 
\begin{equation}
\label{eq:ub1periodicity}
\begin{split}
	u_{\mathbf{k}+\mathbf{k}_0^m+\mathbf{b}_1^m}(\mathbf{r}) =  -e^{-\pi i \frac{a_2^m}{a_1^m}}e^{2\pi i \frac{z_0}{a_1^m}}e^{2\pi i\frac{k}{b_2^m}}e^{-i\mathbf{b}_1^m\cdot\mathbf{r}} u_{\mathbf{k}+\mathbf{k}_0^m}(\mathbf{r})\equiv e^{i\theta_{\mathbf{k},\mathbf{b}_1^{\color{black}{m}}}} e^{-i\mathbf{b}_1^m\cdot\mathbf{r}} u_{\mathbf{k}+\mathbf{k}_0^m}(\mathbf{r}).
\end{split}
\end{equation}
Eqs.~\ref{eq:ub1periodicity} and~\ref{eq:ub2periodicity} imply that
\begin{equation}
\begin{split}
	\frac{N_{\mathbf{k}+\mathbf{b}_i^{\color{black}{m}}}}{N_\mathbf{k}} &= e^{\Im(\theta_{\mathbf{k},\mathbf{b}_i^{\color{black}{m}}})}.\\
	u_{N,\mathbf{k}+\mathbf{k}_0^m+\mathbf{b}_i^{\color{black}{m}}}&=e^{i\Re(\theta_{\mathbf{k},\mathbf{b}_i^{\color{black}{m}}})}e^{-i\mathbf{b}_i^{\color{black}{m}}\cdot \mathbf{r}}u_{N,\mathbf{k}+\mathbf{k}_0^m}.
\end{split}
\end{equation}
Using this, one finds $\mathbf{A}_{\mathbf{k}+\mathbf{b}_i^{\color{black}{m}}} = \mathbf{A}_{\mathbf{k}}-\boldsymbol{\nabla}_\mathbf{k}\Re(\theta_{\mathbf{k},\mathbf{b}_i^{\color{black}{m}}})$. Notice that in our case,
\begin{equation}
\begin{split}
	\boldsymbol{\nabla}_\mathbf{k}\Re(\theta_{\mathbf{k},\mathbf{b}_2^{\color{black}{m}}}) &= 0\\
	\boldsymbol{\nabla}_\mathbf{k}\Re(\theta_{\mathbf{k},\mathbf{b}_1^{\color{black}{m}}}) &= \frac{\hat{z}\cdot(\mathbf{a}_1^m\times\mathbf{a}_2^m)}{a^m}(-\sin\phi_1,\cos\phi_1),
\end{split}
\end{equation}
where we used the expressions $k = k_x+ik_y$ and $b_2^m = \frac{2\pi a^m}{|\mathbf{a}_1^m\times \mathbf{a}_2^m|}e^{i\phi_1+i\pi/2}$. Therefore, using Eq.~\ref{eq:ChernBZBoundary} the Chern number is
\begin{equation}
\begin{split}
	C &= \frac{1}{2\pi}\int_{\mathbf{0}\rightarrow\mathbf{b}_2^m}d\mathbf{k}\cdot \left(- \frac{\hat{z}\cdot(\mathbf{a}_1^m\times\mathbf{a}_2^m)}{a^m}(-\sin\phi_1,\cos\phi_1)\right)\\
	&= \frac{1}{2\pi}\int_0^1 dx \frac{2\pi a^m}{|\mathbf{a}_1^m\times \mathbf{a}_2^m|}(-\sin\phi_1,\cos\phi_1)\cdot \left(- \frac{\hat{z}\cdot(\mathbf{a}_1^m\times\mathbf{a}_2^m)}{a^m}(-\sin\phi_1,\cos\phi_1)\right)\\
	&= -1,
\end{split}
\end{equation}
where we parametrized the line $\mathbf{0}\rightarrow\mathbf{b}_2^m$ by $x$. Therefore, the Chern number is always $|C| = 1$ (whereas the sign of the Chern number depends on the convention).
\subsection{Topology of more than two flatbands -- multiple flat bands on each sub-lattice}
\begin{figure}[h!]
     \centering
\includegraphics[scale=1]{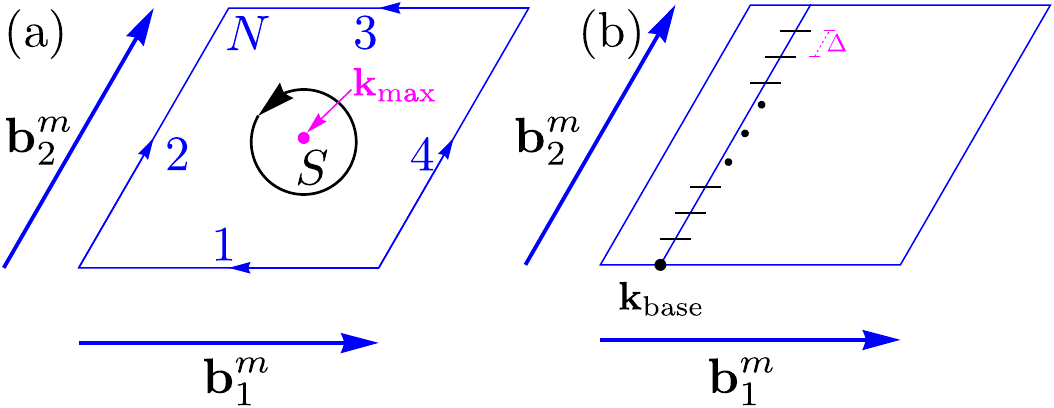}
     \caption{(a) If a WF is smooth at all $\mathbf{k}$ in the Brillouin zone defined by the reciprocal lattice vectors $\mathbf{b}_i^m$, its Chern number can be evaluated from the line integrals of the Berry connections along the paths $1, 2, 3$ and $4$ as shown in Eq.~\ref{eq:ChernBZBoundary}. If, in some gauge, the WF is singular at $\mathbf{k}_\text{max}$, the Brillouin zone needs to be divided into patches $N$ and $S$, where $S$ contains $\mathbf{k}_\text{max}$. A different gauge for the WF is needed in $S$ that is smooth in $S$. (b) A noncontractible loop parallel to $\mathbf{b}_2^m$ with base momentum $\mathbf{k}_\text{base}$. The Wilson loop is evaluated along such loops. For numerical evaluation of Wilson loop, the noncontractible loop needs to be discretized. The Wilson loop is then a function of $\mathbf{k}_\text{base}$, which can be varied parallel to $\mathbf{b}_1^m$.
    }
     \label{fig:path}
\end{figure}
As was described in the main text, if there are multiple zeros of $\psi_{\mathbf{k}_\text{max}}$ at $\mathbf{r}_0^{(1)},\dots,\mathbf{r}_0^{(m)}$, then $m$ FB WFs $\mathbf{f}_{\mathbf{k}}(z;\mathbf{r}_0^{(i)})\psi_{\mathbf{k}_\text{max}}(\mathbf{r})$ can be constructed on one sublattice. These WFs are smooth functions of $\mathbf{k}$, and are independent (but not mutually orthogonal) of each other everywhere except at $\mathbf{k} = 0$ since $\mathbf{f}_{\mathbf{0}}(z;\mathbf{r}_0^{(i)}) = 1$ for all $i$. For concreteness and simplicity, let us focus on $m = 2$, but the procedure can be straightforwardly extended to any integer value of $m$. Similar to what was shown in the main text, after orthogonalization, 
\begin{equation}
\label{eq:PartialOrthogonalization2}
    \tilde{u}_{\mathbf{k}}^{(i)}(\mathbf{r}) = \begin{cases}
        u_\mathbf{k}^{(1)}(\mathbf{r}) &\text{ if } i =1,\\
        u_\mathbf{k}^{(2)}(\mathbf{r}) -\frac{\langle u_\mathbf{k}^{(1)}|u_\mathbf{k}^{(2)}\rangle}{\langle u_\mathbf{k}^{(1)}|u_\mathbf{k}^{(1)}\rangle}u_\mathbf{k}^{(1)}(\mathbf{r}) &\text{ if } i = 2,\\
    \end{cases}
\end{equation}
where $u_{\mathbf{k}+\mathbf{k}_\text{max}}^{(i)}(\mathbf{r})= e^{-i\mathbf{k}\cdot\mathbf{r}}f_{\mathbf{k}}(z;\mathbf{r}_0^{(i)})\psi_{\mathbf{k}_\text{max}}(\mathbf{r}) = \tilde{f}_{\mathbf{k}}(z;\mathbf{r}_0^{(i)})\psi_{\mathbf{k}_\text{max}}(\mathbf{r})$ is the periodic part of the full Bloch function.
The function $\tilde{u}_\mathbf{k}^{(1)}(\mathbf{r})$ is exactly the same as the one discussed for two FB case, and has Chern number -1. The function $\tilde{u}_\mathbf{k}^{(2)}(\mathbf{r})$ is more interesting. It is smooth at any $\mathbf{k}$ except $\mathbf{k}_\text{max}$ due to the reason mentioned above. Furthermore, around $\mathbf{k}_\text{max}$ the function $\tilde{u}_\mathbf{k}^{(2)}(\mathbf{r})$ has the form
\begin{equation}
    \tilde{u}_{\mathbf{k}+\mathbf{k}_\text{max}}^{(2)}(\mathbf{r}) \approx k(\tilde{f}_\mathbf{0}'(\mathbf{r};\mathbf{r}_0^{(2)})-\tilde{f}_\mathbf{0}'(\mathbf{r};\mathbf{r}_0^{(1)})-\langle\psi_{\mathbf{k}_\text{max}}|(\tilde{f}_\mathbf{0}'(\mathbf{r};\mathbf{r}_0^{(2)})-\tilde{f}_\mathbf{0}'(\mathbf{r};\mathbf{r}_0^{(1)}))\psi_{\mathbf{k}_\text{max}}\rangle)\psi_{\mathbf{k}_\text{max}}(\mathbf{r})
\end{equation}
where $k=k_x+ik_y$ and $\tilde{f}_\mathbf{0}'(\mathbf{r};\mathbf{r}_0^{(i)}) = \left.\partial_k \tilde{f}_\mathbf{k}(\mathbf{r};\mathbf{r}_0^{(i)})\right|_{\mathbf{k}=\mathbf{0}}$, where $\partial_k =(\partial_{k_x}-i\partial_{k_y})/2$. Its phase winds around $\mathbf{k}_\text{max}$. However, the gauge $e^{-i\varphi_\mathbf{k}}\tilde{u}_{\mathbf{k}+\mathbf{k}_\text{max}}^{(2)}(\mathbf{r})$ (where $\varphi_\mathbf{k} = \text{arg}(k_x+ik_y)$) is smooth. However, the Berry curvature $F_{xy}(\mathbf{k})$ is gauge independent. So, while calculating the Chern number using the usual formula $C = \frac{1}{2\pi}\int_\text{mBZ}d^2\mathbf{k} F_{xy}(\mathbf{k})$ (where $F_{xy}(\mathbf{k}) = \partial_{k_x}A_y(\mathbf{k})-\partial_{k_y}A_x(\mathbf{k})$), we can break the mBZ into two regions $S$ and $N$ (see {\color{red}Fig.~\ref{fig:path}}), and use $\tilde{u}^{S}_{\mathbf{k}+\mathbf{k}_\text{max}}(\mathbf{r})=e^{-i\varphi_\mathbf{k}}\tilde{u}_{\mathbf{k}+\mathbf{k}_\text{max}}^{(2)}(\mathbf{r})$ and $\tilde{u}^{N}_{\mathbf{k}+\mathbf{k}_\text{max}}(\mathbf{r})=\tilde{u}_{\mathbf{k}+\mathbf{k}_\text{max}}^{(2)}(\mathbf{r})$ in the two regions respectively. Using the normalization factor $N_\mathbf{k}$ ($N_\mathbf{k}^{-2}= \langle\tilde{u}^{S}_{\mathbf{k}+\mathbf{k}_\text{max}}|\tilde{u}^{S}_{\mathbf{k}+\mathbf{k}_\text{max}}\rangle = \langle\tilde{u}^{N}_{\mathbf{k}+\mathbf{k}_\text{max}}|\tilde{u}^{N}_{\mathbf{k}+\mathbf{k}_\text{max}}\rangle$), we define $\mathbf{A}^{S/N}_\mathbf{k}\equiv i\langle N_\mathbf{k} u_{\mathbf{k}+\mathbf{k}_\text{max}}^{S/N}|\boldsymbol{\nabla}_\mathbf{k}N_\mathbf{k} u_{\mathbf{k}+\mathbf{k}_\text{max}}^{S/N}\rangle$. Then, in the overlap region between the $S$ and $N$ regions, we have $\mathbf{A}^{S}_\mathbf{k} = \mathbf{A}^{N}_\mathbf{k}+\boldsymbol{\nabla}_\mathbf{k}\varphi_\mathbf{k}$. Using all of these, we get the expression for the Chern number of the second band to be
\begin{equation}
\begin{split}
    C &= \frac{1}{2\pi}\int_\text{mBZ}d^2\mathbf{k} F_{xy}(\mathbf{k})\\
    &= \frac{1}{2\pi}\left(\int_Sd^2\mathbf{k} F_{xy}^S(\mathbf{k})+\int_Nd^2\mathbf{k} F_{xy}^N(\mathbf{k})\right) \\
    &= \frac{1}{2\pi}\left(\int_{\partial S}d\mathbf{k}\cdot (\mathbf{A}_{\mathbf{k}}^S-\mathbf{A}_{\mathbf{k}}^N)+\int_{\partial \text{mBZ}}d\mathbf{k}\cdot \mathbf{A}_{\mathbf{k}}^N\right)\\
    &= \frac{1}{2\pi}\left(\int_{\partial S}d\mathbf{k}\cdot \boldsymbol{\nabla}_\mathbf{k}\varphi_\mathbf{k}+\int_{\partial \text{mBZ}}d\mathbf{k}\cdot \mathbf{A}_{\mathbf{k}}^N\right)\\
    &= \frac{1}{2\pi}\int_{\partial \text{mBZ}}d\mathbf{k}\cdot \mathbf{A}_{\mathbf{k}}^N+1.
\end{split}
\end{equation}
Now the integral over the edge of the mBZ is very similar to what we did in the previous subsection. We know that 
\begin{equation}
\begin{split}
    &\frac{1}{2\pi}\int_{\partial \text{mBZ}}d\mathbf{k}\cdot \mathbf{A}_{\mathbf{k}}^N\\
    =& \frac{1}{2\pi} (\int_{3}-\int_{1}+\int_4-\int_2)d\mathbf{k}\cdot\mathbf{A}_\mathbf{k}^N\\
	=&\frac{1}{2\pi}\int_{-\mathbf{b}_1^m/2\rightarrow\mathbf{b}_1^m/2}d\mathbf{k}\cdot (\mathbf{A}_\mathbf{k}^N-\mathbf{A}_{\mathbf{k}+\mathbf{b}_2^m}^N)+\frac{1}{2\pi}\int_{-\mathbf{b}_2^m/2\rightarrow\mathbf{b}_2^m/2}d\mathbf{k}\cdot (\mathbf{A}_{\mathbf{k}+\mathbf{b}_1^m}^N-\mathbf{A}_\mathbf{k}^N).
\end{split}
\end{equation}
Now, using Eq.~\ref{eq:ub1periodicity} and Eq.~\ref{eq:ub2periodicity} we get
\begin{equation}
\begin{split}
    &\tilde{u}_{\mathbf{k}+\mathbf{k}_\text{max}+\mathbf{b}_i^m}^{N}(\mathbf{r})\\
    =&\tilde{u}_{\mathbf{k}+\mathbf{k}_\text{max}+\mathbf{b}_i^m}^{(2)}(\mathbf{r})\\
    =& u_{\mathbf{k}+\mathbf{k}_\text{max}+\mathbf{b}_i^m}^{(2)}(\mathbf{r}) -\frac{\langle u_{\mathbf{k}+\mathbf{k}_\text{max}+\mathbf{b}_i^m}^{(1)}|u_{\mathbf{k}+\mathbf{k}_\text{max}+\mathbf{b}_i^m}^{(2)}\rangle}{\langle u_{\mathbf{k}+\mathbf{k}_\text{max}+\mathbf{b}_i^m}^{(1)}|u_{\mathbf{k}+\mathbf{k}_\text{max}+\mathbf{b}_i^m}^{(1)}\rangle}u_{\mathbf{k}+\mathbf{k}_\text{max}+\mathbf{b}_i^m}^{(1)}(\mathbf{r})\\
    =& e^{-i\mathbf{b}_i^m\cdot\mathbf{r}}\left[e^{i\theta_{\mathbf{k},\mathbf{b}_1^{\color{black}{m}}}^{(2)}}u_{\mathbf{k}+\mathbf{k}_\text{max}}^{(2)}(\mathbf{r})- \frac{\langle e^{i\theta_{\mathbf{k},\mathbf{b}_i^{\color{black}{m}}}^{(1)}} u_{\mathbf{k}+\mathbf{k}_\text{max}}^{(1)}| e^{i\theta_{\mathbf{k},\mathbf{b}_i^{\color{black}{m}}}^{(2)}}u_{\mathbf{k}+\mathbf{k}_\text{max}}^{(2)}\rangle}{\langle e^{i\theta_{\mathbf{k},\mathbf{b}_i^{\color{black}{m}}}^{(1)}} u_{\mathbf{k}+\mathbf{k}_\text{max}}^{(1)}|e^{i\theta_{\mathbf{k},\mathbf{b}_i^{\color{black}{m}}}^{(1)}}u_{\mathbf{k}+\mathbf{k}_\text{max}}^{(1)}\rangle}e^{i\theta_{\mathbf{k},\mathbf{b}_i^{\color{black}{m}}}^{(1)}}u_{\mathbf{k}+\mathbf{k}_\text{max}}^{(1)}(\mathbf{r})\right]\\
    =& e^{-i\mathbf{b}_i^m\cdot\mathbf{r}}e^{i\theta_{\mathbf{k},\mathbf{b}_i^{\color{black}{m}}}^{(2)}}\left[u_{\mathbf{k}+\mathbf{k}_\text{max}}^{(2)}(\mathbf{r})- \frac{\langle u_{\mathbf{k}+\mathbf{k}_\text{max}}^{(1)}| u_{\mathbf{k}+\mathbf{k}_\text{max}}^{(2)}\rangle}{\langle u_{\mathbf{k}+\mathbf{k}_\text{max}}^{(1)}|u_{\mathbf{k}+\mathbf{k}_\text{max}}^{(1)}\rangle}u_{\mathbf{k}+\mathbf{k}_\text{max}}^{(1)}(\mathbf{r})\right]\\
    =&e^{-i\mathbf{b}_i^m\cdot\mathbf{r}}e^{i\theta_{\mathbf{k},\mathbf{b}_i^{\color{black}{m}}}^{(2)}}\tilde{u}_{\mathbf{k}+\mathbf{k}_\text{max}}^{(2)}(\mathbf{r})\\
    =&e^{-i\mathbf{b}_i^m\cdot\mathbf{r}}e^{i\theta_{\mathbf{k},\mathbf{b}_i^{\color{black}{m}}}^{(2)}}\tilde{u}_{\mathbf{k}+\mathbf{k}_\text{max}}^{N}(\mathbf{r}),
\end{split}
\end{equation}
which means that the ratio of wave-functions at the two sides of the mBZ related by reciprocal lattice vector remains the same before and after the orthogonalization in Eq.~\ref{eq:PartialOrthogonalization2}. But, we know from the previous subsection that the integral $\frac{1}{2\pi}\int_{\partial \text{mBZ}}d\mathbf{k}\cdot \mathbf{A}_{\mathbf{k}}^N$ only depends on this ratio. Therefore, using our knowledge from the previous subsection, we immediately obtain $\frac{1}{2\pi}\int_{\partial \text{mBZ}}d\mathbf{k}\cdot \mathbf{A}_{\mathbf{k}}^N=-1$. Therefore, the Chern number of the second band is
\begin{equation}
    C= \frac{1}{2\pi}\int_{\partial \text{mBZ}}d\mathbf{k}\cdot \mathbf{A}_{\mathbf{k}}^N+1 = -1+1=0.
\end{equation}
A similar calculation shows that the Chern number of the third band would also be zero in the case of $m=3$. This proves that the total Chern number of all FB WFs polarized on the same sublattice is $C =\pm 1$.

\subsection{Details of Wilson loop spectrum calculation}
To numerically capture the topology of the flat bands, we evaluate Wilson loop spectrum as shown in {\color{red}Figs.~1-2 of main text as well as Figs.~S3-S7, S9}. Here we give details of the Wilson loop calculation. The Wilson loop at point $\mathbf{k}_\text{base}$ along a noncontractible loop in the $\mathbf{b}_2^m$ direction (see Fig.~\ref{fig:path}(b)) is defined as
\begin{equation}
W_{\mathbf{k}_\text{base}+\mathbf{b}_2^m\leftarrow \mathbf{k}_\text{base}}^{mn}= \lim_{N\rightarrow \infty} \sum_{i_1,\dots,i_{N-1}\in \text{FBs}}\langle u_{\mathbf{k}_\text{base}+\mathbf{b}_2^m}^{m}| u_{\mathbf{k}_\text{base}+\frac{N-1}{N}\mathbf{b}_2^m}^{i_{N-1}}\rangle \langle u_{\mathbf{k}_\text{base}+\frac{N-1}{N}\mathbf{b}_2^m}^{i_{N-1}}|\dots | u_{\mathbf{k}_\text{base}+\frac{1}{N}\mathbf{b}_2^m}^{i_{1}}\rangle\langle  u_{\mathbf{k}_\text{base}+\frac{1}{N}\mathbf{b}_2^m}^{i_{1}}| u_{\mathbf{k}_\text{base}}^{n}\rangle.
\end{equation}
In numerical evaluations, $N$ is finite. The quantities $\theta_{i}(\mathbf{k}) = \Im\log \left(\text{eig}_i (W_{\mathbf{k}+\mathbf{b}_2^m\leftarrow \mathbf{k}})\right)$ are plotted as vary $\mathbf{k}$ between $\mathbf{0}\rightarrow\mathbf{b}_1^m$ in the figures mentioned above. Due to the anti-unitary particle-hole symmetry $\mathcal{P}$ ($\mathcal{P}=\mathcal{S}\mathcal{T}$ in QBCP systems and $\mathcal{P} = P\mathcal{C}_{2z}\mathcal{T}$ in TBG (see Sec.~S-I)) that satisfies $\mathcal{P}^2 = -\mathds{1}$, the Wilson loop spectrum $\theta_{i}(\mathbf{k}) $ is (Kramers) doubly degenerate at $\mathbf{k} = \mathbf{0}, \frac{1}{2}\mathbf{b}^m$ and $\mathbf{b}^m$~\citesupp{TBGIIBernevigs}. Furthermore, for $\mathcal{C}_{2z}\mathcal{T}$ symmetric systems, the degeneracies at $\theta_{i}(\mathbf{k}) = 0,\pi$ are protected. In all examples discussed here (except twisted bilayer checkerboard lattice, which is discussed later), the winding of the Wilson loop spectrum is $\pm 1$, which is consistent with the analysis done earlier in this section.

\section{Ideal quantum geometry of the flat-bands}
In this section we prove that the set of FBs polarized on each sublattice satisfy ideal non-Abelian quantum geometry. 

As was described in the main text, if there are multiple zeros of $\psi_{\mathbf{k}_\text{max}}$ at $\mathbf{r}_0^{(1)},\dots,\mathbf{r}_0^{(m)}$, then $m$ FB WFs $\mathbf{f}_{\mathbf{k}}(z;\mathbf{r}_0^{(i)})\psi_{\mathbf{k}_\text{max}}(\mathbf{r})$ can be constructed on one sublattice. These WFs are smooth functions of $\mathbf{k}$, and are independent (but not mutually orthogonal) of each other everywhere except at $\mathbf{k} = 0$ since $\mathbf{f}_{\mathbf{0}}(z;\mathbf{r}_0^{(i)}) = 1$ for all $i$. For concreteness and simplicity, let us focus on $m = 2$, but the procedure can be straightforwardly extended to any integer value of $m$. Similar to what was shown in the main text, after orthogonalization, 
\begin{equation}
\label{eq:PartialOrthogonalization1}
    \tilde{u}_{\mathbf{k}}^{(i)}(\mathbf{r}) = \begin{cases}
        u_\mathbf{k}^{(1)}(\mathbf{r}) &\text{ if } i =1,\\
        u_\mathbf{k}^{(2)}(\mathbf{r}) -\frac{\langle u_\mathbf{k}^{(1)}|u_\mathbf{k}^{(2)}\rangle}{\langle u_\mathbf{k}^{(1)}|u_\mathbf{k}^{(1)}\rangle}u_\mathbf{k}^{(1)}(\mathbf{r}) &\text{ if } i = 2,\\
    \end{cases}
\end{equation}
where $u_{\mathbf{k}+\mathbf{k}_\text{max}}^{(i)}(\mathbf{r})= e^{-i\mathbf{k}\cdot\mathbf{r}}f_{\mathbf{k}}(z;\mathbf{r}_0^{(i)})\psi_{\mathbf{k}_\text{max}}(\mathbf{r}) = \tilde{f}_{\mathbf{k}}(z;\mathbf{r}_0^{(i)})\psi_{\mathbf{k}_\text{max}}(\mathbf{r})$ is the periodic part of the full Bloch function. In the case of single FB per sublattice, the ideal quantum geometry of the FB follows form the fact that $u_\mathbf{k}^{(1)}(\mathbf{r})$ is a holomorphic function of $k$~\citesupp{claassen2015positions,ledwith2020fractionals}. However, here the the orthogonalization of $u_\mathbf{k}^{(2)}(\mathbf{r})$ from $u_\mathbf{k}^{(1)}(\mathbf{r})$ gives factors like $\langle u_\mathbf{k}^{(1)}|u_\mathbf{k}^{(2)}\rangle$, which makes $\tilde{u}_{\mathbf{k}}^{(2)}(\mathbf{r})$ non-holomorphic in $k$. However, below we show that the two WFs $\tilde{u}_{\mathbf{k}}^{(1)}(\mathbf{r})$ and $\tilde{u}_{\mathbf{k}}^{(2)}(\mathbf{r})$ together satisfies ideal non-Abelian quantum geometry.

The non-Abelian Fubiny-Study metric~\citesupp{marzari1997maximallys,resta2011insulatings,marzari2012maximallys,xie2020topologys,ledwith2020fractionals} for the bands polarized on one sublattice is defined as:
\begin{equation}
    g_{\alpha\beta}^{mn}(\mathbf{k}) = \Re\left[\langle \partial_{k_\alpha}\tilde{u}_{N,\mathbf{k}}^{(m)}| \left(\mathds{1} - \sum_{n_1=1}^2 |\tilde{u}_{N,\mathbf{k}}^{(n_1)}\rangle\langle \tilde{u}_{N,\mathbf{k}}^{(n_1)}|\right)| \partial_{k_\beta}\tilde{u}_{N,\mathbf{k}}^{(n)}\rangle\right],
\end{equation}
where the sum over $n_1$ is restricted to the FBs polarized on one sublattice, $\tilde{u}_{N,\mathbf{k}}^{(m)}$ are the normalized FB WFs, and $\langle f|g\rangle \equiv\int_\text{moir\'e unit cell} d^2\mathbf{r}\,f^*(\mathbf{r})g(\mathbf{r})$. Then, the trace of the non-Abelian Fubini-Study metric can be written in terms of unnormalized WFs as the following:
\begin{equation}
    \text{tr}(g_{\alpha\beta}^{mn}(\mathbf{k})) = \sum_{n=1}^2\sum_{\alpha \in \{x,y\}} g_{\alpha\alpha}^{nn}(\mathbf{k}) = \sum_{n=1}^2\sum_{\alpha \in \{x,y\}}\left(\frac{\langle\partial_{k_\alpha}\tilde{u}_{\mathbf{k}}^{(n)}| \partial_{k_\alpha}\tilde{u}_{\mathbf{k}}^{(n)} \rangle}{||\tilde{u}_{\mathbf{k}}^{(n)}||^2}-\sum_{n'=1}^2\frac{\langle\partial_{k_\alpha}\tilde{u}_{\mathbf{k}}^{(n)}|\tilde{u}_{\mathbf{k}}^{(n')} \rangle\langle\tilde{u}_{\mathbf{k}}^{(n')}|\partial_{k_\alpha}\tilde{u}_{\mathbf{k}}^{(n)} \rangle}{||\tilde{u}_{\mathbf{k}}^{(n)}||^2 ||\tilde{u}_{\mathbf{k}}^{(n')}||^2}\right),
\end{equation}
where $||f||^2 = \langle f|f\rangle$.

Now, we can use the expressions in Eq.~\eqref{eq:PartialOrthogonalization1}, and take advantage of the fact that $u_\mathbf{k}^{(i)}(\mathbf{r})$ is holomorphic function in the following way. First, writing, $\partial_{k_x} = (\partial_k +\overline{\partial_k})$ and $\partial_{k_y} = i(\partial_k -\overline{\partial_k})$ (here $\partial_k = \frac{1}{2} (\partial_{k_x}-i\partial_{k_y})$ and $\overline{\partial_k} = \frac{1}{2} (\partial_{k_x}+i\partial_{k_y})$), we find
\begin{equation}
\label{eq:trgComplex}
    \text{tr}(g_{\alpha\beta}^{mn}(\mathbf{k})) = 2\sum_{n=1}^2 \left(\frac{\langle\partial_{k}\tilde{u}_{\mathbf{k}}^{(n)}| \partial_{k}\tilde{u}_{\mathbf{k}}^{(n)} \rangle}{||\tilde{u}_{\mathbf{k}}^{(n)}||^2}+\frac{\langle\overline{\partial_{k}}\tilde{u}_{\mathbf{k}}^{(n)}| \overline{\partial_{k}}\tilde{u}_{\mathbf{k}}^{(n)} \rangle}{||\tilde{u}_{\mathbf{k}}^{(n)}||^2}-\sum_{n'=1}^2\left(\frac{\langle\partial_{k}\tilde{u}_{\mathbf{k}}^{(n)}|\tilde{u}_{\mathbf{k}}^{(n')} \rangle\langle\tilde{u}_{\mathbf{k}}^{(n')}|\partial_{k}\tilde{u}_{\mathbf{k}}^{(n)} \rangle}{||\tilde{u}_{\mathbf{k}}^{(n)}||^2 ||\tilde{u}_{\mathbf{k}}^{(n')}||^2}+\frac{\langle\overline{\partial_{k}}\tilde{u}_{\mathbf{k}}^{(n)}|\tilde{u}_{\mathbf{k}}^{(n')} \rangle\langle\tilde{u}_{\mathbf{k}}^{(n')}|\overline{\partial_{k}}\tilde{u}_{\mathbf{k}}^{(n)} \rangle}{||\tilde{u}_{\mathbf{k}}^{(n)}||^2 ||\tilde{u}_{\mathbf{k}}^{(n')}||^2}\right)\right).
\end{equation}
Furthermore, we have that
\begin{equation}
\label{eq:uIdentities}
\begin{split}
    &\overline{\partial_{k}}\tilde{u}_\mathbf{k}^{(1)}(\mathbf{r}) = 0\\
    &\overline{\partial_{k}}\tilde{u}_\mathbf{k}^{(2)}(\mathbf{r}) = \overline{\partial_{k}} \left(u_\mathbf{k}^{(2)}(\mathbf{r}) -\frac{\langle u_\mathbf{k}^{(1)}|u_\mathbf{k}^{(2)}\rangle}{\langle u_\mathbf{k}^{(1)}|u_\mathbf{k}^{(1)}\rangle}u_\mathbf{k}^{(1)}(\mathbf{r})\right) = -\frac{\langle \partial_k u_\mathbf{k}^{(1)}|u_\mathbf{k}^{(2)}\rangle}{\langle u_\mathbf{k}^{(1)}|u_\mathbf{k}^{(1)}\rangle}u_\mathbf{k}^{(1)}(\mathbf{r})+\frac{\langle  u_\mathbf{k}^{(1)}|u_\mathbf{k}^{(2)}\rangle}{\langle u_\mathbf{k}^{(1)}|u_\mathbf{k}^{(1)}\rangle^2}\langle \partial_ku_\mathbf{k}^{(1)} |u_\mathbf{k}^{(1)}\rangle u_\mathbf{k}^{(1)}(\mathbf{r})\\
    &\phantom{\overline{\partial_{k}}\tilde{u}_\mathbf{k}^{(2)}(\mathbf{r})}= -\frac{\langle \partial_k \tilde{u}_\mathbf{k}^{(1)}|\tilde{u}_\mathbf{k}^{(2)}\rangle}{\langle \tilde{u}_\mathbf{k}^{(1)}|\tilde{u}_\mathbf{k}^{(1)}\rangle}\tilde{u}_\mathbf{k}^{(1)}(\mathbf{r}).
\end{split}
\end{equation}
The last line also means that $\langle\tilde{u}_\mathbf{k}^{(2)}|\overline{\partial_{k}}\tilde{u}_\mathbf{k}^{(2)}\rangle = 0$ (sine $\tilde{u}_\mathbf{k}^{(2)}$ and $\tilde{u}_\mathbf{k}^{(1)}$ are orthogonal) and $\langle\tilde{u}_\mathbf{k}^{(1)}|\overline{\partial_{k}}\tilde{u}_\mathbf{k}^{(2)}\rangle = -\langle \partial_k \tilde{u}_\mathbf{k}^{(1)}|\tilde{u}_\mathbf{k}^{(2)}\rangle$. Moreover, since $\partial_k \tilde{u}_\mathbf{k}^{(2)} = \partial_k\left(u_\mathbf{k}^{(2)} -\frac{\langle u_\mathbf{k}^{(1)}|u_\mathbf{k}^{(2)}\rangle}{\langle u_\mathbf{k}^{(1)}|u_\mathbf{k}^{(1)}\rangle}u_\mathbf{k}^{(1)}\right) = \partial_k u_\mathbf{k}^{(2)} - \frac{\langle u_\mathbf{k}^{(1)}|u_\mathbf{k}^{(2)}\rangle}{\langle u_\mathbf{k}^{(1)}|u_\mathbf{k}^{(1)}\rangle}\partial_ku_\mathbf{k}^{(1)} - \frac{\langle u_\mathbf{k}^{(1)}|\partial_k u_\mathbf{k}^{(2)}\rangle}{\langle u_\mathbf{k}^{(1)}|u_\mathbf{k}^{(1)}\rangle}u_\mathbf{k}^{(1)}+\frac{\langle u_\mathbf{k}^{(1)}|u_\mathbf{k}^{(2)}\rangle}{\langle u_\mathbf{k}^{(1)}|u_\mathbf{k}^{(1)}\rangle^2}\langle u_\mathbf{k}^{(1)}|\partial_k u_\mathbf{k}^{(1)}\rangle u_\mathbf{k}^{(1)}$, we have $\langle \tilde{u}_\mathbf{k}^{(1)}| \partial_k \tilde{u}_\mathbf{k}^{(2)}\rangle = 0$.  Plugging all of these into the expression in Eq.~\eqref{eq:trgComplex}, we find the following simplified expression
\begin{equation}
\label{eq:trgFinal}
    \text{tr}(g_{\alpha\beta}^{mn}(\mathbf{k})) = 2\left(\frac{||\partial_k \tilde{u}_\mathbf{k}^{(1)}||^2}{|| \tilde{u}_\mathbf{k}^{(1)}||^2}+\frac{||\partial_k \tilde{u}_\mathbf{k}^{(2)}||^2}{|| \tilde{u}_\mathbf{k}^{(2)}||^2} - \frac{|\langle \tilde{u}_\mathbf{k}^{(1)}|\partial_k \tilde{u}_\mathbf{k}^{(1)}\rangle|^2}{|| \tilde{u}_\mathbf{k}^{(1)}||^4} - \frac{|\langle \tilde{u}_\mathbf{k}^{(2)}|\partial_k \tilde{u}_\mathbf{k}^{(2)}\rangle|^2}{|| \tilde{u}_\mathbf{k}^{(2)}||^4}- \frac{|\langle \tilde{u}_\mathbf{k}^{(2)}|\partial_k \tilde{u}_\mathbf{k}^{(1)}\rangle|^2}{|| \tilde{u}_\mathbf{k}^{(1)}||^2 || \tilde{u}_\mathbf{k}^{(2)}||^2}\right).
\end{equation}
Clearly, $\text{tr}(g_{\alpha\beta}^{mn}(\mathbf{k})) >0$.

Similarly, the expression for the trace of the non-Abelian Berry curvature is 
\begin{equation}
\begin{split}
    \text{tr}(F_{xy}^{mn}(\mathbf{k})) &= \sum_{n=1}^2 F_{xy}^{nn}(\mathbf{k}) = \sum_{n=1}^2i \left(\langle\partial_{k_x}\tilde{u}_{\mathbf{k}}^{(n)}| \partial_{k_y}\tilde{u}_{\mathbf{k}}^{(n)}\rangle- \langle\partial_{k_y}\tilde{u}_{{\mathbf{k}}}^{(n)}| \partial_{k_x}\tilde{u}_{{\mathbf{k}}}^{(n)}\rangle\right)\\
    &=\sum_{n=1}^2 i\left[\left(\frac{\langle\partial_{k_x}\tilde{u}_{\mathbf{k}}^{(n)}| \partial_{k_y}\tilde{u}_{\mathbf{k}}^{(n)} \rangle}{||\tilde{u}_{\mathbf{k}}^{(n)}||^2}- (x\leftrightarrow y)\right) - \left(\frac{\langle\partial_{k_x}\tilde{u}_{\mathbf{k}}^{(n)}|\tilde{u}_{\mathbf{k}}^{(n)}\rangle\langle \tilde{u}_{\mathbf{k}}^{(n)}| \partial_{k_y}\tilde{u}_{\mathbf{k}}^{(n)} \rangle}{||\tilde{u}_{\mathbf{k}}^{(n)}||^4}- (x\leftrightarrow y)\right)\right]\\
    &=\sum_{n=1}^2 2i\left[\left(\frac{\langle\partial_{k}\tilde{u}_{\mathbf{k}}^{(n)}| \partial_{k}\tilde{u}_{\mathbf{k}}^{(n)} \rangle}{||\tilde{u}_{\mathbf{k}}^{(n)}||^2}- \frac{\langle\overline{\partial_{k}}\tilde{u}_{\mathbf{k}}^{(n)}| \overline{\partial_{k}}\tilde{u}_{\mathbf{k}}^{(n)} \rangle}{||\tilde{u}_{\mathbf{k}}^{(n)}||^2}\right) - \left(\frac{\langle\partial_{k}\tilde{u}_{\mathbf{k}}^{(n)}|\tilde{u}_{\mathbf{k}}^{(n)}\rangle\langle \tilde{u}_{\mathbf{k}}^{(n)}| \partial_{k}\tilde{u}_{\mathbf{k}}^{(n)} \rangle}{||\tilde{u}_{\mathbf{k}}^{(n)}||^4}- \frac{\langle\overline{\partial_{k}}\tilde{u}_{\mathbf{k}}^{(n)}|\tilde{u}_{\mathbf{k}}^{(n)}\rangle\langle \tilde{u}_{\mathbf{k}}^{(n)}| \overline{\partial_{k}}\tilde{u}_{\mathbf{k}}^{(n)} \rangle}{||\tilde{u}_{\mathbf{k}}^{(n)}||^4}\right)\right].
\end{split}
\end{equation}
Now, using the identities in Eq.~\eqref{eq:uIdentities} and below it, we simplify the expression to get the following
\begin{equation}
\label{eq:trFFinal}
    \text{tr}(F_{xy}^{mn}(\mathbf{k})) =2i\left(\frac{||\partial_k \tilde{u}_\mathbf{k}^{(1)}||^2}{|| \tilde{u}_\mathbf{k}^{(1)}||^2}+\frac{||\partial_k \tilde{u}_\mathbf{k}^{(2)}||^2}{|| \tilde{u}_\mathbf{k}^{(2)}||^2} - \frac{|\langle \tilde{u}_\mathbf{k}^{(1)}|\partial_k \tilde{u}_\mathbf{k}^{(1)}\rangle|^2}{|| \tilde{u}_\mathbf{k}^{(1)}||^4} - \frac{|\langle \tilde{u}_\mathbf{k}^{(2)}|\partial_k \tilde{u}_\mathbf{k}^{(2)}\rangle|^2}{|| \tilde{u}_\mathbf{k}^{(2)}||^4}- \frac{|\langle \tilde{u}_\mathbf{k}^{(2)}|\partial_k \tilde{u}_\mathbf{k}^{(1)}\rangle|^2}{|| \tilde{u}_\mathbf{k}^{(1)}||^2 || \tilde{u}_\mathbf{k}^{(2)}||^2}\right).
\end{equation}
Comparing Eq.~\eqref{eq:trgFinal} and Eq.~\eqref{eq:trFFinal}, we have
\begin{equation}
    \text{tr}(g_{\alpha\beta}^{mn}(\mathbf{k}))=|\text{tr}(F_{xy}^{mn}(\mathbf{k}))|.
\end{equation}

\section{Examples}
Below, we show 5 new examples listed in the Fig.~3(b) of the main text: 2 FBs in systems with QBCP under periodic strain having $p3$ and $p4$ space group symmetries, 4 FBs in systems with QBCP under periodic strain having $p4$, $p4mm$ and $p4gm$ space group symmetry.

\subsection{2 flat bands in single layer system with QBCP under moir\'e potential with space group symmetry $p3$}
\begin{figure}[h!]
     \centering
\includegraphics[scale=1.1]{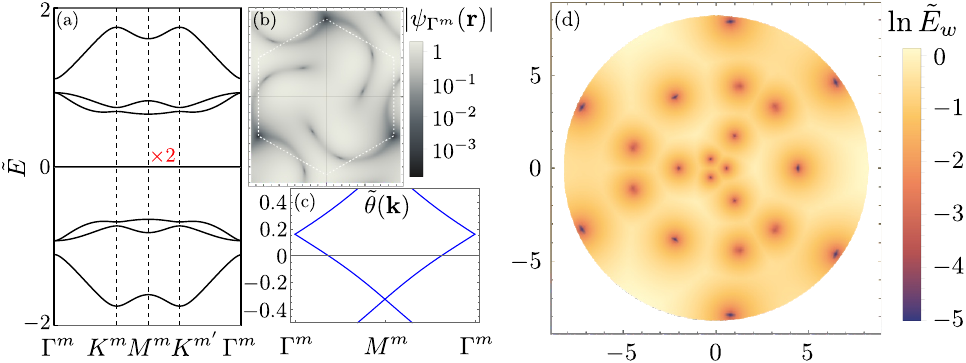}
     \caption{
     Flat bands in system with QBCP under moir\'e potential $\mathcal{D}_U(\mathbf{r};\boldsymbol{\alpha}=\alpha)= \frac{\alpha}{2}\sum_{n=1}^{3} e^{i(1-n)\phi} \;\;\exp \left(-i (\mathbf{b}_n^m \cdot \mathbf{r}+\phi_{1}) \right)$.
     Here, $\phi=2\pi /3$, $\mathbf{b}^m_{1} = \frac{4\pi}{\sqrt{3}a^m}(0,1)$ and $\mathbf{b}^m_{2,3} = \frac{4\pi}{\sqrt{3}a^m}(\mp\sqrt{3}/2,-1/2)$ are the reciprocal lattice vectors and $a^m$ is the lattice constant of the superlattice. The system for $\phi_1 \neq 2\pi m/3$ ($m \in \mathds{Z}$) has $p3$ space group symmetry. For $\phi_1 = 2\pi m/3$ ($m \in \mathds{Z}$), it has $p31m$ symmetry.
     (a) Band structure showing 2 exact flat bands at $\tilde{\alpha}=\frac{\alpha}{|\mathbf{b}^m| ^{2}}=4.58$ and $\phi_{1}=0.7962$ along the high symmetry path in the Moir\'e Brillouin zone. The eigen-energy in the vertical axis is normalized as $\tilde{E}=\frac{E}{ |\mathbf{b}^m|^2}$.
     (b) Density plots of $|\psi_{{\Gamma}^{m}}(\mathbf{r})|$. The white dashed line marks the boundary of the moir\'e unit cell. The dark points indicate position of zeros of $\psi_{\Gamma^{m}}(\mathbf{r})$. Clearly, there is only one zero of $\psi_{\Gamma^m}(\mathbf{r})$ at an HSP, namely the corner, in the unit cell. (c) Wilson loop spectrum $\tilde{\theta}({\mathbf{k}})=\frac{\theta({\mathbf{k}})}{2\pi}$ of the flat bands in (a). 
     (d) Bandwidth of the middle two bands $\ln \tilde{E}_w$ as a function of $\phi_{1}$ and $\tilde{\alpha}$ in polar coordinate of $\tilde{\alpha}^{2}$ (radius) and $\phi_{1}$ (polar angle). The dark points in the plot imply flat bands. Since the system has $p3$ symmetry (except the special lines $\phi_1 = 2\pi m/3$), the co-dimension of the tuning parameter to obtain flat bands is 2; hence we see flat-bands occurring at isolated points in the $\tilde{\alpha}^2-\phi_1$ plane.
     }
     \label{fig:p3}
\end{figure}
\newpage
\subsection{2 flat bands in single layer system with QBCP under moir\'e potential with space group symmetry $p4$}
\begin{figure}[h!]
     \centering
\includegraphics[scale=2]{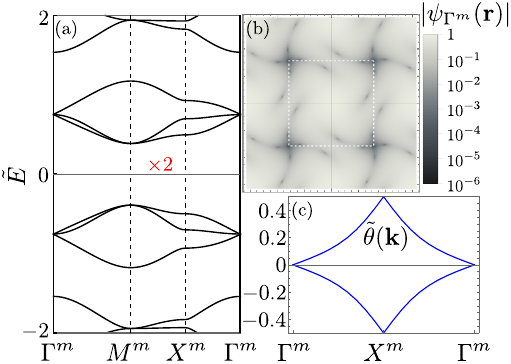}
     \caption{
     Flat bands in system with QBCP under moir\'e potential $\mathcal{D}_U(\mathbf{r};\boldsymbol{\alpha}=\alpha)= \alpha \exp(-i\phi) \sum_{n=1}^{2} (-1)^{1-n} \;\;\cos \left(\mathbf{b}_n^m \cdot \mathbf{r} \right)$ having space group symmetry $p4$. Here,  $\mathbf{b}^m_{1} = \frac{2\pi}{a^m}(1,0)$ and $\mathbf{b}^m_{2} = \frac{2\pi}{a^m}(0,1)$ are the reciprocal lattice vectors and $a^m$ is the lattice constant of the superlattice.
     (a) Band structure showing 2 exact flat bands at $\tilde{\alpha}=\frac{\alpha}{|\mathbf{b}^m| ^{2}}=4.83$ and $\phi=1.203067$ along the high symmetry path in the Moir\'e Brillouin zone. The eigen-energy in the vertical axis is normalized as $\tilde{E}=\frac{E}{ |\mathbf{b}^m|^2}$.
     (b) Density plots of $|\psi_{{\Gamma}^{m}}(\mathbf{r})|$ (normalized by its maximum). The white dashed line marks the boundary of the moir\'e unit cell. The dark points indicate position of zeros of $\psi_{\Gamma^{m}}(\mathbf{r})$. Clearly, there is only one zero of $\psi_{\Gamma^m}(\mathbf{r})$ at an HSP, namely the corner, in the unit cell. (c) Wilson loop spectrum $\tilde{\theta}({\mathbf{k}})=\frac{\theta({\mathbf{k}})}{2\pi}$ of the flat bands in (a). 
     }
     \label{fig:p4_2FB}
\end{figure}
\newpage
\subsection{4 flat bands in single layer system with QBCP under moir\'e potential with space group symmetry $p4$}
\begin{figure}[h!]
     \centering
\includegraphics[scale=2]{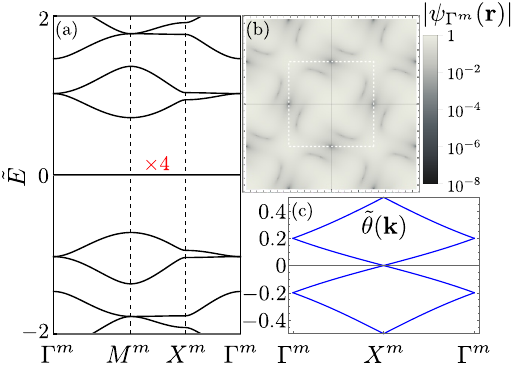}
     \caption{
     Flat bands in system with QBCP under moir\'e potential $\mathcal{D}_U(\mathbf{r};\boldsymbol{\alpha}=\alpha)= \alpha \exp(-i\phi) \sum_{n=1}^{2} (-1)^{1-n} \;\;\cos \left(\mathbf{b}_n^m \cdot \mathbf{r} \right)$ having space group symmetry $p4$. Here,  $\mathbf{b}^m_{1} = \frac{2\pi}{a^m}(1,0)$ and $\mathbf{b}^m_{2} = \frac{2\pi}{a^m}(0,1)$ are the reciprocal lattice vectors and $a^m$ is the lattice constant of the superlattice.
     (a) Band structure showing 4 exact flat bands at $\tilde{\alpha}=\frac{\alpha}{|\mathbf{b}^m| ^{2}}=8.41$ and $\phi=0.98130229$ along the high symmetry path in the Moir\'e Brillouin zone. The eigen-energy in the vertical axis is normalized as $\tilde{E}=\frac{E}{ |\mathbf{b}^m|^2}$.
     (b) Density plots of $|\psi_{{\Gamma}^{m}}(\mathbf{r})|$ (normalized by its maximum). The white dashed line marks the boundary of the moir\'e unit cell. The dark points indicate position of zeros of $\psi_{\Gamma^{m}}(\mathbf{r})$. Clearly, there are two zeros of $\psi_{\Gamma^m}(\mathbf{r})$ at HSPs, namely center of the edges, in the unit cell. (c) Wilson loop spectrum $\tilde{\theta}({\mathbf{k}})=\frac{\theta({\mathbf{k}})}{2\pi}$ of the flat bands in (a). 
     }
     \label{fig:p4_4FB}
\end{figure}
\newpage
\subsection{4 flat bands in single layer system with QBCP under moir\'e potential with space group symmetry $p4gm$}
\begin{figure}[h!]
     \centering
\includegraphics[scale=2]{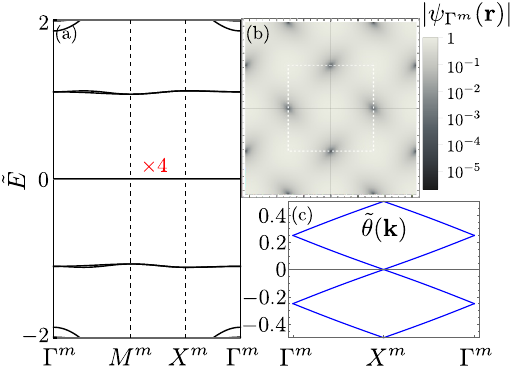}
     \caption{
     Flat bands in system with QBCP under moir\'e potential $\mathcal{D}_U(\mathbf{r};\boldsymbol{\alpha}=\alpha)= i\alpha \sum_{n=1}^{2} (-1)^{1-n} \;\;\cos \left(\mathbf{b}_n^m \cdot \mathbf{r} \right)$ having space group symmetry $p4gm$. Here,  $\mathbf{b}^m_{1} = \frac{2\pi}{a^m}(1,0)$ and $\mathbf{b}^m_{2} = \frac{2\pi}{a^m}(0,1)$ are the reciprocal lattice vectors and $a^m$ is the lattice constant of the superlattice. Notice that in addition to $\mathcal{C}_{4z}$, this system has glide symmetry $\mathcal{G}_{10}=\{\mathcal{M}_{10}|\frac{1}{2},\frac{1}{2}\}$: $\mathcal{D}_U(\mathcal{G}_{10}\mathbf{r};\boldsymbol{\alpha}=\alpha) = -\mathcal{D}_U(\mathbf{r};\boldsymbol{\alpha}=\alpha) = \mathcal{D}_U^*(\mathbf{r};\boldsymbol{\alpha}=\alpha)$ for $\alpha \in \mathds{R}$ (by $\mathcal{M}_{10}$ we mean the mirror whose normal is in the direction of the lattice vector $\mathbf{a}_1^m$, the translation part of the glide is $(\mathbf{a}_1^m+\mathbf{a}_2^m)/2$).
     (a) Band structure showing 4 exact flat bands at $\tilde{\alpha}=\frac{\alpha}{|\mathbf{b}^m| ^{2}}=2.24$ along the high symmetry path in the Moir\'e Brillouin zone. The eigen-energy in the vertical axis is normalized as $\tilde{E}=\frac{E}{ |\mathbf{b}^m|^2}$.
     (b) Density plots of $|\psi_{{\Gamma}^{m}}(\mathbf{r})|$ (normalized by its maximum). The white dashed line marks the boundary of the moir\'e unit cell. The dark points indicate position of zeros of $\psi_{\Gamma^{m}}(\mathbf{r})$. Clearly, there are two zeros of $\psi_{\Gamma^m}(\mathbf{r})$ at HSPs, namely center of the edges, in the unit cell. (c) Wilson loop spectrum $\tilde{\theta}({\mathbf{k}})=\frac{\theta({\mathbf{k}})}{2\pi}$ of the flat bands in (a). 
	}
     \label{fig:p4gm}
\end{figure}
\newpage
\subsection{4 flat bands in single layer system with QBCP under moir\'e potential with space group symmetry $p4mm$}
\begin{figure}[h!]
     \centering
\includegraphics[scale=2]{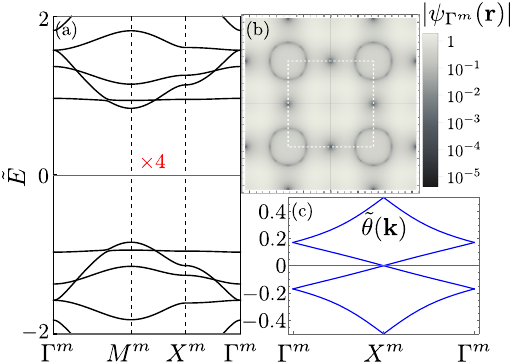}
     \caption{
     Flat bands in system with QBCP under moir\'e potential $\mathcal{D}_U(\mathbf{r};\boldsymbol{\alpha}=\alpha)= \alpha \sum_{n=1}^{2} (-1)^{1-n}(\cos \left(\mathbf{b}_n^m \cdot \mathbf{r} \right)- \cos \left(2\mathbf{b}_n^m \cdot \mathbf{r} \right))$ having space group symmetry $p4mm$. Here,  $\mathbf{b}^m_{1} = \frac{2\pi}{a^m}(1,0)$ and $\mathbf{b}^m_{2} = \frac{2\pi}{a^m}(0,1)$ are the reciprocal lattice vectors and $a^m$ is the lattice constant of the superlattice. Notice that in addition to $\mathcal{C}_{4z}$, this system has mirror symmetry $\mathcal{M}_{10}$: $\mathcal{D}_U(\mathcal{M}_{10}\mathbf{r};\boldsymbol{\alpha}=\alpha) = \mathcal{D}_U(\mathbf{r};\boldsymbol{\alpha}=\alpha) = \mathcal{D}_U^*(\mathbf{r};\boldsymbol{\alpha}=\alpha)$ for $\alpha \in \mathds{R}$.
     (a) Band structure showing 4 exact flat bands at $\tilde{\alpha}=\frac{\alpha}{|\mathbf{b}^m| ^{2}}=-2.62$ along the high symmetry path in the Moir\'e Brillouin zone. The eigen-energy in the vertical axis is normalized as $\tilde{E}=\frac{E}{ |\mathbf{b}^m|^2}$.
     (b) Density plots of $|\psi_{{\Gamma}^{m}}(\mathbf{r})|$. The white dashed line marks the boundary of the moir\'e unit cell. The dark points indicate position of zeros of $\psi_{\Gamma^{m}}(\mathbf{r})$. Clearly, there are two zeros of $\psi_{\Gamma^m}(\mathbf{r})$ at HSPs, namely center of the edges, in the unit cell. (c) Wilson loop spectrum $\tilde{\theta}({\mathbf{k}})=\frac{\theta({\mathbf{k}})}{2\pi}$ of the flat bands in (a). 
     }
     \label{fig:p4mm}
\end{figure}

\section{Twisted bilayer checkerboard lattice (TBCL) is two uncoupled copies of single layer QBCP system under periodic strain field}

\begin{figure}[h!]
     \centering
\includegraphics[scale=1]{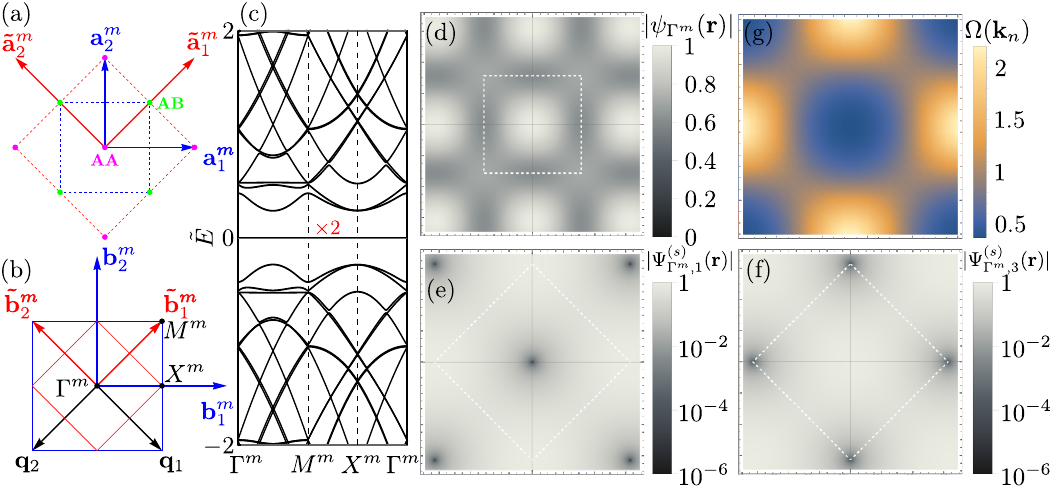}
     \caption{TBCL.
      (a) Moir\'e unit cell of TBCL system is plotted in blue dashed line. Moir\'e unit cell of single layer QBCP system is plotted in red dashed line. $\mathbf{a}^{m}_{i}$ denotes the Moir\'e lattice vector of TBCL system. AA region is marked by magenta disk and AB region is marked by green disk. $\tilde{\mathbf{a}}^{m}_{i}$ denotes the Moir\'e lattice vector of the single layer QBCP Hamiltonians that the TBCL Hamiltonian can be decomposed into (see Eq.~\eqref{eq:HTBCLDecomposition}).
     (b) Moir\'e BZ of TBCL system is plotted in blue solid line. Moir\'e BZ of single layer QBCP system is plotted in red solid line. $\mathbf{b}^{m}_{i}$ denotes the Moir\'e reciprocal lattice vector of TBCL system. $\tilde{\mathbf{b}}^{m}_{i}$ denotes the Moir\'e reciprocal lattice vector of the single layer QBCP system.
     (c) Band structure of TBCL system with 2 exact flat bands at $\tilde{\alpha}=\frac{\alpha}{|\mathbf{b}^{m}|^{2}}=0.13$.
     (d) Density plot of $ |\psi_{\Gamma^{m}}(\mathbf{r})|$ (normalized by its maximum).
     $|\psi_{\Gamma^{m}}(\mathbf{r})|$ has no zero in the unit cell.
     (e), (f) Density plot of $|\Psi_{\Gamma^{m},1}^{(s)}(\mathbf{r})|$ and $|\Psi_{\Gamma^{m},3}^{(s)}(\mathbf{r})|$ (normalized by their respective maximum). See the text below Eq.~\eqref{eq:HTBCLUtranslation} for the definition of these two functions.
     The zero of $|\Psi_{\Gamma^m,3}^{(s)}(\mathbf{r})|$ in (f) is shifted by $(\tilde{\mathbf{a}_1^{m}}+\tilde{\mathbf{a}_2^{m}})/2 = \mathbf{a}_2^m$ from the zero of $|\Psi_{\Gamma^m,1}^{(s)}(\mathbf{r})|$ in (e).
     (g) Berry Curvature distribution $\Omega(\mathbf{k}_n)$ (normalized by its average) of the FBs of TBCL Hamiltonian plotted within the moir\'e BZ of TBCL Hamiltonian. Clearly, the Berry Curvature distribution actually has a smaller periodicity than the moir\'e BZ of TBCL system.
    }
     \label{fig:TBCL}
\end{figure}
Checkerboard lattice has $\mathcal{C}_{(n=4)z}$ and $\mathcal{T}$ protected QBCP at the corner, $\mathbf{k}_0=M$ point of the Brillouin zone. Upon twisting the the two layers on top of each other, the $M$ point of one layer gets mapped to $\mathbf{k}_0^m=\Gamma^m$, the $M$ point of the other layer gets mapped of $M^m$ of the mBZ. Following the derivation of {\color{red}Sec.~SM.1D}, we have $\rho(\mathcal{C}_{4z}) = \text{Diag}\{i,i,-i,-i\}$, $\rho(\mathcal{T}) = \sigma_x\otimes\mathds{1}$. Furthermore, the moir\'e potential $U_1(\mathbf{r})$ in Eq.~\eqref{eq:HTB} satisfies (Eqs.~\eqref{eq:Urotation} and~\eqref{eq:Utranslation})
\begin{equation}
	U_1(\mathcal{C}_{4z}\mathbf{r}) = - U_1(\mathbf{r}),\; U_1(\mathbf{r}) = U_2^*(\mathbf{r}),\; U_1(\mathbf{r}) = \sum_{\mathbf{b}^m} a_{\mathbf{b}^m}e^{-i(\mathbf{q}_1+\mathbf{b}^m)\cdot \mathbf{r}}, 
\end{equation}
 where $\mathbf{q}_1 = (\mathbf{b}_1^m-\mathbf{b}_2^m)/2$ as shown in Fig.~\ref{fig:TBCL}. Together, we have
 \begin{equation}
 \begin{split}
 	U_1(\mathcal{C}_{4z}\mathbf{r}) = - U_1(\mathbf{r}) &\Rightarrow \sum_{\mathbf{b}^m} a_{\mathbf{b}^m}e^{-i(\mathbf{q}_1+\mathbf{b}^m)\cdot \mathcal{C}_{4z}\mathbf{r}} = - \sum_{\mathbf{b}^m} a_{\mathbf{b}^m}e^{-i(\mathbf{q}_1+\mathbf{b}^m)\cdot \mathbf{r}}\\ 
&\Rightarrow \sum_{\mathbf{b}^m} a_{\mathbf{b}^m}e^{-i( \mathcal{C}_{4z}^{-1}\mathbf{q}_1+ \mathcal{C}_{4z}^{-1} \mathbf{b}^m)\cdot\mathbf{r}} = - \sum_{\mathbf{b}^m} a_{\mathbf{b}^m}e^{-i(\mathbf{q}_1+\mathbf{b}^m)\cdot \mathbf{r}}\\ 
&\Rightarrow \sum_{\mathbf{b}^m} a_{\mathbf{b}^m}e^{-i(\mathbf{q}_1-\mathbf{b}_1^m+ \mathcal{C}_{4z}^{-1} \mathbf{b}^m)\cdot\mathbf{r}} = - \sum_{\mathbf{b}^m} a_{\mathbf{b}^m}e^{-i(\mathbf{q}_1+\mathbf{b}^m)\cdot \mathbf{r}}\\ 
&\Rightarrow \sum_{\mathbf{b}^m} a_{\mathcal{C}_{4z}(\mathbf{b}^m+\mathbf{b}^m_1)}e^{-i(\mathbf{q}_1 +\mathbf{b}^m)\cdot\mathbf{r}} = - \sum_{\mathbf{b}^m} a_{\mathbf{b}^m}e^{-i(\mathbf{q}_1+\mathbf{b}^m)\cdot \mathbf{r}},\\ 
&\Rightarrow a_{\mathcal{C}_{4z}(\mathbf{b}^m+\mathbf{b}^m_1)} = -a_{\mathbf{b}^m} \text{ for all reciprocal lattice vector }\mathbf{b}^m. 
 \end{split}
 \end{equation}
 Using above equation, starting from $a_\mathbf{0}\equiv\alpha$, we get 
\begin{equation} 
\begin{split} 
&a_{\mathcal{C}_{4z}\mathbf{b}^m_1} = a_{\mathbf{b}^m_2} =-a_{\mathbf{0}} = -\alpha, \\ 
&a_{\mathcal{C}_{4z}(\mathbf{b}^m_1+\mathbf{b}^m_2)} = a_{\mathbf{b}^m_2-\mathbf{b}^m_1} = -a_{\mathbf{b}^m_2} = a_{\mathbf{0}} = \alpha,\\
&a_{\mathcal{C}_{4z}(\mathbf{b}^m_2)} = a_{-\mathbf{b}^m_1} = -a_{\mathbf{b}^m_2-\mathbf{b}^m_1} = -a_{\mathbf{0}} = -\alpha.
\end{split} 
\end{equation} 
Hence, if we keep only lowest harmonics the expression for $U_1(\mathbf{r})$ becomes 
\begin{equation} 
\begin{split} 
U_1(\mathbf{r}) &= a_\mathbf{0} (e^{-i\mathbf{q}_1\cdot \mathbf{r}}- e^{-i(\mathbf{q}_1+\mathbf{b}_2^m)\cdot \mathbf{r}}+ e^{-i(\mathbf{q}_1+\mathbf{b}_2^m-\mathbf{b}^m_1)\cdot \mathbf{r}}-e^{-i(\mathbf{q}_1-\mathbf{b}^m_1)\cdot \mathbf{r}})\\ 
&= \alpha(e^{-i\mathbf{q}_1\cdot \mathbf{r}}- e^{i\mathbf{q}_2\cdot \mathbf{r}}+e^{i\mathbf{q}_1\cdot \mathbf{r}}- e^{-i\mathbf{q}_2\cdot \mathbf{r}})\\
&= 2\alpha(\cos(\mathbf{q}_1\cdot \mathbf{r})-\cos(\mathbf{q}_2\cdot \mathbf{r})). 
\end{split} 
\end{equation} 
An additional mirror symmetry $\mathcal{M}_x$ with representation $\rho(\mathcal{M}_x) = \sigma_x\otimes\mathds{1}$, would result in
\begin{equation}
\begin{split}
	&U_1(\mathcal{M}_x\mathbf{r}) = U_2(\mathbf{r}) =U_1^*(\mathbf{r})\\
	\Rightarrow & 2\alpha(\cos(\mathbf{q}_1\cdot \mathcal{M}_x\mathbf{r})-\cos(\mathbf{q}_2\cdot \mathcal{M}_x\mathbf{r})) = 2\alpha^*(\cos(\mathbf{q}_1\cdot \mathbf{r})-\cos(\mathbf{q}_2\cdot \mathbf{r}))\\
	\Rightarrow & 2\alpha(\cos((\mathcal{M}_x\mathbf{q}_1)\cdot \mathbf{r})-\cos((\mathcal{M}_x\mathbf{q}_2)\cdot\mathbf{r})) = 2\alpha^*(\cos(\mathbf{q}_1\cdot \mathbf{r})-\cos(\mathbf{q}_2\cdot \mathbf{r}))\\
	\Rightarrow & 2\alpha(\cos(\mathbf{q}_2\cdot \mathbf{r})-\cos(\mathbf{q}_1\cdot\mathbf{r})) = 2\alpha^*(\cos(\mathbf{q}_1\cdot \mathbf{r})-\cos(\mathbf{q}_2\cdot \mathbf{r}))\\
	\Rightarrow &\alpha = -\alpha^*
\end{split}
\end{equation}
Then, replace $\alpha\rightarrow i\alpha$ such that $\alpha \in \mathbb{R}$. Lastly, performing the following transformation $\splitatcommas{\text{Diag}\{e^{\pi i/4},e^{\pi i/4},e^{-\pi i/4},e^{-\pi i/4}\}\mathcal{H}_{TB}(\mathbf{r})\text{Diag}\{e^{-\pi i/4},e^{-\pi i/4},e^{\pi i/4},e^{\pi i/4}\}}$ on the Hamiltonian in Eq.~\ref{eq:HTB}, we obtain
\begin{equation}\label{eq:HTBCL}
\mathcal{H}_{TBCL}(\mathbf{r}) = \begin{pmatrix} 
0 & 0 & i(-2i\partial_z)^2 & iU_1^*(\mathbf{r})\\ 
0 & 0 & iU_1^*(\mathbf{r}) & i(-2i\partial_z)^2\\ 
-i(-2i\overline{\partial_{z}})^2 & -iU_1(\mathbf{r}) & 0 & 0\\ 
-iU_1(\mathbf{r}) & -i(-2i\overline{\partial_{z}})^2\ & 0 & 0\ 
\end{pmatrix}  \text{, }U_1(\mathbf{r}) = 2\alpha i (\cos(\mathbf{q}_1\cdot \mathbf{r})-\cos(\mathbf{q}_2\cdot \mathbf{r})),\; \alpha \in \mathbb{R},
\end{equation}
this is the Hamiltonian considered in~\citesupp{li2022magics}. Notice that there It was shown in~\citesupp{li2022magics} (see also Fig.~\ref{fig:TBCL}(c)), that for some magic values of $\alpha$, 2 exact flat bands appear at the charge neutrality point. However, 
\begin{itemize}
\item these bands have Chern number $C =\pm 2$, which suggest that the FB WFs are not simply $f_\mathbf{k}(z;\mathbf{r}_0)\psi_{\mathbf{k}_0^m}(\mathbf{r})$, because if it were, then the Chern number would be $C = \pm 1$
\item It is also clear from Fig.~\ref{fig:TBCL}(d), that the WF $\psi_{\Gamma^m}$ does not have a zero at the magic value of $\alpha$ in the moir\'e unit cell.
\end{itemize}
These two points makes the FBs in TBCL intriguing. A major clue to solving this problem comes from the Berry curvature distribution of the sublattice polarized FB WF in Fig.~\ref{fig:TBCL}(g). The periodicity of the Berry curvature distribution in the reciprocal space is smaller than the reciprocal lattice vectors. This alludes to the possibility that this model is two copies of Chern number $C = \pm 1$ bands unfolded to a larger Brillouin zone. Below we show that this is indeed the case. 
Consider the following transformation
\begin{equation}
\label{eq:HTBCLDecomposition}
\mathcal{H}^{(s)}_{TBCL}(\mathbf{r}) =U \mathcal{H}_{TBCL}(\mathbf{r}) U^\dagger = \begin{pmatrix} 
0 &  i(-2i\partial_z)^2+iU_1^*(\mathbf{r}) & 0  & 0\\ 
-i(-2i\overline{\partial_{z}})^2-iU_1(\mathbf{r}) & 0 & 0 & 0\\ 
0  & 0 & 0 & i(-2i\partial_z)^2-iU_1^*(\mathbf{r})\\ 
0  & 0 & -i(-2i\overline{\partial_{z}})^2+iU_1(\mathbf{r}) & 0\ 
\end{pmatrix},
\end{equation}
where 
\begin{equation}
U = \frac{1}{\sqrt{2}}\begin{pmatrix}
1 & 1 & 0 & 0\\
0 & 0 & 1 & 1\\
-1 & 1 & 0 & 0\\
0 & 0 & -1 & 1
\end{pmatrix}.
\end{equation}
Each of the diagonal blocks clearly corresponds to a single layer QBCP system with $p4mm$ space group symmetry and moir\'e lattice vectors $\tilde{\mathbf{a}}_i^m$ (and corresponding reciprocal lattice vector $\tilde{\mathbf{b}}_i^m$) as shown in Fig.~\ref{fig:TBCL}(a-b). However, the lattice vectors $\mathbf{a}_i^m$ of $\mathcal{H}_{TBCL}$ are smaller (see Fig.~\ref{fig:TBCL}(a-b)) because $U_1(\mathbf{r}+\mathbf{a}_i^m) = -U_1(\mathbf{r})$, hence
\begin{equation}
\label{eq:HTBCLUtranslation}
\mathcal{H}^{(s)}_{TBCL}(\mathbf{r}+\mathbf{a}_i^m) = \sigma_x\otimes\mathds{1} \mathcal{H}^{(s)}_{TBCL}(\mathbf{r})\sigma_x\otimes\mathds{1}.
\end{equation}
In fact, the diagonal blocks are exactly the same as the Hamiltonian that was reported to host exact FBs for magic values of $\alpha$  in~\citesupp{eugenio2022twisteds} (the Hamiltonian in~\citesupp{eugenio2022twisteds} is written in a $\pi/4$ rotated coordinate system than the one here, which results in a difference in factor of imaginary $i$ between the two). The magic value reported in~\citesupp{eugenio2022twisteds} are the same as that  reported in~\citesupp{li2022magics} after nondimensionalization. Indeed, when we transform the TBCL WF $\Psi_{\Gamma^m}(\mathbf{r})=\{\psi_{\Gamma^m}(\mathbf{r}),\mathbf{0}\}^T$ to $\Psi_{\Gamma^m}^{(s)}(\mathbf{r})=U\Psi_{\Gamma^m}(\mathbf{r})$, the nonzero components of it, $\Psi_{\Gamma^m,1}^{(s)}(\mathbf{r})=(\psi_{\Gamma^m,1}(\mathbf{r})+\psi_{\Gamma^m,2}(\mathbf{r}))/\sqrt{2}$ and $\Psi_{\Gamma^m,3}^{(s)}(\mathbf{r})=(-\psi_{\Gamma^m,1}(\mathbf{r})+\psi_{\Gamma^m,2}(\mathbf{r}))/\sqrt{2}$) (recall $\psi_{\Gamma^m}(\mathbf{r})=\{\psi_{\Gamma^m,1}(\mathbf{r}),\psi_{\Gamma^m,2}(\mathbf{r})\}$ is a two component function for twisted bilayer systems) have zeros in the unit cell defined by vectors $\tilde{\mathbf{a}}_i^m$ shifted from each other by $\mathbf{a}_1^m$ for magic values of $\alpha$ as shown in Fig.~\ref{fig:TBCL}(e-f). 

Clearly, from the Eq.~\ref{eq:WFtranslation} and subsequent discussion, we know that $\psi_{\Gamma^m,1}(\mathbf{r})$ has periodicity $\mathbf{a}_i^m$, whereas $\psi_{\Gamma^m,2}(\mathbf{r})$ has periodicity $\tilde{\mathbf{a}}_i^m$; hence $\Psi_{\Gamma^m,1/3}^{(s)}(\mathbf{r})$ have periodicity $\tilde{\mathbf{a}}_i^m$. Also, since $\psi_{\Gamma^m,2}(\mathbf{r}-\mathbf{a}_1^m)=-\psi_{\Gamma^m,2}(\mathbf{r})$, we have $\Psi_{\Gamma^m,1}^{(s)}(\mathbf{r}-\mathbf{a}_1^m)=\Psi_{\Gamma^m,3}^{(s)}(\mathbf{r})$. Using the zeros of $\Psi_{\Gamma^m,1}^{(s)}(\mathbf{r})$ and $\Psi_{\Gamma^m,3}^{(s)}(\mathbf{r})$ at $\mathbf{r}_0 = \mathbf{0}$ and $\mathbf{r}_0 = -\mathbf{a}_1^m$ respectively (see Fig.~\ref{fig:TBCL}(e-f)), we can construct the FB WF of $\mathcal{H}^{(s)}(\mathbf{r})$ as
\begin{equation}\label{eq:WFsTBCL}
\Psi_\mathbf{k}^{(s)}(\mathbf{r})  =\begin{Bmatrix}
f_\mathbf{k}(z;\mathbf{0})\psi_{\Gamma^m,1}^{(s)}(\mathbf{r})\\
0\\
e^{-i\mathbf{k}\cdot\mathbf{a}_1^m}f_\mathbf{k}(z+a_1^m;\mathbf{0})\psi_{\Gamma^m,1}^{(s)}(\mathbf{r}+\mathbf{a}_1^m)\\
0
\end{Bmatrix}\text{, } f_\mathbf{k}(z;\mathbf{0}) = e^{i (\mathbf{k}\cdot\tilde{\mathbf{a}}_1^m) z/\tilde{a}_1^m}\frac{\vartheta\left(\frac{z}{\tilde{a}_1^m}-\frac{k}{\tilde{b}_2^m},\tilde{a}_2^m/\tilde{a}_1^m\right)}{\vartheta\left(\frac{z}{\tilde{a}_1^m},,\tilde{a}_2^m/\tilde{a}_1^m\right)}
\end{equation}
where  $f_\mathbf{k}(z;\mathbf{r}_0)$ satisfies $f_\mathbf{k}(z+\tilde{a}_i^m;\mathbf{r}_0) = e^{i\mathbf{k}\cdot\tilde{\mathbf{a}}_i^m}f_\mathbf{k}(z;\mathbf{r}_0)$, and $f_\mathbf{k}(z+a_1^m;\mathbf{0}) = f_\mathbf{k}(z;-\mathbf{a}_1^m)$. It can be easily checked that $\Psi_\mathbf{k}^{(s)}(\mathbf{r})$ satisfies the Bloch periodicity $\Psi_\mathbf{k}^{(s)}(\mathbf{r}+\mathbf{a}_i^m) = e^{i\mathbf{k}\cdot\mathbf{a}_i^m}\sigma_x\otimes\mathds{1}\Psi_\mathbf{k}^{(s)}(\mathbf{r})$ corresponding to Eq.~\eqref{eq:HTBCLUtranslation}:
\begin{subequations}
\begin{align}
    &\Psi_\mathbf{k}^{(s)}(\mathbf{r}+\mathbf{a}_1^m)\nonumber\\ 
    =&\begin{Bmatrix}
    f_\mathbf{k}(z+a_1^m;\mathbf{0})\psi_{\Gamma^m,1}^{(s)}(\mathbf{r}+\mathbf{a}_1^m)\\
    0\\
    e^{-i\mathbf{k}\cdot\mathbf{a}_1^m}f_\mathbf{k}(z+2a_1^m;\mathbf{0})\psi_{\Gamma^m,1}^{(s)}(\mathbf{r}+2\mathbf{a}_1^m)\\
    0
    \end{Bmatrix}\nonumber\\
    =&\sigma_x\otimes\mathds{1}\begin{Bmatrix}
    e^{-i\mathbf{k}\cdot\mathbf{a}_1^m}f_\mathbf{k}(z+2a_1^m;\mathbf{0})\psi_{\Gamma^m,1}^{(s)}(\mathbf{r}+2\mathbf{a}_1^m)\\
    0\\
    f_\mathbf{k}(z+a_1^m;\mathbf{0})\psi_{\Gamma^m,1}^{(s)}(\mathbf{r}+\mathbf{a}_1^m)\\
    0
    \end{Bmatrix}\nonumber\\
    =&\sigma_x\otimes\mathds{1}\begin{Bmatrix}
    e^{-i\mathbf{k}\cdot\mathbf{a}_1^m}f_\mathbf{k}(z+\tilde{a}_1^m-\tilde{a}_2^m;\mathbf{0})\psi_{\Gamma^m,1}^{(s)}(\mathbf{r}+\tilde{\mathbf{a}}_1^m-\tilde{\mathbf{a}}_2^m)\\
    0\\
    f_\mathbf{k}(z+a_1^m;\mathbf{0})\psi_{\Gamma^m,1}^{(s)}(\mathbf{r}+\mathbf{a}_1^m)\\
    0
    \end{Bmatrix}\nonumber\\
    =&\sigma_x\otimes\mathds{1}\begin{Bmatrix}
    e^{-i\mathbf{k}\cdot\mathbf{a}_1^m}e^{i\mathbf{k}\cdot(\tilde{\mathbf{a}}_1^m-\tilde{\mathbf{a}}_2^m)}f_\mathbf{k}(z;\mathbf{0})\psi_{\Gamma^m,1}^{(s)}(\mathbf{r})\\
    0\\
    f_\mathbf{k}(z+a_1^m;\mathbf{0})\psi_{\Gamma^m,1}^{(s)}(\mathbf{r}+\mathbf{a}_1^m)\\
    0
    \end{Bmatrix},\begin{array}{c}
         \text{ since } f_\mathbf{k}(z+\tilde{a}_1^m-\tilde{a}_2^m;\mathbf{0}) = e^{i\mathbf{k}\cdot(\tilde{\mathbf{a}}_1^m-\tilde{\mathbf{a}}_2^m)}f_\mathbf{k}(z;\mathbf{0})  \\
         \text{ and }\psi_{\Gamma^m,1}^{(s)}(\mathbf{r}+\tilde{\mathbf{a}}_1^m-\tilde{\mathbf{a}}_2^m) = \psi_{\Gamma^m,1}^{(s)}(\mathbf{r}) 
    \end{array}\nonumber\\
    =&e^{i\mathbf{k}\cdot\mathbf{a}_1^m}\sigma_x\otimes\mathds{1}\begin{Bmatrix}
    f_\mathbf{k}(z;\mathbf{0})\psi_{\Gamma^m,1}^{(s)}(\mathbf{r})\\
    0\\
    e^{-i\mathbf{k}\cdot\mathbf{a}_1^m}f_\mathbf{k}(z+a_1^m;\mathbf{0})\psi_{\Gamma^m,1}^{(s)}(\mathbf{r}+\mathbf{a}_1^m)\\
    0
    \end{Bmatrix},\text{ since }e^{-i\mathbf{k}\cdot\mathbf{a}_1^m}e^{i\mathbf{k}\cdot(\tilde{\mathbf{a}}_1^m-\tilde{\mathbf{a}}_2^m)} = e^{-i\mathbf{k}\cdot\mathbf{a}_1^m}e^{i\mathbf{k}\cdot 2\tilde{\mathbf{a}}_1^m} =e^{i\mathbf{k}\cdot\mathbf{a}_1^m}\nonumber\\
    =&e^{i\mathbf{k}\cdot\mathbf{a}_1^m}\sigma_x\otimes\mathds{1}\Psi_\mathbf{k}^{(s)}(\mathbf{r}),\nonumber\\
    &\Psi_\mathbf{k}^{(s)}(\mathbf{r}+\mathbf{a}_2^m)\nonumber\\ 
    =&\begin{Bmatrix}
    f_\mathbf{k}(z+a_2^m;\mathbf{0})\psi_{\Gamma^m,1}^{(s)}(\mathbf{r}+\mathbf{a}_2^m)\\
    0\\
    e^{-i\mathbf{k}\cdot\mathbf{a}_1^m}f_\mathbf{k}(z+a_1^m+a_2^m;\mathbf{0})\psi_{\Gamma^m,1}^{(s)}(\mathbf{r}+\mathbf{a}_1^m+\mathbf{a}_2^m)\\
    0
    \end{Bmatrix}\nonumber\\
    =&\sigma_x\otimes\mathds{1}\begin{Bmatrix}
    e^{-i\mathbf{k}\cdot\mathbf{a}_1^m}f_\mathbf{k}(z+a_1^m+a_2^m;\mathbf{0})\psi_{\Gamma^m,1}^{(s)}(\mathbf{r}+\mathbf{a}_1^m+\mathbf{a}_2^m)\\
    0\\
    f_\mathbf{k}(z+a_2^m;\mathbf{0})\psi_{\Gamma^m,1}^{(s)}(\mathbf{r}+\mathbf{a}_2^m)\\
    0
    \end{Bmatrix}\nonumber\\
    =&\sigma_x\otimes\mathds{1}\begin{Bmatrix}
    e^{-i\mathbf{k}\cdot\mathbf{a}_1^m}f_\mathbf{k}(z+\tilde{a}_1^m;\mathbf{0})\psi_{\Gamma^m,1}^{(s)}(\mathbf{r}+\tilde{\mathbf{a}}_1^m)\\
    0\\
    f_\mathbf{k}(z+a_1^m+a_2^m-a_1^m;\mathbf{0})\psi_{\Gamma^m,1}^{(s)}(\mathbf{r}+\mathbf{a}_1^m+\mathbf{a}_2^m-\mathbf{a}_1^m)\\
    0
    \end{Bmatrix}\nonumber\\
    =&\sigma_x\otimes\mathds{1}\begin{Bmatrix}
    e^{-i\mathbf{k}\cdot\mathbf{a}_1^m}e^{i\mathbf{k}\cdot\tilde{\mathbf{a}}_1^m}f_\mathbf{k}(z;\mathbf{0})\psi_{\Gamma^m,1}^{(s)}(\mathbf{r})\\
    0\\
    f_\mathbf{k}(z+a_1^m+\tilde{a}_2^m;\mathbf{0})\psi_{\Gamma^m,1}^{(s)}(\mathbf{r}+\mathbf{a}_1^m+\tilde{\mathbf{a}}_2^m)\\
    0
    \end{Bmatrix},\begin{array}{c}
         \text{ since } f_\mathbf{k}(z+\tilde{a}_1^m;\mathbf{0}) = e^{i\mathbf{k}\cdot\tilde{\mathbf{a}}_1^m}f_\mathbf{k}(z;\mathbf{0})  \\
         \text{ and }\psi_{\Gamma^m,1}^{(s)}(\mathbf{r}+\tilde{\mathbf{a}}_1^m) = \psi_{\Gamma^m,1}^{(s)}(\mathbf{r}) 
    \end{array}\nonumber\\
    =&\sigma_x\otimes\mathds{1}\begin{Bmatrix}
    e^{i\mathbf{k}\cdot\mathbf{a}_2^m}f_\mathbf{k}(z;\mathbf{0})\psi_{\Gamma^m,1}^{(s)}(\mathbf{r})\\
    0\\
    e^{i\mathbf{k}\cdot\tilde{\mathbf{a}}_2^m}f_\mathbf{k}(z+a_1^m;\mathbf{0})\psi_{\Gamma^m,1}^{(s)}(\mathbf{r}+\mathbf{a}_1^m)\\
    0
    \end{Bmatrix},\begin{array}{c}
         \text{ since } f_\mathbf{k}(z+a_1^m+\tilde{a}_2^m;\mathbf{0}) = e^{i\mathbf{k}\cdot\tilde{\mathbf{a}}_2^m}f_\mathbf{k}(z+a_1^m;\mathbf{0})  \\
         \text{, }\psi_{\Gamma^m,1}^{(s)}(\mathbf{r}+\mathbf{a}_2^m+\tilde{\mathbf{a}}_1^m) = \psi_{\Gamma^m,1}^{(s)}(\mathbf{r}+\mathbf{a}_2^m) \text{, and } e^{-i\mathbf{k}\cdot\mathbf{a}_1^m}e^{i\mathbf{k}\cdot\tilde{\mathbf{a}}_1^m} = e^{i\mathbf{k}\cdot\mathbf{a}_2^m}
    \end{array}\nonumber\\
    =&e^{i\mathbf{k}\cdot\mathbf{a}_2^m}\sigma_x\otimes\mathds{1}\begin{Bmatrix}
    f_\mathbf{k}(z;\mathbf{0})\psi_{\Gamma^m,1}^{(s)}(\mathbf{r})\\
    0\\
    e^{-i\mathbf{k}\cdot\mathbf{a}_1^m}f_\mathbf{k}(z+a_1^m;\mathbf{0})\psi_{\Gamma^m,1}^{(s)}(\mathbf{r}+\mathbf{a}_1^m)\\
    0
    \end{Bmatrix},\text{ since }e^{i\mathbf{k}\cdot\tilde{\mathbf{a}}_2^m}  =e^{i\mathbf{k}\cdot\mathbf{a}_2^m}e^{-i\mathbf{k}\cdot\mathbf{a}_1^m}\nonumber\\
    =&e^{i\mathbf{k}\cdot\mathbf{a}_2^m}\sigma_x\otimes\mathds{1}\Psi_\mathbf{k}^{(s)}(\mathbf{r}).
\end{align}
\end{subequations}
Hence the WF
\begin{equation} \label{eq:WFTBCL}
    \Psi_\mathbf{k}(\mathbf{r}) = U^\dagger\Psi_\mathbf{k}^{(s)}(\mathbf{r}) = \frac{1}{\sqrt{2}}\begin{Bmatrix}
        f_\mathbf{k}(z;\mathbf{0})\psi_{\Gamma^m,1}^{(s)}(\mathbf{r})-e^{-i\mathbf{k}\cdot\mathbf{a}_1^m}f_\mathbf{k}(z+a_1^m;\mathbf{0})\psi_{\Gamma^m,1}^{(s)}(\mathbf{r}+\mathbf{a}_1^m)\\
        f_\mathbf{k}(z;\mathbf{0})\psi_{\Gamma^m,1}^{(s)}(\mathbf{r})+e^{-i\mathbf{k}\cdot\mathbf{a}_1^m}f_\mathbf{k}(z+a_1^m;\mathbf{0})\psi_{\Gamma^m,1}^{(s)}(\mathbf{r}+\mathbf{a}_1^m)\\
        0\\
        0
    \end{Bmatrix}
\end{equation}
satisfies $\mathcal{H}_{TBCL}(\mathbf{r})\Psi_\mathbf{k}(\mathbf{r}) =\mathbf{0}$ and have the correct Bloch periodicity. The FB WF polarized on the other sublattice can be obtained as $\sigma_x\otimes\mathds{1}\Psi_\mathbf{k}^*(\mathbf{r})$. 

Lastly, we can calculate Chern number of the FB WF in the gauge of $\Psi^{(s)}_\mathbf{k}(\mathbf{r})$. The two nonzero components both have $f_\mathbf{k}(z;\mathbf{0})$, and hence would give Chern number $C=-1$ if we integrate over the BZ given by $\tilde{\mathbf{b}}_1^m$ and $\tilde{\mathbf{b}}_2^m$. But, due to Eq.~\eqref{eq:HTBCLUtranslation}, the BZ over which we have to integrate is $\mathbf{b}_1^m$ and $\mathbf{b}_2^m$, which is twice as big as the one given by $\tilde{\mathbf{b}}_1^m$ and $\tilde{\mathbf{b}}_2^m$. This is why the Chern number of these two FBs are $C=\pm 2$.

\subsection{TBCL type Hamiltonian with 4 flat bands}
\begin{figure}[h!]
     \centering
\includegraphics[scale=1]{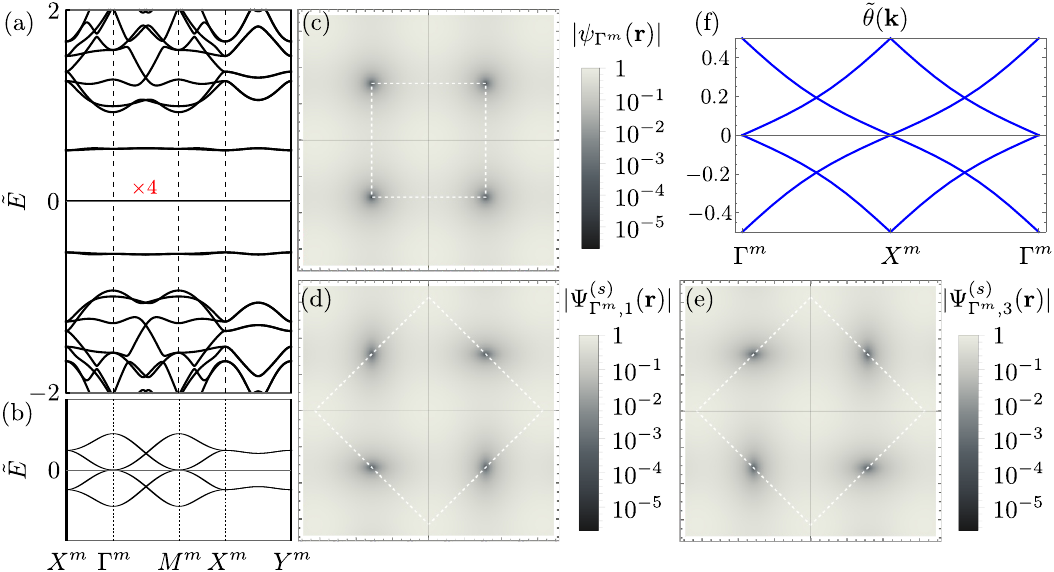}
     \caption{TBCL-4FBs. (a)Band structure of TBCL type system with 4 exact FBs at $\tilde{\alpha}=\frac{\alpha}{|\mathbf{b}^{m}|^{2}}=0.56i$.
     (b) Zoom in of the 4 FBs in (a), which shows "band sticking" along $X^{m}-Y^{m}$.
     (c) Density plot of $ |\psi_{\Gamma^{m}}(\mathbf{r})|$ (normalized by its maximum). Clearly, there is only one zero of $\psi_{\Gamma^m}(\mathbf{r})$ at an HSP, namely the corner, in the unit cell. Moir\'e unit cell of TBCL system is plotted in white dashed line.
     (d),(e) Density plot of $|\Psi_{\Gamma^{m},1}^{(s)}(\mathbf{r})|$ and $|\Psi_{\Gamma^{m},3}^{(s)}(\mathbf{r})|$ (normalized by their respective maximum). See the text below Eq.~\eqref{eq:TBCLg1} for the definition of these two functions. Moir\'e unit cell of single layer QBCP system is plotted in white dashed line. Clearly, there are two zeros of $|\Psi_{\Gamma^{m},1}^{(s)}(\mathbf{r})|$/$|\Psi_{\Gamma^{m},3}^{(s)}(\mathbf{r})|$ at HSPs, namely center of the edges, in the unit cell.
     (f) Wilson loop spectrum $\tilde{\theta}({\mathbf{k}})=\frac{\theta({\mathbf{k}})}{2\pi}$ of the flat bands in (a). 
    }
     \label{fig:4FBC2}
\end{figure}
We end the section by showing the example of a closely related Hamiltonian to that of TBCL. Consider the Hamiltonian that is obtained by replacing $U_1(\mathbf{r}) \rightarrow iU_1(\mathbf{r})$ in Eq.~\eqref{eq:HTBCL}
\begin{equation}\label{eq:TBCLg}
	\mathcal{H}(\mathbf{r}) = \begin{pmatrix} 
0 & 0 & i(-2i\partial_z)^2 & U_1^*(\mathbf{r})\\ 
0 & 0 & U_1^*(\mathbf{r}) & i(-2i\partial_z)^2\\ 
-i(-2i\overline{\partial_{z}})^2 & U_1(\mathbf{r}) & 0 & 0\\ 
U_1(\mathbf{r}) & -i(-2i\overline{\partial_{z}})^2\ & 0 & 0\ 
\end{pmatrix}  \text{, }U_1(\mathbf{r}) = 2\alpha i (\cos(\mathbf{q}_1\cdot \mathbf{r})-\cos(\mathbf{q}_2\cdot \mathbf{r})),\; \alpha \in \mathbb{R}.
\end{equation}
Remarkably, this Hamiltonian at a ``magic'' value of $\alpha$ has 4 FBs (see {\color{red}Fig.~\ref{fig:4FBC2}(a-b)}) with bands polarized to each sublattice possessing Chern number $C = \pm 2$ as can be seen the Wilson loop spectrum in {\color{red}Fig.~\ref{fig:4FBC2}(f)}. Even more curious is the the fact that the wave-function $\psi_{\Gamma^m}(\mathbf{r})$ has a single zero at corner of the unit cell ({\color{red}Fig.~\ref{fig:4FBC2}(c)}). The Chern number as well as the number of zeros are, once again, seemingly in disagreement with the construction of FB WFs discussed in the main text. This can be understood using the decomposition that was used for TBCL and symmetries of the decomposed single layer Hamiltonian. 
Once we transform the Hamiltonian to
\begin{equation}\label{eq:TBCLg1}
\mathcal{H}^{(s)}(\mathbf{r}) =U \mathcal{H}_{TBCL}(\mathbf{r}) U^\dagger = \begin{pmatrix} 
0 &  i(-2i\partial_z)^2+U_1^*(\mathbf{r}) & 0  & 0\\ 
-i(-2i\overline{\partial_{z}})^2+U_1(\mathbf{r}) & 0 & 0 & 0\\ 
0  & 0 & 0 & i(-2i\partial_z)^2-U_1^*(\mathbf{r})\\ 
0  & 0 & -i(-2i\overline{\partial_{z}})^2-U_1(\mathbf{r}) & 0\ 
\end{pmatrix}.
\end{equation}
We find a new symmetry for each diagonal block. Same as in TBCL, each of the diagonal blocks clearly corresponds to a single layer QBCP system with moir\'e lattice vectors $\tilde{\mathbf{a}}_i^m$, whereas the lattice vectors $\mathbf{a}_i^m$ of $\mathcal{H}$ are smaller (see Fig.~\ref{fig:TBCL}(a-b)) because $\mathcal{H}^{(s)}(\mathbf{r}+\mathbf{a}_i^m) = \sigma_x\otimes\mathds{1} \mathcal{H}^{(s)}(\mathbf{r})\sigma_x\otimes\mathds{1}$. Notice that each block has a glide symmetry $\mathcal{G}_{10}=\{\mathcal{M}_{10}|\frac{1}{2},\frac{1}{2}\}$, where $\mathcal{G}_{10}\mathbf{r} = \mathcal{G}_{10}(x,y) =  \mathcal{M}_{10}(x, y) + \frac{1}{2}\tilde{\mathbf{a}}_1^m+\frac{1}{2} \tilde{\mathbf{a}}_2^m = (-y,-x)+ \frac{1}{2}\tilde{\mathbf{a}}_1^m+\frac{1}{2} \tilde{\mathbf{a}}_2^m$ (we denote the mirror as  $\mathcal{M}_{10}$ since its normal is the direction $\tilde{\mathbf{a}}_1^m$ direction): $\mathcal{H}^{(s)}(\mathcal{G}_{10}\mathbf{r}) = \mathds{1}\otimes \sigma_x \mathcal{H}^{(s)}(\mathbf{r}) \mathds{1}\otimes \sigma_x$ (which is due to the fact that $U_1(\mathcal{G}_{10}\mathbf{r}) =  U_1^*(\mathbf{r})$ for $\alpha = \alpha^*$). Therefore, each block forms a QBCP system under periodic potential with $p4gm$ symmetry. However, it is known that in $p4gm$ lattice, high symmetry points have multiplicity of 2 (see, for example, BCS~\citesupp{aroyo2011crystallographys,aroyo2006bilbaoIs,aroyo2006bilbaoIIs} for a list of high symmetry points in the unit cell for any space group) in the unit cell. Therefore, if the components of $\Psi_{\Gamma^m} ^{(s)} = U  \Psi_{\Gamma^m} $, $\Psi_{\Gamma^m,1}^{(s)}(\mathbf{r})=(\psi_{\Gamma^m,1}(\mathbf{r})+\psi_{\Gamma^m,2}(\mathbf{r}))/\sqrt{2}$ and $\Psi_{\Gamma^m,3}^{(s)}(\mathbf{r})=(-\psi_{\Gamma^m,1}(\mathbf{r})+\psi_{\Gamma^m,2}(\mathbf{r}))/\sqrt{2}$, have zeros, they come in pairs in the unit cell defined by lattice vectors $\tilde{\mathbf{a}}_i^m$. Indeed, there are two zeros in each of $\Psi_{\Gamma^m,1}^{(s)}(\mathbf{r}) $ and $\Psi_{\Gamma^m,3}^{(s)}(\mathbf{r})$ at ``magic'' value of $\alpha$ as can be seen from {\color{red}Fig.~\ref{fig:4FBC2}(d-e)}. Because of this, for each diagonal block of $\mathcal{H}^{(s)}(\mathbf{r})$, once can construct 2 FB WFs polarized to each sublattice (indeed each block of $\mathcal{H}^{(s)}(\mathbf{r})$ is equivalent to the Hamiltonian we consider in Fig.~\ref{fig:p4gm} where we find 4 FBs at ``magic'' $\alpha$). This degeneracy can also be understood from ``band sticking'' effect in nonsymmorphic lattices. The two zeros allow for defining two independent holomorphic functions $f_\mathbf{k}(z;\mathbf{r}_0^{(1)})$ and $f_\mathbf{k}(z;\mathbf{r}_0^{(2)})$; this, together with the construction of Eqs.~\eqref{eq:WFsTBCL} and~\eqref{eq:WFTBCL} for each holomorphic function, give the 2 FB wFs polarized to each sublattice (considering two sublattices, total 4 FBs) for the Hamiltonian $\mathcal{H}(\mathbf{r})$ in Eq.~\eqref{eq:TBCLg}. Furthermore, we know ( from {\color{red}Sec.~S-IVB}) that the total Chern number of the multiple bands polarized on the same sublattice is $C=1$ when evaluated over Brillouin zone defined by $\tilde{\mathbf{b}}_i^m$. However, due to $\mathcal{H}^{(s)}(\mathbf{r}+\mathbf{a}_i^m) = \sigma_x\otimes\mathds{1} \mathcal{H}^{(s)}(\mathbf{r})\sigma_x\otimes\mathds{1}$, the Brillouin zone is now twice large (defined by reciprocal lattice $\mathbf{b}_i^m$). This is why the Chern number of two bands polarized on each sublattice is $C = \pm 2$.

\section{Details of the topological heavy fermion (THF) model shown in Fig.~4 of the main text}
As was discussed in the main text, due to the antiunitary particle-hole symmetry $\mathcal{P}$, any set of bands symmetric about the charge neutrality point is topological~\citesupp{TBGIIBernevigs}. As a consequence, a tight binding description of these bands is never possible. However, in the case of TBG, due to the fact that the Berry curvature distribution is peaked at $\Gamma^m$ point in the mBZ, it was shown in~\citesupp{song2022magics} that hybridization of 2 atomic limit HF bands with 4 topological conduction bands (having nontrivial winding) at $\Gamma^m$ point can describe the 2 topological FBs of TBG. This THF model keeps all the relevant symmetries of TBG and captures the correct topology of the bands. We find that similar THF description of the high number of FBs discussed in this article is possible as long as the Berry curvature distribution has a pronounced peak around some point in the mBZ. We showed an example of this in Fig.~4 of the main text for the system (space group $G = p6mm$) with 6 FBs in Fig.~4(c). The irreps of the 6 FBs at the HSMs are $\Gamma_5 \oplus 2\Gamma_6 - 2M_1 \oplus 2M_2 \oplus M_3 \oplus M_4 - K_1\oplus K_2 \oplus 2K_3$, which is not a linear combination of elementary band representations (EBR)~\citesupp{bradlyn2017topologicals}. On the other hand, the two lowest higher energy bands have representations $\Gamma_1$ and $\Gamma_2$ at $\Gamma$. Furthermore, replacement of one $\Gamma_6$ with $\Gamma_1\oplus\Gamma_2$ allows for band representation $BR = (A_1\uparrow G)_{1a}\oplus (A_2\uparrow G)_{1a} \oplus (E_1\uparrow G)_{1a} \oplus (E_2\uparrow G)_{1a}$ (we use the same notation as Topological Quantum Chemistry section of Bilbao Crystallography Server (BCS)~\citesupp{aroyo2011crystallographys,aroyo2006bilbaoIs,aroyo2006bilbaoIIs}). This and the fact that the Berry curvature distribution of the 6 FBs being peaked at $\Gamma^m$ (Fig~4(c)) suggest a THF model composed of local orbitals having band representation $BR$ (we will refer to them as $f$-electrons) and topological conduction bands (we will refer to them as $c$-bands) with representation $\Gamma_6$. The details of the construction of localized Wannier functions as well as the single particle THF Hamiltonian are shown below. This section follows~\citesupp{song2022magics}.

\subsection{Maximally localized Wannier functions for the $f$-electrons}
We start by recalling the basis states $|\mathbf{k}_0+\mathbf{k},A/B\rangle$ of the QBCP described in {\color{red}Sec. S-I}.  In systems with 3-fold rotation symmetry, QBCP can only appear at the $\Gamma$ point, hence $\mathbf{k}_0 = \mathbf{0}$ in our example. The basis in the real space then is
\begin{equation}
	|\mathbf{r}, \alpha\rangle = \sum_\mathbf{k}e^{-i\mathbf{k}\cdot\mathbf{r}}|\mathbf{k}, \alpha\rangle =  \sum_{\mathbf{k}\in \text{mBZ}}\sum_{\mathbf{b}^m}e^{-i(\mathbf{k}+\mathbf{b}^m)\cdot\mathbf{r}}|\mathbf{k},\mathbf{b}^m, \alpha\rangle, \text{, }\alpha \in\{A,B\},
\end{equation}
where we broke down the sum over the $\mathbf{k}$ into sum over $\mathbf{k}$ in the mBZ and sum over moir\'e reciprocal lattice vectors and $|\mathbf{k},\mathbf{b}^m, \alpha\rangle \equiv |\mathbf{k}+\mathbf{b}^m, \alpha\rangle$. Recall that these basis states have the following transformation properties
\begin{equation}
\begin{split}
	\mathcal{C}_{3z}\{|\mathbf{r}, A\rangle,|\mathbf{r}, B\rangle\} &= \{|\mathcal{C}_{3z}\mathbf{r}, A\rangle,|\mathcal{C}_{3z}\mathbf{r}, B\rangle\}\rho(\mathcal{C}_{3z}),\, \rho(\mathcal{C}_{3z}) = \text{Diag}\{e^{4\pi i/3},e^{2\pi i/3}\},\\
	\mathcal{C}_{2z}\{|\mathbf{r}, A\rangle,|\mathbf{r}, B\rangle\} &= \{|\mathcal{C}_{2z}\mathbf{r}, A\rangle,|\mathcal{C}_{2z}\mathbf{r}, B\rangle\}\rho(\mathcal{C}_{2z}),\, \rho(\mathcal{C}_{2z}) = \mathds{1},\\
	\mathcal{M}_{x}\{|\mathbf{r}, A\rangle,|\mathbf{r}, B\rangle\} &= \{|\mathcal{M}_{x}\mathbf{r}, A\rangle,|\mathcal{M}_{x}\mathbf{r}, B\rangle\}\rho(\mathcal{M}_{x}),\, \rho(\mathcal{M}_{x}) = \sigma_x,\\
	\mathcal{T}\{|\mathbf{r}, A\rangle,|\mathbf{r}, B\rangle\} &= \{|\mathbf{r}, A\rangle,|\mathbf{r}, B\rangle\}\rho(\mathcal{M}_{x}),\, \rho(\mathcal{T}) = \sigma_x,\\
	T_\mathbf{R}\{|\mathbf{r}, A\rangle,|\mathbf{r}, B\rangle\} &= \{|\mathbf{r}+\mathbf{R}, A\rangle,|\mathbf{r}+\mathbf{R}, B\rangle\}
\end{split}
\end{equation}
where we chose $\rho(\mathcal{C}_{2z}) = \mathds{1}$ to specify that the irrep label of the QBCP is $\Gamma_5$ in the notation of BCS (we could have just as easily chose $\rho(\mathcal{C}_{2z}) = -\mathds{1}$, then irrep would have been $\Gamma_6$). Also, here, $\mathbf{R} = n_1\mathbf{a}_1^m+n_2\mathbf{a}_2^m$ is a moir\'e lattice vector. 

We want to construct trial Wannier functions that transform as $A_1$ ($s$-type orbital), $A_2$, $E_1$ ($p$-type orbitals) and $E_2$ ($d$-type orbitals) representations of $\mathcal{C}_{6v}$  at $1a$ Wyckoff position or the center of the Wigner-Seitz unit cell. The simplest one of these is the construction of the $E_2$ reps:
\begin{equation}
\begin{split}
	|W'_{\mathbf{R},E_2,d_{x^2-y^2}+id_{2xy}}\rangle &=\frac{1}{\Omega\sqrt{2\pi\lambda_0^2}} \int d^2\mathbf{r} e^{-(\mathbf{r}-\mathbf{R})^2/2\lambda_0^2} |\mathbf{r}, A\rangle = \frac{1}{{\color{black}\Omega}\sqrt{2\pi\lambda_0^2}} \int d^2\mathbf{r} e^{-(\mathbf{r}-\mathbf{R})^2/2\lambda_0^2} \sum_{\mathbf{k}\in \text{mBZ}}\sum_{\mathbf{b}^m}e^{-i(\mathbf{k}+\mathbf{b}^m)\cdot\mathbf{r}}|\mathbf{k},\mathbf{b}^m, A\rangle\\
	&=\frac{\sqrt{2\pi\lambda_0^2}}{\Omega} \sum_{\mathbf{k}\in \text{mBZ}}\sum_{\mathbf{b}^m} e^{-i\mathbf{k}\cdot\mathbf{R}-\frac{1}{2}\lambda_0^2(\mathbf{k}+\mathbf{b}^m)^2}|\mathbf{k},\mathbf{b}^m, A\rangle,\\
	|W'_{\mathbf{R},E_2,d_{x^2-y^2}-id_{2xy}}\rangle &=\frac{1}{\Omega\sqrt{2\pi\lambda_0^2}} \int d^2\mathbf{r} e^{-(\mathbf{r}-\mathbf{R})^2/2\lambda_0^2} |\mathbf{r}, B\rangle = \frac{1}{{\color{black}\Omega}\sqrt{2\pi\lambda_0^2}} \int d^2\mathbf{r} e^{-(\mathbf{r}-\mathbf{R})^2/2\lambda_0^2} \sum_{\mathbf{k}\in \text{mBZ}}\sum_{\mathbf{b}^m}e^{-i(\mathbf{k}+\mathbf{b}^m)\cdot\mathbf{r}}|\mathbf{k},\mathbf{b}^m, B\rangle\\
	&=\frac{\sqrt{2\pi\lambda_0^2}}{\Omega} \sum_{\mathbf{k}\in \text{mBZ}}\sum_{\mathbf{b}^m} e^{-i\mathbf{k}\cdot\mathbf{R}-\frac{1}{2}\lambda_0^2(\mathbf{k}+\mathbf{b}^m)^2}|\mathbf{k},\mathbf{b}^m, B\rangle,
\end{split}
\end{equation}
because the basis states $|\mathbf{r},\alpha\rangle$ transform as $\Gamma_5$ rep, which are also $d$-type. This part is exactly the same as constructing $p_x\pm ip_y$ orbitals in TBG~\citesupp{song2022magics}. However, constructing the other 4 Wannier functions is new in this system compared to TBG. To create the $E_1$ rep or the $p$-orbitals, all we need is to have an extra negative sign under rotation, that can be done in the following way
\begin{equation}
\begin{split}
	|W'_{\mathbf{R},E_1,p_{x}+ip_{y}}\rangle &=\frac{\sqrt{2\pi\lambda_0^2}}{\Omega} \sum_{\mathbf{k}\in \text{mBZ}}\sum_{\mathbf{b}^m} ie^{i\theta_{\mathbf{k}+\mathbf{b}^m}}e^{-i\mathbf{k}\cdot\mathbf{R}-\frac{1}{2}\lambda_0^2(\mathbf{k}+\mathbf{b}^m)^2}|\mathbf{k},\mathbf{b}^m, A\rangle,\\
	|W'_{\mathbf{R},E_1,p_{x}-ip_{y}}\rangle &=\frac{\sqrt{2\pi\lambda_0^2}}{\Omega} \sum_{\mathbf{k}\in \text{mBZ}}\sum_{\mathbf{b}^m} ie^{-i\theta_{\mathbf{k}+\mathbf{b}^m}}e^{-i\mathbf{k}\cdot\mathbf{R}-\frac{1}{2}\lambda_0^2(\mathbf{k}+\mathbf{b}^m)^2}|\mathbf{k},\mathbf{b}^m, B\rangle,
\end{split}
\end{equation}
where $\theta_{\mathbf{k}+\mathbf{b}^m} = \text{arg}((k_x+b^m_x)+i(k_y+b^m_y))$. We can easily verify
\begin{subequations}
\begin{align}
	&\mathcal{C}_{3z}\{|W'_{\mathbf{R},E_1,p_{x}+ip_{y}}\rangle,|W'_{\mathbf{R},E_1,p_{x}-ip_{y}}\rangle\} \nonumber\\
	&= i\frac{\sqrt{2\pi\lambda_0^2}}{\Omega} \sum_{\mathbf{k}\in \text{mBZ}}\sum_{\mathbf{b}^m}e^{-i\mathbf{k}\cdot\mathbf{R}-\frac{1}{2}\lambda_0^2(\mathbf{k}+\mathbf{b}^m)^2}\{ e^{i\theta_{\mathbf{k}+\mathbf{b}^m}}\mathcal{C}_{3z}|\mathbf{k},\mathbf{b}^m, A\rangle, e^{-i\theta_{\mathbf{k}+\mathbf{b}^m}}\mathcal{C}_{3z}|\mathbf{k},\mathbf{b}^m, B\rangle\}\nonumber\\
	&= i\frac{\sqrt{2\pi\lambda_0^2}}{\Omega} \sum_{\mathbf{k}\in \text{mBZ}}\sum_{\mathbf{b}^m}e^{-i\mathbf{k}\cdot\mathbf{R}-\frac{1}{2}\lambda_0^2(\mathbf{k}+\mathbf{b}^m)^2}\{ e^{i\theta_{\mathbf{k}+\mathbf{b}^m}}|\mathcal{C}_{3z}\mathbf{k},\mathcal{C}_{3z}\mathbf{b}^m, A\rangle, e^{-i\theta_{\mathbf{k}+\mathbf{b}^m}}|\mathcal{C}_{3z}\mathbf{k},\mathcal{C}_{3z}\mathbf{b}^m, B\rangle\}\text{Diag}\{e^{4\pi i/3},e^{2\pi i/3}\}\nonumber\\
	&= i\frac{\sqrt{2\pi\lambda_0^2}}{\Omega} \sum_{\mathbf{k}\in \text{mBZ}}\sum_{\mathbf{b}^m}e^{-i\mathcal{C}_{3z}^{-1}\mathbf{k}\cdot\mathbf{R}-\frac{1}{2}\lambda_0^2(\mathcal{C}_{3z}^{-1}\mathbf{k}+\mathcal{C}_{3z}^{-1}\mathbf{b}^m)^2}\{ e^{i\theta_{\mathcal{C}_{3z}^{-1}(\mathbf{k}+\mathbf{b}^m)}}|\mathbf{k},\mathbf{b}^m, A\rangle, e^{-i\theta_{\mathcal{C}_{3z}^{-1}(\mathbf{k}+\mathbf{b}^m)}}|\mathbf{k},\mathbf{b}^m, B\rangle\}\text{Diag}\{e^{4\pi i/3},e^{2\pi i/3}\}\nonumber\\
	&= i\frac{\sqrt{2\pi\lambda_0^2}}{\Omega} \sum_{\mathbf{k}\in \text{mBZ}}\sum_{\mathbf{b}^m}e^{-i\mathbf{k}\cdot\mathcal{C}_{3z}\mathbf{R}-\frac{1}{2}\lambda_0^2(\mathbf{k}+\mathbf{b}^m)^2}\{ e^{i(\theta_{\mathbf{k}+\mathbf{b}^m}-2\pi/3)}|\mathbf{k},\mathbf{b}^m, A\rangle, e^{-i(\theta_{\mathbf{k}+\mathbf{b}^m}-2\pi/3)}|\mathbf{k},\mathbf{b}^m, B\rangle\}\text{Diag}\{e^{4\pi i/3},e^{2\pi i/3}\}\nonumber\\
	&= i\frac{\sqrt{2\pi\lambda_0^2}}{\Omega} \sum_{\mathbf{k}\in \text{mBZ}}\sum_{\mathbf{b}^m}e^{-i\mathbf{k}\cdot\mathcal{C}_{3z}\mathbf{R}-\frac{1}{2}\lambda_0^2(\mathbf{k}+\mathbf{b}^m)^2}\{ e^{i\theta_{\mathbf{k}+\mathbf{b}^m}}|\mathbf{k},\mathbf{b}^m, A\rangle, e^{-i\theta_{\mathbf{k}+\mathbf{b}^m}}|\mathbf{k},\mathbf{b}^m, B\rangle\}\text{Diag}\{e^{4\pi i/3}e^{-2\pi i/3},e^{2\pi i/3}e^{2\pi i/3}\}\nonumber\\
	&= \{|W'_{\mathcal{C}_{3z}\mathbf{R},E_1,p_{x}+ip_{y}}\rangle,|W'_{\mathcal{C}_{3z}\mathbf{R},E_1,p_{x}-ip_{y}}\rangle\}\text{Diag}\{e^{2\pi i/3},e^{4\pi i/3}\},\\
	\nonumber\\
	&\mathcal{C}_{2z}\{|W'_{\mathbf{R},E_1,p_{x}+ip_{y}}\rangle,|W'_{\mathbf{R},E_1,p_{x}-ip_{y}}\rangle\} \nonumber\\
	&= i\frac{\sqrt{2\pi\lambda_0^2}}{\Omega} \sum_{\mathbf{k}\in \text{mBZ}}\sum_{\mathbf{b}^m}e^{-i\mathbf{k}\cdot\mathbf{R}-\frac{1}{2}\lambda_0^2(\mathbf{k}+\mathbf{b}^m)^2}\{ e^{i\theta_{\mathbf{k}+\mathbf{b}^m}}\mathcal{C}_{2z}|\mathbf{k},\mathbf{b}^m, A\rangle, e^{-i\theta_{\mathbf{k}+\mathbf{b}^m}}\mathcal{C}_{2z}|\mathbf{k},\mathbf{b}^m, B\rangle\}\nonumber\\
	&= i\frac{\sqrt{2\pi\lambda_0^2}}{\Omega}\sum_{\mathbf{k}\in \text{mBZ}}\sum_{\mathbf{b}^m}e^{-i\mathbf{k}\cdot\mathbf{R}-\frac{1}{2}\lambda_0^2(\mathbf{k}+\mathbf{b}^m)^2}\{ e^{i\theta_{\mathbf{k}+\mathbf{b}^m}}|\mathcal{C}_{2z}\mathbf{k},\mathcal{C}_{2z}\mathbf{b}^m, A\rangle, e^{-i\theta_{\mathbf{k}+\mathbf{b}^m}}|\mathcal{C}_{2z}\mathbf{k},\mathcal{C}_{2z}\mathbf{b}^m, B\rangle\}\nonumber\\
	&= i\frac{\sqrt{2\pi\lambda_0^2}}{\Omega} \sum_{\mathbf{k}\in \text{mBZ}}\sum_{\mathbf{b}^m}e^{-i\mathcal{C}_{2z}^{-1}\mathbf{k}\cdot\mathbf{R}-\frac{1}{2}\lambda_0^2(\mathcal{C}_{2z}^{-1}\mathbf{k}+\mathcal{C}_{2z}^{-1}\mathbf{b}^m)^2}\{ e^{i\theta_{\mathcal{C}_{2z}^{-1}(\mathbf{k}+\mathbf{b}^m)}}|\mathbf{k},\mathbf{b}^m, A\rangle, e^{-i\theta_{\mathcal{C}_{2z}^{-1}(\mathbf{k}+\mathbf{b}^m)}}|\mathbf{k},\mathbf{b}^m, B\rangle\}\nonumber\\
	&= i\frac{\sqrt{2\pi\lambda_0^2}}{\Omega}\sum_{\mathbf{k}\in \text{mBZ}}\sum_{\mathbf{b}^m}e^{-i\mathbf{k}\cdot\mathcal{C}_{2z}\mathbf{R}-\frac{1}{2}\lambda_0^2(\mathbf{k}+\mathbf{b}^m)^2}\{ e^{i(\theta_{\mathbf{k}+\mathbf{b}^m}-\pi)}|\mathbf{k},\mathbf{b}^m, A\rangle, e^{-i(\theta_{\mathbf{k}+\mathbf{b}^m}-\pi)}|\mathbf{k},\mathbf{b}^m, B\rangle\}\nonumber\\
	&= i\frac{\sqrt{2\pi\lambda_0^2}}{\Omega}\sum_{\mathbf{k}\in \text{mBZ}}\sum_{\mathbf{b}^m}e^{-i\mathbf{k}\cdot\mathcal{C}_{2z}\mathbf{R}-\frac{1}{2}\lambda_0^2(\mathbf{k}+\mathbf{b}^m)^2}\{ e^{i\theta_{\mathbf{k}+\mathbf{b}^m}}|\mathbf{k},\mathbf{b}^m, A\rangle, e^{-i\theta_{\mathbf{k}+\mathbf{b}^m}}|\mathbf{k},\mathbf{b}^m, B\rangle\}\text{Diag}\{e^{-\pi i},e^{\pi i}\}\nonumber\\
	&= \{|W'_{\mathcal{C}_{2z}\mathbf{R},E_1,p_{x}+ip_{y}}\rangle,|W'_{\mathcal{C}_{2z}\mathbf{R},E_1,p_{x}-ip_{y}}\rangle\}\text{Diag}\{-1,-1\},\\
	\nonumber\\
	&\mathcal{M}_{x}\{|W'_{\mathbf{R},E_1,p_{x}+ip_{y}}\rangle,|W'_{\mathbf{R},E_1,p_{x}-ip_{y}}\rangle\} \nonumber\\
	&= i\frac{\sqrt{2\pi\lambda_0^2}}{\Omega} \sum_{\mathbf{k}\in \text{mBZ}}\sum_{\mathbf{b}^m}e^{-i\mathbf{k}\cdot\mathbf{R}-\frac{1}{2}\lambda_0^2(\mathbf{k}+\mathbf{b}^m)^2}\{ e^{i\theta_{\mathbf{k}+\mathbf{b}^m}}\mathcal{M}_{x}|\mathbf{k},\mathbf{b}^m, A\rangle, e^{-i\theta_{\mathbf{k}+\mathbf{b}^m}}\mathcal{M}_{x}|\mathbf{k},\mathbf{b}^m, B\rangle\}\nonumber\\
	&= i\frac{\sqrt{2\pi\lambda_0^2}}{\Omega} \sum_{\mathbf{k}\in \text{mBZ}}\sum_{\mathbf{b}^m}e^{-i\mathbf{k}\cdot\mathbf{R}-\frac{1}{2}\lambda_0^2(\mathbf{k}+\mathbf{b}^m)^2}\{ e^{i\theta_{\mathbf{k}+\mathbf{b}^m}}|\mathcal{M}_{x}\mathbf{k},\mathcal{M}_{x}\mathbf{b}^m, B\rangle, e^{-i\theta_{\mathbf{k}+\mathbf{b}^m}}|\mathcal{M}_{x}\mathbf{k},\mathcal{M}_{x}\mathbf{b}^m, A\rangle\}\nonumber\\
	&= i\frac{\sqrt{2\pi\lambda_0^2}}{\Omega} \sum_{\mathbf{k}\in \text{mBZ}}\sum_{\mathbf{b}^m}e^{-i\mathcal{M}_{x}^{-1}\mathbf{k}\cdot\mathbf{R}-\frac{1}{2}\lambda_0^2(\mathcal{M}_{x}\mathbf{k}+\mathcal{M}_{x}\mathbf{b}^m)^2}\{ e^{i\theta_{\mathcal{M}_{x}^{-1}(\mathbf{k}+\mathbf{b}^m)}}|\mathbf{k},\mathbf{b}^m, B\rangle, e^{-i\theta_{\mathcal{M}_{x}^{-1}(\mathbf{k}+\mathbf{b}^m)}}|\mathbf{k},\mathbf{b}^m, A\rangle\}\nonumber\\
	&= i\frac{\sqrt{2\pi\lambda_0^2}}{\Omega} \sum_{\mathbf{k}\in \text{mBZ}}\sum_{\mathbf{b}^m}e^{-i\mathbf{k}\cdot\mathcal{M}_{x}\mathbf{R}-\frac{1}{2}\lambda_0^2(\mathbf{k}+\mathbf{b}^m)^2}\{ e^{i(\pi-\theta_{\mathbf{k}+\mathbf{b}^m})}|\mathbf{k},\mathbf{b}^m, B\rangle, e^{-i(\pi-\theta_{\mathbf{k}+\mathbf{b}^m})}|\mathbf{k},\mathbf{b}^m, A\rangle\}\nonumber\\
	&= i\frac{\sqrt{2\pi\lambda_0^2}}{\Omega} \sum_{\mathbf{k}\in \text{mBZ}}\sum_{\mathbf{b}^m}e^{-i\mathbf{k}\cdot\mathcal{M}_{x}\mathbf{R}-\frac{1}{2}\lambda_0^2(\mathbf{k}+\mathbf{b}^m)^2}\{ e^{-i\theta_{\mathbf{k}+\mathbf{b}^m}}|\mathbf{k},\mathbf{b}^m, B\rangle, e^{i\theta_{\mathbf{k}+\mathbf{b}^m}}|\mathbf{k},\mathbf{b}^m, A\rangle\}\text{Diag}\{-1,-1\}\nonumber\\
	&= \{|W'_{\mathcal{M}_{x}\mathbf{R},E_1,p_{x}-ip_{y}}\rangle,|W'_{\mathcal{M}_{x}\mathbf{R},E_1,p_{x}+ip_{y}}\rangle\}\text{Diag}\{-1,-1\}\nonumber\\
	&= \{|W'_{\mathcal{M}_{x}\mathbf{R},E_1,p_{x}+ip_{y}}\rangle,|W'_{\mathcal{M}_{x}\mathbf{R},E_1,p_{x}-ip_{y}}\rangle\}(-\sigma_x),\\
	\nonumber\\
	&\mathcal{T}\{|W'_{\mathbf{R},E_1,p_{x}+ip_{y}}\rangle,|W'_{\mathbf{R},E_1,p_{x}-ip_{y}}\rangle\} \nonumber\\
	&= -i\frac{\sqrt{2\pi\lambda_0^2}}{\Omega} \sum_{\mathbf{k}\in \text{mBZ}}\sum_{\mathbf{b}^m}e^{i\mathbf{k}\cdot\mathbf{R}-\frac{1}{2}\lambda_0^2(\mathbf{k}+\mathbf{b}^m)^2}\{ e^{-i\theta_{\mathbf{k}+\mathbf{b}^m}}\mathcal{T}|\mathbf{k},\mathbf{b}^m, A\rangle, e^{i\theta_{\mathbf{k}+\mathbf{b}^m}}\mathcal{T}|\mathbf{k},\mathbf{b}^m, B\rangle\}\nonumber\\
	&= -i\frac{\sqrt{2\pi\lambda_0^2}}{\Omega}\sum_{\mathbf{k}\in \text{mBZ}}\sum_{\mathbf{b}^m}e^{i\mathbf{k}\cdot\mathbf{R}-\frac{1}{2}\lambda_0^2(\mathbf{k}+\mathbf{b}^m)^2}\{ e^{-i\theta_{\mathbf{k}+\mathbf{b}^m}}|-\mathbf{k},-\mathbf{b}^m, B\rangle, e^{i\theta_{\mathbf{k}+\mathbf{b}^m}}|-\mathbf{k},-\mathbf{b}^m, A\rangle\}\nonumber\\
	&= -i\frac{\sqrt{2\pi\lambda_0^2}}{\Omega} \sum_{\mathbf{k}\in \text{mBZ}}\sum_{\mathbf{b}^m}e^{i(-\mathbf{k})\cdot\mathbf{R}-\frac{1}{2}\lambda_0^2(-\mathbf{k}-\mathbf{b}^m)^2}\{ e^{-i\theta_{-(\mathbf{k}+\mathbf{b}^m)}}|\mathbf{k},\mathbf{b}^m, B\rangle, e^{i\theta_{-(\mathbf{k}+\mathbf{b}^m)}}|\mathbf{k},\mathbf{b}^m, A\rangle\}\nonumber\\
	&= -i\frac{\sqrt{2\pi\lambda_0^2}}{\Omega}\sum_{\mathbf{k}\in \text{mBZ}}\sum_{\mathbf{b}^m}e^{-i\mathbf{k}\cdot\mathbf{R}-\frac{1}{2}\lambda_0^2(\mathbf{k}+\mathbf{b}^m)^2}\{ e^{-i(\pi+\theta_{\mathbf{k}+\mathbf{b}^m})}|\mathbf{k},\mathbf{b}^m, B\rangle, e^{i(\pi+\theta_{\mathbf{k}+\mathbf{b}^m})}|\mathbf{k},\mathbf{b}^m, A\rangle\}\nonumber\\
	&= i\frac{\sqrt{2\pi\lambda_0^2}}{\Omega} \sum_{\mathbf{k}\in \text{mBZ}}\sum_{\mathbf{b}^m}e^{-i\mathbf{k}\cdot\mathbf{R}-\frac{1}{2}\lambda_0^2(\mathbf{k}+\mathbf{b}^m)^2}\{ e^{-i\theta_{\mathbf{k}+\mathbf{b}^m}}|\mathbf{k},\mathbf{b}^m, B\rangle, e^{i\theta_{\mathbf{k}+\mathbf{b}^m}}|\mathbf{k},\mathbf{b}^m, A\rangle\}\nonumber\\
	&= \{|W'_{\mathbf{R},E_1,p_{x}+ip_{y}}\rangle,|W'_{\mathbf{R},E_1,p_{x}-ip_{y}}\rangle\}\sigma_x.
\end{align}
\end{subequations}
Similiarly, one can check that the following trial Wannier functions transform as $A_1$ and $A_2$ rep of $C_{6v}$
\begin{equation}
\begin{split}
	|W'_{\mathbf{R},A_1}\rangle &=\frac{\sqrt{\pi\lambda_0^2}}{\Omega} \sum_{\mathbf{k}\in \text{mBZ}}\sum_{\mathbf{b}^m} e^{-i\mathbf{k}\cdot\mathbf{R}-\frac{1}{2}\lambda_0^2(\mathbf{k}+\mathbf{b}^m)^2}(e^{2i\theta_{\mathbf{k}+\mathbf{b}^m}}|\mathbf{k},\mathbf{b}^m, A\rangle+e^{-2i\theta_{\mathbf{k}+\mathbf{b}^m}}|\mathbf{k},\mathbf{b}^m, B\rangle),\\
	|W'_{\mathbf{R},A_2}\rangle &=-\frac{\sqrt{\pi\lambda_0^2}}{\Omega} \sum_{\mathbf{k}\in \text{mBZ}}\sum_{\mathbf{b}^m} e^{-i\mathbf{k}\cdot\mathbf{R}-\frac{1}{2}\lambda_0^2(\mathbf{k}+\mathbf{b}^m)^2}i(e^{2i\theta_{\mathbf{k}+\mathbf{b}^m}}|\mathbf{k},\mathbf{b}^m, A\rangle-e^{-2i\theta_{\mathbf{k}+\mathbf{b}^m}}|\mathbf{k},\mathbf{b}^m, B\rangle).
\end{split}
\end{equation}
In the basis $\{|W'_{\mathbf{R},A_1}\rangle,|W'_{\mathbf{R},A_2}\rangle,|W'_{\mathbf{R},E_1,p_x+ip_y}\rangle,|W'_{\mathbf{R},E_1,p_x-ip_y}\rangle,|W'_{\mathbf{R},E_2,d_{x^2-y^2}+d_{2xy}}\rangle,|W'_{\mathbf{R},E_2,d_{x^2-y^2}-d_{2xy}}\rangle\}$ the representations of the symmetries are
\begin{equation}
\begin{split}
	\rho^f(\mathcal{C}_{3z}) &=\begin{pmatrix}
1 & 0 & 0 & 0 & 0 & 0\\
0 & 1 & 0 & 0 & 0 & 0\\
0 & 0 & e^{2\pi i/3} & 0 & 0 & 0\\
0 & 0 & 0 & e^{4\pi i/3} & 0 & 0\\
0 & 0 & 0 & 0 & e^{4\pi i/3} & 0\\
0 & 0 & 0 & 0 & 0 & e^{2\pi i/3}\\
\end{pmatrix}, \rho^f(\mathcal{C}_{2z}) =\begin{pmatrix}
1 & 0 & 0 & 0 & 0 & 0\\
0 & 1 & 0 & 0 & 0 & 0\\
0 & 0 & -1 & 0 & 0 & 0\\
0 & 0 & 0 & -1 & 0 & 0\\
0 & 0 & 0 & 0 & 1 & 0\\
0 & 0 & 0 & 0 & 0 & 1\\
\end{pmatrix}, \rho^f(\mathcal{M}_{x}) =\begin{pmatrix}
1 & 0 & 0 & 0 & 0 & 0\\
0 & -1 & 0 & 0 & 0 & 0\\
0 & 0 & 0 & -1 & 0 & 0\\
0 & 0 & -1 & 0 & 0 & 0\\
0 & 0 & 0 & 0 & 0 & 1\\
0 & 0 & 0 & 0 & 1 & 0\\
\end{pmatrix}\\
\rho^f(\mathcal{T}) &=\begin{pmatrix}
1 & 0 & 0 & 0 & 0 & 0\\
0 & 1 & 0 & 0 & 0 & 0\\
0 & 0 & 0 & 1 & 0 & 0\\
0 & 0 & 1 & 0 & 0 & 0\\
0 & 0 & 0 & 0 & 0 & 1\\
0 & 0 & 0 & 0 & 1 & 0\\
\end{pmatrix},
 \rho^f(\mathcal{S}) =\begin{pmatrix}
0 & -i & 0 & 0 & 0 & 0\\
i & 0 & 0 & 0 & 0 & 0\\
0 & 0 & 1 & 0 & 0 & 0\\
0 & 0 & 0 & -1 & 0 & 0\\
0 & 0 & 0 & 0 & 1 & 0\\
0 & 0 & 0 & 0 & 0 & -1\\
\end{pmatrix},
\rho^f(\mathcal{P})=\rho^f(\mathcal{ST}) =\begin{pmatrix}
0 & -i & 0 & 0 & 0 & 0\\
i & 0 & 0 & 0 & 0 & 0\\
0 & 0 & 0 & 1 & 0 & 0\\
0 & 0 & -1 & 0 & 0 & 0\\
0 & 0 & 0 & 0 & 0 & 1\\
0 & 0 & 0 & 0 & -1 & 0\\
\end{pmatrix},
\end{split}
\end{equation}

Next we calculate the overlap between these trial Wannier functions and the energy eigenstates. Denoting the numerically obtained energy eigenstates as $|\psi_{\mathbf{k},n}\rangle$, we define the overlap matrix as
\begin{align}\label{eq:Overlap}
	A_{n,1}(\mathbf{k}) &\equiv \langle \psi_{\mathbf{k},n}|W'_{\mathbf{0},A_1}\rangle = \frac{\sqrt{\pi\lambda_0^2}}{\Omega}\sum_{\mathbf{k}\in \text{mBZ}}\sum_{\mathbf{b}^m} e^{-\frac{1}{2}\lambda_0^2(\mathbf{k}+\mathbf{b}^m)^2}(e^{2i\theta_{\mathbf{k}+\mathbf{b}^m}}\langle \psi_{\mathbf{k},n}|\mathbf{k},\mathbf{b}^m, A\rangle+e^{-2i\theta_{\mathbf{k}+\mathbf{b}^m}}\langle \psi_{\mathbf{k},n}|\mathbf{k},\mathbf{b}^m, B\rangle)\nonumber\\
	A_{n,2}(\mathbf{k}) &\equiv \langle \psi_{\mathbf{k},n}|W'_{\mathbf{0},A_2}\rangle = \frac{\sqrt{\pi\lambda_0^2}}{\Omega}\sum_{\mathbf{k}\in \text{mBZ}}\sum_{\mathbf{b}^m} e^{-\frac{1}{2}\lambda_0^2(\mathbf{k}+\mathbf{b}^m)^2}i(e^{2i\theta_{\mathbf{k}+\mathbf{b}^m}}\langle \psi_{\mathbf{k},n}|\mathbf{k},\mathbf{b}^m, A\rangle-e^{-2i\theta_{\mathbf{k}+\mathbf{b}^m}}\langle \psi_{\mathbf{k},n}|\mathbf{k},\mathbf{b}^m, B\rangle)\nonumber\nonumber\\
	A_{n,3}(\mathbf{k}) &\equiv \langle \psi_{\mathbf{k},n}|W'_{\mathbf{0},E_1,p_x+ip_y}\rangle =\frac{\sqrt{2\pi\lambda_0^2}}{\Omega} \sum_{\mathbf{k}\in \text{mBZ}}\sum_{\mathbf{b}^m} e^{-\frac{1}{2}\lambda_0^2(\mathbf{k}+\mathbf{b}^m)^2}ie^{i\theta_{\mathbf{k}+\mathbf{b}^m}}\langle \psi_{\mathbf{k},n}|\mathbf{k},\mathbf{b}^m, A\rangle\nonumber\\
	A_{n,4}(\mathbf{k}) &\equiv \langle \psi_{\mathbf{k},n}|W'_{\mathbf{0},E_1,p_x-ip_y}\rangle = \frac{\sqrt{2\pi\lambda_0^2}}{\Omega}\sum_{\mathbf{k}\in \text{mBZ}}\sum_{\mathbf{b}^m} e^{-\frac{1}{2}\lambda_0^2(\mathbf{k}+\mathbf{b}^m)^2}ie^{i\theta_{\mathbf{k}+\mathbf{b}^m}}\langle \psi_{\mathbf{k},n}|\mathbf{k},\mathbf{b}^m, B\rangle\nonumber\\
	A_{n,5}(\mathbf{k}) &\equiv \langle \psi_{\mathbf{k},n}|W'_{\mathbf{0},E_2,d_{x^2-y^2}+id_{2xy}}\rangle =\frac{\sqrt{2\pi\lambda_0^2}}{\Omega} \sum_{\mathbf{k}\in \text{mBZ}}\sum_{\mathbf{b}^m} e^{-\frac{1}{2}\lambda_0^2(\mathbf{k}+\mathbf{b}^m)^2}\langle \psi_{\mathbf{k},n}|\mathbf{k},\mathbf{b}^m, A\rangle\nonumber\\
	A_{n,6}(\mathbf{k}) &\equiv \langle \psi_{\mathbf{k},n}|W'_{\mathbf{0},E_2,d_{x^2-y^2}-id_{2xy}}\rangle = \frac{\sqrt{2\pi\lambda_0^2}}{\Omega}\sum_{\mathbf{k}\in \text{mBZ}}\sum_{\mathbf{b}^m} e^{-\frac{1}{2}\lambda_0^2(\mathbf{k}+\mathbf{b}^m)^2}\langle \psi_{\mathbf{k},n}|\mathbf{k},\mathbf{b}^m, B\rangle
\end{align}
\begin{figure}[h!]
     \centering
\includegraphics[scale=1]{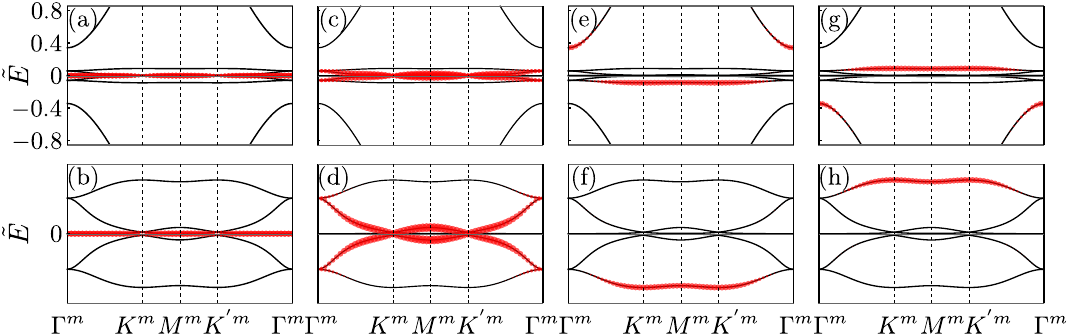}
     \caption{Overlap $|A_{n,\alpha}(\mathbf{k})|$ (Eq.~\eqref{eq:Overlap}) between the trial Wannier functions and the energy eigenstates are plotted in red circles on the energy bands. The top row shows all the 8 lowest bands, whereas the bottom row zooms into the the lowest 6 bands for better visualization. (a-b) show the overlap $|A_{n,5}(\mathbf{k})|=|A_{n,6}(\mathbf{k})|$ (the equality is due to chiral symmetry $\mathcal{S}$) of the Wannier functions corresponding to $E_2$ ($d$-type) irrep. (c-d) show the overlap $|A_{n,3}(\mathbf{k})|=|A_{n,4}(\mathbf{k})|$ (the equality is due to chiral symmetry $\mathcal{S}$) of the Wannier functions corresponding to $E_1$ ($p$-type) irrep. (e-f) show the overlap $|A_{n,1}(\mathbf{k})|$ of the Wannier function corresponding to $A_1$ ($s$-type) irrep. (g-h) show the overlap $|A_{n,2}(\mathbf{k})|$ of the Wannier function corresponding to $A_2$ irrep.} 
     \label{fig:hfoverlap}
\end{figure}
We plot these overlap  functions for $n=\pm1,\pm2,\pm3,\pm4$ (the bands are numbered away from the charge neutrality as $\pm1,\pm2,\dots$) in {\color{red}Fig.~\ref{fig:hfoverlap}}. Clearly the Wannier functions are completely supported by the middle six bands everywhere in the mBZ except at $\Gamma^m$, where the $A_1$ and $A_2$ type Wannier functions are supported by the lowest higher bands at $\Gamma^m$. We feed the overlap matrix $A_{n,\alpha}(\mathbf{k})$ ($n=\pm1,\dots,\pm4, \alpha = 1,\dots,6$) into the machinary of Wannier90~\citesupp{marzari1997maximallys,souza2001maximallys,pizzi2020wannier90s} to construct the Maximally localized Wannier functions (MLWFs). We chose $\lambda_0 = a^m/10$ for the numerical calculation, and used a $20\times 20$ grid to discretize the mBZ, and chose an energy window such that only the lowest 8 bands fall inside the window for the disentanglement and Wannierization procedure. Wannier90 returns MLWFs in the plane wave basis $|\mathbf{k},\mathbf{b}^m,\beta\rangle$ ($\beta = A,B$) as
\begin{equation}
	|W_{\mathbf{R},\alpha}\rangle =\frac{1}{\Omega\sqrt{N}} \sum_{\mathbf{k}\in \text{mBZ}}\sum_{\mathbf{b}^m} |\mathbf{k},\mathbf{b}^m,\beta\rangle e^{-i\mathbf{k}\cdot\mathbf{R}} \tilde{v}_{\mathbf{b}^m\beta,\alpha}(\mathbf{k}),
\end{equation}
where $N$ is the number of moir\'e unit cell. We can write the MLWFs in the real space basis
\begin{equation}
	w_{\beta,\alpha}(\mathbf{r}-\mathbf{R})= \langle\mathbf{r},\beta|W_{\mathbf{R},\alpha}\rangle = \frac{1}{\Omega\sqrt{N}} \sum_{\mathbf{k}\in \text{mBZ}}\sum_{\mathbf{b}^m}  e^{i(\mathbf{k}+\mathbf{b}^m)\cdot(\mathbf{r}-\mathbf{R})} \tilde{v}_{\mathbf{b}^m\beta,\alpha}(\mathbf{k}).
\end{equation}
The density plots of $\sum_\beta|w_{\beta,\alpha}|^2$ are shown in {\color{red}Fig.~4(d)} of the main text. Note that since the $|W_{\mathbf{R},A_1}\rangle$ and $|W_{\mathbf{R},A_2}\rangle$ are related by chiral symmetry, their density plot look the same. Also, since $p_x\pm ip_y$ are related by $\mathcal{T}$,  $\sum_\beta|w_{\beta,3}|^2 = \sum_\beta|w_{\beta,4}|^2$; this is why we only plotted $\sum_\beta|w_{\beta,3}|^2$. Similarly, $\sum_\beta|w_{\beta,5}|^2 = \sum_\beta|w_{\beta,6}|^2$ since $d_{x^2-y^2}\pm id_{2xy}$ are related by $\mathcal{T}$, and we plotted $\sum_\beta|w_{\beta,5}|^2$. Clearly, they are well localized within the unit cell.

A creation operators of the Wannier states can be introduced as
\begin{equation}
\begin{split}
	f_{\mathbf{R},\alpha}^\dagger &= \sum_\beta \int d^2\mathbf{r} \langle\mathbf{r},\beta|W_{\mathbf{R},\alpha}\rangle c_{\beta}^\dagger(\mathbf{r}) = \sum_\beta \int d^2\mathbf{r}\, w_{\beta,\alpha}(\mathbf{r}-\mathbf{R}) c_{\beta}^\dagger(\mathbf{r})  = \frac{1}{\sqrt{N}}\sum_{\mathbf{k},\mathbf{b}^m,\beta}  e^{-i\mathbf{k}\cdot\mathbf{R}} \tilde{v}_{\mathbf{b}^m\beta,\alpha}(\mathbf{k}) c_{\mathbf{k},\mathbf{b}^m,\beta}^\dagger,\\
	f_{\mathbf{k},\alpha}^\dagger &= \sum_{\mathbf{b}^m,\beta} \tilde{v}_{\mathbf{b}^m\beta,\alpha}(\mathbf{k}) c_{\mathbf{k},\mathbf{b}^m,\beta}^\dagger,
\end{split}
\end{equation}
where $c_{\mathbf{k},\mathbf{b}^m,\beta}^\dagger$ is the creation operator of the plane-wave state $|\mathbf{k},\mathbf{b}^m,\beta\rangle$.

\subsection{The $c$-electrons}
The Wannier functions span $\Gamma_1 \oplus\Gamma_2\oplus\Gamma_5\oplus\Gamma_6$ representations among $\Gamma_1 \oplus\Gamma_2\oplus\Gamma_5\oplus2\Gamma_6$ representation formed by the middle 8 bands. However, the middle six bands actually have representations $\Gamma_5\oplus2\Gamma_6$. Thus to get the correct band structure at the $\Gamma^m$ point, we need two additional degrees of freedom that form $\Gamma_6$ representation. Following~\citesupp{song2022magics}, they can be formally written as 
\begin{equation}
	c_{\mathbf{k},a}^\dagger = \sum_{\mathbf{b}^m,\beta} \tilde{u}_{\mathbf{b}^m\beta,a}(\mathbf{k})c_{\mathbf{k},\mathbf{b}^m,\beta}^\dagger,
\end{equation}
where $ \tilde{u}_{\mathbf{b}^m\beta,a}(\mathbf{k})$ will be determined below. Note that in the plane wave basis the single layer QBCP Hamiltonian in Eq.~\eqref{eq:QBCPS}, can be written as
\begin{equation}
\label{eq:HQBCPTable}
\begin{split}
	\hat{H} &= \sum_{\mathbf{k}\in \text{mBZ}}\sum_{\mathbf{b}^m,{\mathbf{b}^m}'}\sum_{\alpha,\beta \in \{A,B\}} [h_{\mathbf{b}^m,{\mathbf{b}^m}'}(\mathbf{k})]_{\alpha,\beta} c_{\mathbf{k},\mathbf{b}^m,\alpha}^\dagger c_{\mathbf{k},{\mathbf{b}^m}',\beta},\\
	[h_{\mathbf{b}^m,{\mathbf{b}^m}'}(\mathbf{k})] & = \begin{bmatrix} 0 & (k^*+{b^m}^*)^2 \delta_{\mathbf{b}^m,{\mathbf{b}^m}'} + \tilde{A}^*(\mathbf{b}^m -{\mathbf{b}^m}')\\ (k+b^m)^2 \delta_{\mathbf{b}^m,{\mathbf{b}^m}'} + \tilde{A}(\mathbf{b}^m -{\mathbf{b}^m}') & 0\end{bmatrix},
\end{split}
\end{equation}
where $ \tilde{A}(\mathbf{b}^m)$  are the Fourier components of the periodic field $\mathcal{D}_U(\mathbf{r};\mathbf{\alpha}) \equiv \tilde{A}(\mathbf{r})$ in Eq.~\eqref{eq:QBCPS}, $k = k_x+ik_y$ and $b^m = b^m_x+ib^m_y$. We can diagonalize the Hamiltonian 
\begin{equation}
	\sum_{{\mathbf{b}^m}',\beta} [h_{\mathbf{b}^m,{\mathbf{b}^m}'}(\mathbf{k})]_{\alpha,\beta} u_{\mathbf{k},{\mathbf{b}^m}',\beta,n} = \epsilon_n(\mathbf{k})u_{\mathbf{k},{\mathbf{b}^m},\alpha,n},
\end{equation}
where $\{u_{\mathbf{k},n}\}$ is the $n$-th eigenvector of matrix $[h(\mathbf{k})]$ with eigenvalue $\epsilon_n(\mathbf{k})$. We denote the eigenvalues as $\dots\leq\epsilon_{-2}(\mathbf{k})\leq\epsilon_{-1}(\mathbf{k})\leq\epsilon_{1}(\mathbf{k})\leq\epsilon_{2}(\mathbf{k})\leq\dots$, where $\epsilon_{-1}(\mathbf{k})$ and $\epsilon_{1}(\mathbf{k})$ are the eigenvalues with lowest magnitude. With this, we can define the projector to the lowest 8 bands
\begin{equation}
\label{eq:projection81}
	P_{{\mathbf{b}^m}\alpha,{\mathbf{b}^m}'\beta} (\mathbf{k})= \sum_{n = \pm1,\pm2,\pm3,\pm4} u_{\mathbf{k},\mathbf{b}^m,\alpha,n} u_{\mathbf{k},{\mathbf{b}^m}',\beta,n}^*.
\end{equation}
On the other hand, the projector to the 6 Wannier states are 
\begin{equation}
	Q_{{\mathbf{b}^m}\alpha,{\mathbf{b}^m}'\beta} (\mathbf{k})= \sum_{\gamma = 1,\dots,6} \tilde{v}_{\mathbf{b}^m\alpha,\gamma}(\mathbf{k}) \tilde{v}_{{\mathbf{b}^m}'\beta,\gamma}^*(\mathbf{k}).
\end{equation}
Since, by construction, the Wannier states are linear combinations of the 8 lowest bands, we have $P(\mathbf{k})Q(\mathbf{k})P(\mathbf{k}) = Q(\mathbf{k})$. Then, the projector to the remaining states is given by $P(\mathbf{k}) - Q(\mathbf{k})$. The eigenvectors of $P(\mathbf{k}) - Q(\mathbf{k})$ with eigenvalue 1 are $ \tilde{u}_{\mathbf{b}^m\beta,a}(\mathbf{k})$. We fix the gauge of these two vectors by requiring the representations of the symmetries to be the following
\begin{equation}
	\rho^c(\mathcal{C}_{3z})=\begin{pmatrix}e^{4\pi i/3} & 0\\ 0 & e^{2\pi i/3}\end{pmatrix}, \rho^c(\mathcal{C}_{2z})=\begin{pmatrix}-1 & 0\\ 0 & -1\end{pmatrix}, \rho^c(\mathcal{M}_{x})=\begin{pmatrix}0 & -1\\ -1 & 0\end{pmatrix}, \rho^c(\mathcal{T})=\begin{pmatrix}0 & 1\\ 1 & 0\end{pmatrix}, \rho^c(\mathcal{S})=\begin{pmatrix}1 & 0\\ 0 & -1\end{pmatrix}. 
\end{equation}

\subsection{The single particle Hamiltonian}
After obtaining the $f$-electron and the $c$-band basis, we are at a position to obtain the single-particle effective topological heavy fermion Hamiltonian. To this end, we define two matrices
\begin{equation}
\begin{split}
	[U(\mathbf{k})] &= [\{\tilde{v}_{1}(\mathbf{k})\},\{\tilde{v}_{2}(\mathbf{k})\}, \{\tilde{v}_{3}(\mathbf{k})\}, \{\tilde{v}_{4}(\mathbf{k})\}, \{\tilde{v}_{5}(\mathbf{k})\}, \{\tilde{v}_{6}(\mathbf{k})\}, \{\tilde{u}_{1}(\mathbf{k})\}, \{\tilde{u}_{2}(\mathbf{k})\}],\\
	[U_C(\mathbf{k})] &= [\dots, \{u_{\mathbf{k},-7}\}, \{u_{\mathbf{k},-6}\},\{u_{\mathbf{k},-5}\},\{u_{\mathbf{k},5}\},\{u_{\mathbf{k},6}\},\{u_{\mathbf{k},7}\},\dots]
\end{split}
\end{equation}
Then, we project the hamiltonian matrix $[h(\mathbf{k})]$ into the lowest 8 bands for small $|\mathbf{k}|$ (accurate to the second order in expansion w.r.t. $|\mathbf{k}|$) in the following way 
\begin{equation}
\begin{split}
[h_P(\mathbf{k})] &= [\tilde{h}(\mathbf{k})]-[C(\mathbf{k})]^\dagger[\tilde{h}_C(\mathbf{k})] [C(\mathbf{k})] = \left[\begin{array}{c|c}
H^{(f)}(\mathbf{k}) & H^{(fc)}(\mathbf{k})\\
\hline
 H^{(cf)}(\mathbf{k}) &  H^{(c)}(\mathbf{k})
\end{array}\right],\\
[\tilde{h}(\mathbf{k})] &= [U(\mathbf{0})]^\dagger [h(\mathbf{k})][U(\mathbf{0})],\\
[\tilde{h}_C(\mathbf{k})] &= [U_C(\mathbf{0})]^\dagger [h(\mathbf{k})][U_C(\mathbf{0})],\\
[C(\mathbf{k})] &= [U_C(\mathbf{0})]^\dagger [h(\mathbf{k})][U(\mathbf{0})].
\end{split}
\end{equation}
Due to the choice of the gauge for the $c$-bands in the previous subsection, $H^{(c)}(\mathbf{k})$ has the form
\begin{equation}
H^{(c)}(\mathbf{k}) = \begin{pmatrix}
0 & c_1 (k^*)^2\\ c_1k^2 & 0
\end{pmatrix},
\end{equation}
where $k= k_x+ik_y$ and $c_1$ is a real constant. For the $f$-electrons, since they are localized, we get
\begin{equation}
	H^{(f)}(\mathbf{k}) \approx \begin{pmatrix}
         m\sigma_z & \mathbf{0}_{2\times 4}\\
         \mathbf{0}_{4\times 2} & \mathbf{0}_{4\times 4}
     	\end{pmatrix},
\end{equation}
where $2|m|$ sets the gap between the $\Gamma_1$ and $\Gamma_2$ reps (see {\color{red}Fig.~4(b)}). The coupling between the $c$-bands and the $f$-electrons (keeping only the lowest order terms)
\begin{equation}
	H^{(cf)}(\mathbf{k}) \approx \begin{pmatrix}
         -ic_2(k_x-ik_y) & c_2(k_x-ik_y) & 0 & \gamma & 0 & ic_3(k_x+ik_y)\\
         -ic_2(k_x+ik_y) & -c_2(k_x+ik_y) & \gamma & 0 &  ic_3(k_x-ik_y) & 0\\
     \end{pmatrix},
\end{equation}
where $\gamma$, $c_2$ and $c_3$ are real constants. For the Flat bands, we find $c_3$ to be negligible. Furthermore, $2|\gamma|$ sets the gap between the two $\Gamma_6$ reps. Since the $f$-electrons are localized, the integral $\langle \mathbf{k},a|\hat{H}|W_{\mathbf{0},\alpha}\rangle$ ($\hat{H}$ is the QBCP Hamiltonian in Eq.~\eqref{eq:HQBCPTable}) should decay exponentially with $|\mathbf{k}|$. For simplicity, we choose the decay factor to be $e^{-\lambda^2|\mathbf{k}|^2/2}$ with $\lambda$ being the spread of the Wannier functions; this is same as what was done in the case of TBG~\citesupp{song2022magics}. Lastly, since at large $\mathbf{k}$ has huge kinetic energy, we put a cutoff $|\mathbf{k}|<\Lambda_c$ for the $c$-electron momentum. All these considerations together give the single particle THF Hamiltonian ({\color{red}Eq.~(8) of main text})
\begin{equation}
\begin{split}
     &\hat{\mathcal{H}} = \sum_{|\mathbf{k}|< \Lambda_c} H^{(c)}_{ab}(\mathbf{k}) c^\dagger_a(\mathbf{k}) c_b(\mathbf{k}) + \sum_{\mathbf{R}} H^{(f)}_{\alpha\beta} f^\dagger_\alpha(\mathbf{R}) f_\beta(\mathbf{R})+\\
     &\phantom{\hat{\mathcal{H}}}\sum_{|\mathbf{k}|< \Lambda_c,\mathbf{R}} \left(H^{(cf)}_{a\alpha}(\mathbf{k}) e^{-i\mathbf{k}\cdot\mathbf{R}-|\mathbf{k}|^2\lambda^2/2}c^\dagger_a(\mathbf{k})f_\alpha(\mathbf{R})+\text{h.c.}\right),\\
     &H^{(c)}(\mathbf{k}) \approx c_1 (k_x^2-k_y^2)\sigma_x - 2c_1 k_xk_y\sigma_y,\\
     &H^{(f)} \approx \begin{pmatrix}
         m\sigma_z & \mathbf{0}_{2\times 4}\\
         \mathbf{0}_{4\times 2} & \mathbf{0}_{4\times 4}
     \end{pmatrix},\\
     &H^{(cf)}(\mathbf{k}) \approx \begin{pmatrix}
         -ic_2(k_x-ik_y) & c_2(k_x-ik_y) & 0 & \gamma & 0 & 0\\
         -ic_2(k_x+ik_y) & -c_2(k_x+ik_y) & \gamma & 0 & 0 & 0\\
     \end{pmatrix}.
\end{split}
\end{equation}

\bibliographystylesupp{apsrev4-1}
\bibliographysupp{ref.bib}
\end{document}